\documentclass[12pt]{article}
\renewcommand{\thefootnote}{\fnsymbol{footnote}}

\newcommand{\NP}{{Nucl.\ Phys.\ }}
\newcommand{\PL}{{Phys.\ Lett.\ }}
\newcommand{\PR}{{Phys.\ Rev.\ }}

\begin{document}
\vspace*{1cm} 
\setcounter{footnote}{1}
\begin{center}
{\Large\bf Chemical equilibration of quarks and gluons 
at RHIC and LHC energies}
\\[1cm]
Duncan M.\ Elliott $^1$ and Dirk H.\ Rischke $^2$
\\ ~~ \\
{\it $^1$Department of Physics} \\
{\it University of Cape Town, Rondebosch 7701, South Africa} \\
{\it email: elliott@physci.uct.ac.za} 
\\ ~~ \\
{\it $^2$ RIKEN-BNL Research Center} \\
{\it Brookhaven National Laboratory, Upton, New York 11973, U.S.A.}\\
{\it email: rischke@bnl.gov} 
\\ ~~ \\ ~~ \\
\end{center}
\begin{abstract} 
We study chemical equilibration of quarks and gluons in central
nuclear collisions at RHIC and LHC energies. 
The initial quark and gluon densities are taken from earlier studies
as well as from recent perturbative QCD
estimates and are then evolved via rate equations coupled to
longitudinally boost-invariant fluid dynamics. We find that, for
RHIC initial conditions,
the lifetime of quark-gluon matter is too short in order for
the quark and gluon number densities to chemically equilibrate prior
to hadronization. In contrast, at LHC energies
chemical equilibration is complete before the system hadronizes.
Entropy production due to chemical 
equilibration can be as large as $30 \%$.
\end{abstract}
\renewcommand{\thefootnote}{\arabic{footnote}}
\setcounter{footnote}{0}

\section{Introduction}

Nucleus-nucleus collisions at ultrarelativistic energies probe the properties
of nuclear matter under extreme conditions \cite{QM99}. Of particular
interest is the question whether ordinary nuclear matter undergoes
a (phase) transition to quark-gluon matter, as predicted by lattice
calculations of quantum chromodynamics (QCD) \cite{lattice}. 

A nuclear collision can be viewed as a sequence of 
nucleon-nucleon collisions. At sufficiently high energies, 
multi-particle production leads to the formation of a 
region of high energy and particle number density.
With increasing beam energy, multi-particle production through
processes where the partons inside the nucleons directly
interact with each other, 
becomes more and more important. The dominant partonic
particle-production mechanism is so-called
mini-jet production \cite{eskola}. 

If the momentum transfer in these interactions is sufficiently large, 
mini-jet production is reliably computable within
perturbative QCD (pQCD). It was estimated \cite{eskola2} that
about half of the total transverse energy created 
in $Au+Au$ collisions at the Relativistic Heavy-Ion Collider
(RHIC) at BNL resides in mini-jets, while at CERN's 
Large Hadron Collider (LHC) the transverse energy created is
almost exclusively due to mini-jets.

The transverse energy can be used to 
estimate the energy densities created in the initial stage of the collision. 
If matter is in thermodynamical equilibrium,
both for RHIC and LHC energies the corresponding 
energy densities are found to be 
large enough for nuclear matter to be in the 
quark-gluon phase.
The remaining question is whether high-energy density matter formed
in ultrarelativistic nuclear collisions lives   
sufficiently long enough to actually reach thermodynamical equilibrium, so
that it can be identified with the quark-gluon phase seen in
lattice QCD calculations.

Thermodynamical equilibrium means that matter is in thermal, mechanical,
and chemical equilibrium. In general, the question of thermodynamical
equilibration can only be decided with {\em microscopic\/} transport models
\cite{micromodels}. In this paper, we study a simpler problem which
allows us to use a {\em macroscopic\/} transport model.
We assume that matter reaches thermal and mechanical equilibrium
after a proper time $\tau_0 \sim 0.1 - 0.25$ fm/c.
This is justified given the fact that the rate for elastic
collisions between quarks and gluons, which establish thermal and
mechanical equilibrium, is much larger than for inelastic
collisions which establish chemical equilibrium. 
Thermal and mechanical equilibrium will be referred to as
{\em kinetic equilibrium\/} in the following.

Under these assumptions, and given initial values for
the energy density and the quark, antiquark and gluon number densities,
we can then employ {\em ideal
fluid dynamics\/} to study the subsequent evolution of the kinetically
equilibrated quark-gluon phase, coupled to {\em rate equations\/}
which determine the chemical composition of the system away
from chemical equilibrium.
This problem has been previously studied by Bir\'{o} {\it et al.} \cite{Biro},
and by Srivastava {\it et al.} \cite{MMS,woundedmms}.
As in these previous studies, we assume that
chemical equilibration is driven mainly by the two-body reaction
$gg \leftrightarrow q{\bar q}$, and by gluon multiplication as well as
fusion, $gg \leftrightarrow ggg$. We also
terminate the evolution when the plasma reaches the
hadronization energy density, 
$\epsilon_h \equiv 1.45$ GeV/fm$^{3}$. For a completely
equilibrated plasma, this corresponds to a hadronization temperature of
$T_h \sim 0.17$ GeV. We do not study the evolution in the mixed and 
purely hadronic phases.

Our motivation to reinvestigate this subject is the following.
First of all, the question whether the quark-gluon 
phase created in ultrarelativistic heavy-ion collisions is chemically
equilibrated is highly relevant for the experiments 
commencing at the RHIC collider in the fall of this year. 
Since \cite{Biro,MMS,woundedmms} were published, newer results on
mini-jet production in nuclear collisions at RHIC and LHC energies
have been obtained \cite{eskola2}. We therefore decided to
compare the evolution computed with initial conditions obtained from
the so-called ``self-screened parton cascade'' (SSPC) model 
(used in \cite{MMS,woundedmms}) with that employing the more recent pQCD
estimates of \cite{eskola2}.

Second, the authors of \cite{Biro,MMS,woundedmms} simplified 
the rate equations using an approximate, so-called ``factorized'' 
phase-space distribution function for the quarks and gluons (see below).
In our treatment we use the full distribution function in the rate equations,
and so are in a position to assess the validity of the factorization
assumption.

Third, baryon stopping is non-negligible in nuclear collisions at
SPS energies \cite{Stop0}, and there is mounting evidence 
from theoretical studies that, even at RHIC energies, 
the midrapidity region is not completely net-baryon free 
\cite{RossiVeneziano}. Our analysis therefore also accounts for non-zero 
net-baryon number.

Finally, the numerical algorithms used here are different from
the ones employed before \cite{MMS,woundedmms}. Thus, they
constitute an independent check on the validity of the conclusions
reached previously.

This paper is organized as follows. In section 2 we discuss the
macroscopic transport equations. Starting
from a single-particle phase-space distribution in kinetic
equilibrium, we show that these equations are given
by ideal fluid dynamics, coupled to rate equations which determine
the densities of the individual particle species. We also prove that
the entropy never decreases during the time evolution of the system.
In section 3 we discuss the initial conditions for the transport
equations. Section 4 is devoted to purely longitudinal boost-invariant
expansion. We show that the difference in the equilibration process
is small when using the full phase-space distribution function instead
of the factorized distribution function of \cite{Biro,MMS,woundedmms}.
We compare results obtained with the initial conditions of the SSPC model
\cite{MMS,woundedmms} with those when using pQCD estimates \cite{eskola2}.
We also study the sensitivity of the equilibration process on the
initial time, $\tau_0$, and the strong coupling constant, $\alpha_s$.
In section 5 we include (cylindrically symmetric) transverse expansion 
as well, and present results for the SSPC model. We conclude in section
6 with a summary of our results and an outlook.
Our units are $\hbar = c = k_B =1$, and the metric tensor is
$g^{\mu \nu} = {\rm diag}\, (+,-,-,-)$.

\section{Macroscopic transport equations}

In this section, we discuss the macroscopic transport equations on which
our results are based. 
We assume that elastic collisions between quarks (antiquarks)
and gluons are sufficiently frequent to establish
kinetic equilibrium. Ideal fluid dynamics can then be invoked to
follow the evolution of the energy and momentum densities in the
system, while rate equations for the inelastic reactions
$gg \leftrightarrow ggg$ and $gg \leftrightarrow q \bar{q}$
determine the number density of quarks, antiquarks, and gluons.

\subsection{Phase-space distribution}

Once kinetic equilibrium is achieved, all particles in an (infinitesimal)
volume element at space-time point $x\equiv x^\mu = (t, {\bf x})$ 
have the same temperature, $T(x)$, 
and move with a common average 4-velocity, $u(x) \equiv u^\mu(x)$.
In that case, the single-particle phase-space
distribution for particles of species $i$ 
assumes the same form as in (local) thermodynamical
equilibrium, except that the fugacity (or the chemical potential) 
is not equal to its equilibrium value:
\begin{equation} \label{1}
f_i(x, k \cdot u) = \lambda_i(x)\, \frac{d_i}{(2 \pi)^3} \,
\frac{1}{\exp \left[ k \cdot u(x)/T(x) \right] + \lambda_i(x) 
\, \theta_i } \,\, ,
\end{equation}
where $k \equiv k^\mu = (E_i,{\bf k})$ is the 4-momentum; $E_i =
\sqrt{{\bf k}^2 + m_i^2}$ is the on-shell energy of particles of species 
$i$ with 3-momentum ${\bf k}$ and rest mass $m_i$. The fugacity
$\lambda_i(x) = \exp[\mu_i(x)/T(x)]$ controls the number density of
particle species $i$, $\mu_i$ is the chemical potential. 
$\theta_i = \pm 1$ for
fermions or bosons, respectively. $d_i$ denotes the number of
internal degrees of freedom for particles of species $i$ (like spin,
isospin, color, {\em etc.}). 

In \cite{Biro,MMS,woundedmms}, instead of (\ref{1}) a so-called
{\em factorized\/} distribution function was used, in which the
fugacity $\lambda_i$ in the denominator is approximated by 1,
\begin{equation} \label{fac}
f_i^{\rm fac}(x,k \cdot u) \equiv \lambda_i(x) \, \frac{d_i}{(2 \pi)^3} \,
\frac{1}{\exp \left[ k \cdot u(x)/T(x) \right] +  \theta_i } \,\, .
\end{equation}
We will comment on the validity of this approximation below.

\subsection{Energy-momentum conservation}

With a distribution function of the type (\ref{1}), 
the energy-momentum tensor assumes the so-called {\em
ideal-fluid\/} form
\begin{equation} \label{tmunu}
T^{\mu \nu}(x) \equiv  \sum_i \int \frac{{\rm d}^3{\bf k}}{E_i} \,
k^\mu \, k^\nu \, f_i(x,k \cdot u) = [\epsilon(x) + p(x)]\, u^\mu(x)\,
u^\nu(x) - p(x)\, g^{\mu \nu}\,\, ,
\end{equation}
where 
\begin{equation}
\epsilon(x) \equiv \sum_i \int {\rm d}^3{\bf k}
\, E_i\, f_i(x,E_i) \,\,\, , \,\,\,\,
p(x) \equiv \sum_i \int {\rm d}^3{\bf k}
\, \frac{{\bf k}^2}{3 E_i}\, f_i(x,E_i) 
\end{equation}
are the energy density and pressure in the local
rest frame of a fluid element moving with 4-velocity $u^\mu$.

While it is clear that thermal equilibrium requires all particle
species to have the same temperature, it is less obvious
why the phase-space distribution (\ref{1}) ensures mechanical
equilibrium as well. To see this, consider the tensor
decomposition $T^{\mu \nu}$ as given by (\ref{tmunu}). In the rest
frame of the fluid element, the pressure is completely isotropic,
$p\equiv T^{ii}/3 = T^{xx} = T^{yy} = T^{zz}$, which is synonymous
to mechanical equilibrium.  If
(\ref{1}) depends on more than {\em one\/} 4-vector $u$, 
additional tensors would appear on the right-hand side of (\ref{tmunu}),
such as $u^\mu_k u^\nu_l$, and the diagonal components of the
stress tensor $T^{ij}$ would no longer be identical.
Hence, the common 4-velocity $u$ in (\ref{1}) ensures
mechanical equilibrium.

In our case, matter consists of massless gluons, quarks, and
antiquarks, $i= g,\, q,\, \bar{q}$, with energy density
\begin{eqnarray} \label{e}
\epsilon & = &   \epsilon_g + \epsilon_q + \epsilon_{\bar{q}}  \,\, ,\\
\epsilon_g & = & \lambda_g\, T^4\,  \frac{d_g }{2\pi^2} 
\int_0^\infty {\rm d} z \, \frac{z^3}{e^z - \lambda_g}  \,\, , \label{e1} \\
\epsilon_q & = & \lambda_q\, T^4\, \frac{d_q}{2\pi^2} 
\int_0^\infty {\rm d} z \, \frac{z^3}{e^z + \lambda_q}  \,\, , \\
\epsilon_{\bar{q}} & = & \lambda_{\bar q}\, T^4\, \frac{d_q}{2\pi^2} 
\int_0^\infty {\rm d} z \, \frac{z^3}{e^z + \lambda_{\bar{q}}} \,\, ,
\label{e3}
\end{eqnarray}
where $d_g \equiv 2 (N_c^2-1)$ is the number of internal degrees
of freedom for gluons, $N_c = 3 $ is the number of colors, and
$d_q \equiv 2 N_c  N_f$, is the number of internal degrees of freedom
for quarks and antiquarks, with $N_f$ being the number of 
massless flavors. 
Throughout our analysis, we use $N_f = 2.5$, mimicking the effect
of the nonzero mass of the strange quark by taking $d_s = 0.5\, d_q$.

This approximation interpolates between the region of high temperature,
$T \gg m_s \simeq 150$ MeV, where $N_f \simeq 3$, and the region of 
low temperature, $m_{u,d} \ll T \ll m_s$, where $N_f \simeq 2$.
Since we follow the time evolution of the temperature from very
high $T$ down to $T \simeq T_c \sim m_s$, taking $N_f = 2.5$ is not
a particularly good approximation. We nevertheless make this choice
to be able to compare our results to earlier work \cite{Biro,MMS,woundedmms}.
Also, the chemical reaction rates, cf.\ eqs.\ (\ref{Rs}) below, which
appear on the right-hand side of the rate equations (\ref{r1}) -- (\ref{r3}), 
are more complicated in the case of massive particles \cite{pkoch}. 
The extension of our present study to nonzero strange quark mass is, however,
important and will be pursued in a subsequent publication \cite{dmedhr2}.

For massless bosons, the equilibrium value of the fugacity, 
$\lambda_i^{\rm eq}$, is always equal to 1. The reason is the
following. For bosons,
the equilibrium value of the chemical potential, $\mu_i^{\rm eq}$,
has to be smaller than the rest mass, 
$\mu_i^{\rm eq} \leq m_i$. On the other hand, bosons which carry a
conserved charge, $\mu_i^{\rm eq} \neq 0$, always come in pairs with
their own antiparticles, {\it e.g.}, $\pi^+$ and $\pi^-$, $K$ and
$\bar{K}$, {\it etc.}. Without loss of generality, we can therefore choose
$\mu_i^{\rm eq} \geq 0$ (with the chemical potential of the
associated antiparticle $\mu_{\bar{i}}^{\rm eq}= - \mu_i^{\rm eq} \leq 0$). 
Then, for massless bosons, $\mu_i^{\rm eq} \equiv 0$, such that 
$\lambda_i^{\rm eq} = 1$.

For massless fermions, there is no such constraint, although
$\lambda_q^{\rm eq} \equiv 1/ \lambda_{\bar{q}}^{\rm eq}$ in all cases
and $\lambda_q^{\rm eq} \equiv \lambda_{\bar{q}}^{\rm eq} = 1$ 
for vanishing net-baryon density.

The pressure is related to the energy density through 
\begin{equation} \label{pe3}
p \equiv \frac{\epsilon}{3}\,\, ,
\end{equation}
the well-known equation of state for an ultrarelativistic ideal gas. 
Note that, in order to derive this relation, we {\em only\/} required
the system to be in {\em kinetic\/} equilibrium, {\it i.e.},
eq.\ (\ref{pe3}) is valid even when the system is {\em not\/}
in chemical equilibrium! 

With the factorized distribution function (\ref{fac}),
the integrals in (\ref{e1}) -- (\ref{e3}) 
can be performed analytically, as they
no longer depend on the fugacities.
The energy densities simplify to
\begin{equation} \label{efac}
\epsilon^{\rm fac}_g = a_2\, \lambda_g  \, T^4 \;\;\;\; , \;\;\;\;\;
\epsilon^{\rm fac}_q = b_2\, \lambda_q\,  T^4 \;\;\;\; , \;\;\;\;\;
\epsilon^{\rm fac}_{\bar{q}} = b_2\, \lambda_{\bar{q}}\, T^4 
\,\, ,
\end{equation}
where $a_2 \equiv 8\pi^2/15$ and $b_2 \equiv 7\pi^2 N_f/40$.
The equation of state (\ref{pe3}) remains valid.

\begin{figure}
\vspace{7cm}
\caption{The ratio $\epsilon^{\rm fac}_i/\epsilon_i$ for
gluons (solid line) and quarks (dashed line) as a function of
the fugacity $\lambda_i$.}
\includegraphics{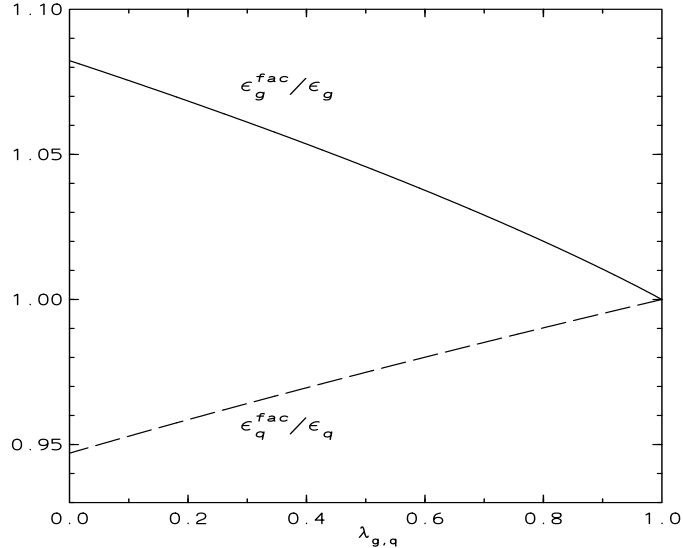}
\label{fig1}
\end{figure}

In Fig.\ \ref{1} we show the ratio $\epsilon^{\rm fac}_i/\epsilon_i$
for quarks and gluons as a function of the fugacity $\lambda_i$.
This ratio does not depend on the temperature.
For bosons, the factorized expression tends to
overestimate the correct result, for fermions, it underestimates it.
In the interval $0 \leq \lambda_i \leq 1$, 
the error is maximized at $\lambda_i = 0$, and decreases monotonically
as $\lambda_i \rightarrow 1$. At $\lambda_i = 0$, it is
$8 \% $ for bosons, and $5 \% $ for fermions.

Energy and momentum is locally conserved,
\begin{equation} \label{2}
\partial_\mu T^{\mu \nu}(x) = 0 \,\, .
\end{equation}
With (\ref{tmunu}),
these are the equations of {\em ideal fluid dynamics}, which can
be used to compute the evolution of the energy density and the fluid
4-velocity of the system. The fluid-dynamical equations (\ref{2}) are
closed by specifying the equation of state of matter under
consideration, {\it i.e.}, the pressure as a function of energy density,
$p= p(\epsilon)$. In our case, this equation of state is simple,
see eq.\ (\ref{pe3}). 

\subsection{Rate equations}

The 4-current of the number of particles of species $i$ is
\begin{equation} \label{nmu}
N_i^\mu(x) \equiv \int \frac{{\rm d}^3{\bf k}}{ E_i} \,
k^\mu \, f_i(x,k \cdot u) = n_i(x)\, u^\mu(x) \,\,  ,
\end{equation}
where
\begin{equation} 
n_i(x) \equiv \int {\rm d}^3{\bf k}\, f_i(x,E_i) 
\end{equation} 
is the number density of particle species $i$ 
in the rest frame of a fluid element.
This density is controlled by the value of the fugacity, $\lambda_i$.
In general, the fugacities for
the different particle species are independent
thermodynamic variables, and one
has to specify additional equations of motion which determine
the number densities. These are the {\em rate equations}.
To close the coupled set of fluid-dynamical equations and rate equations,
one has to be specify an
equation of state which, in general, also depends on the number densities of
the various particle species, $p= p(\epsilon,n_1,n_2, \ldots)$.

The reason why there is actually no such dependence for the system considered
here, eq.\ (\ref{pe3}), is that the quarks and gluons are
considered to be massless. This has the important consequence that,
in our case,
chemical non-equilibrium does {\em not\/} affect the evolution of the
energy and momentum densities of the fluid!
This certainly changes in the case of a nonzero strange quark mass, 
although, for large temperatures $T \gg m_s$, the dependence of the 
pressure $p$ on $n_s$ is relatively weak.

The number densities of gluons, quarks, and antiquarks
in the rest frame of a fluid element are
\begin{eqnarray} \label{ng}
n_g & = & \lambda_g \, T^3\,  \frac{d_g}{2\pi^2} \int_0^{\infty} {\rm d} z\,
\frac{z^2}{e^z - \lambda_g} \,\, , \\
n_q & = & \lambda_q \, T^3 \, \frac{d_q}{2\pi^2} \int_0^{\infty} 
{\rm d} z\, \frac{z^2}{e^z + \lambda_q}     \,\, , \label{nq} \\
n_{\bar q} & = & \lambda_{\bar{q}}\, T^3\, 
\frac{d_q}{2\pi^2} \int_0^{\infty} 
{\rm d} z\, \frac{z^2}{e^z + \lambda_{\bar q}}   \,\, . \label{nbarq}
\end{eqnarray}
With the factorized distribution function (\ref{fac}) these expressions
simplify to
\begin{equation} \label{nfac}
n_g^{\rm fac} = a_1 \, \lambda_g \, T^3 \,\,\, , \,\,\,\, 
n_q^{\rm fac} = b_1\, \lambda_q \,T^3 \,\,\, , \,\,\,\,
n_{\bar{q}}^{\rm fac} = b_1 \, \lambda_{\bar{q}}\, T^3 \,\, ,
\end{equation}
where $a_1 \equiv 16\, \zeta(3)/\pi^2$ and $b_1 \equiv 9\, \zeta(3)\, N_f/
(2\pi^2)$.

\begin{figure}
\vspace{7cm}
\caption{The ratio $n^{\rm fac}_i/n_i$ for
gluons (solid line) and quarks (dashed line) as a function of
the fugacity $\lambda_i$.}
\includegraphics{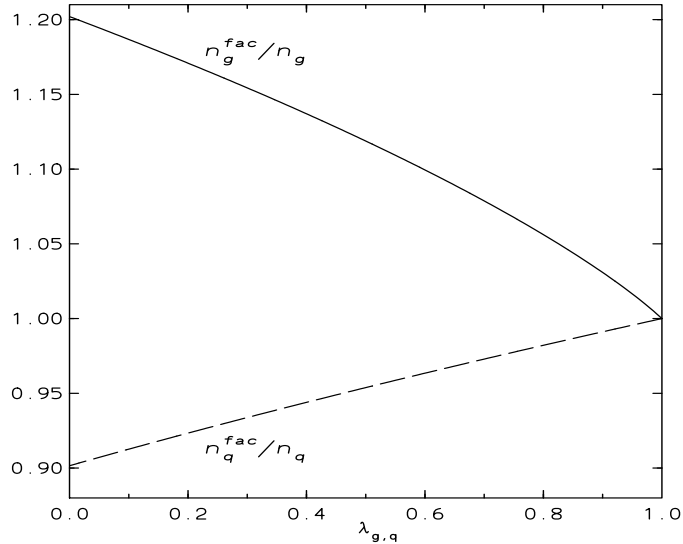}
\label{fig2}
\end{figure}

In Fig.\ \ref{fig2}, we show the ratio $n_i^{\rm fac}/n_i$ as a function of
fugacity. As for the ratio of energy densities, Fig.\ \ref{fig1}, 
this ratio does not depend on the temperature, and
the factorized expression overestimates the correct result for bosons,
while it underestimates it for fermions. Again, the error decreases as
a function of $\lambda_i$, but it is about twice as large as for
the energy densities. The maximum error (at $\lambda_i =0$) is about
$20 \%$ for gluons and $10 \%$ for quarks.

In the absence of chemical equilibrium, the number of particles
of species $i$ is determined from the {\em rate equation}
\begin{equation} \label{5}
\partial_\mu N_i^\mu(x) = R_i(x) \,\, .
\end{equation}
In chemical equilibrium, $R_i$ vanishes.
In our case of a quark-gluon system, we 
assume that the $R_i$ are determined
by the reactions $gg \leftrightarrow ggg$ and
$gg \leftrightarrow q{\bar q}$.
The rate equations (\ref{5}) can then be written in the form \cite{Biro} 
\begin{eqnarray} \label{r1}
{\partial }_{\mu}( n_g u^{\mu} ) & = &  {\rm R}_3 \, n_g 
\left( 1 - \frac{n_g}{\tilde{n}_g} \right) - 
2 \, {\rm R}_2\, n_g \left( 1 - \frac{n_q \,n_{\bar{q}} \, \tilde{n}_g^2}{ 
\tilde{n}_q \, \tilde{n}_{\bar{q}}\, n_g^2} \right)  \,\,, \\
{\partial }_{\mu}( n_q u^{\mu} )& = & 
{\rm R}_2 \, n_g 
\left( 1 - \frac{n_q \, n_{\bar{q}} \, \tilde{n}_g^2 }{\tilde{n}_q\,
\tilde{n}_{\bar{q}} \, n_g^2} \right)   \,\, , \\
{\partial }_{\mu}( n_{\bar q} u^{\mu} )& = &
{\rm R}_2\, n_g \left(1 - \frac{n_q\, n_{\bar{q}}\, \tilde{n}_g^2 }
{\tilde{n}_q\, \tilde{n}_{\bar{q}}\, n_g^2} \right)  \,\,, \label{r3}
\end{eqnarray}
where the $\tilde{n}_i$ are the number densities computed at the
same temperature $T$, but for $\lambda_i = 1$.
The terms ${\rm R}_2$ and ${\rm R}_3$ are given by \cite{Biro}
\begin{equation} \label{Rs}
{\rm R}_2 \simeq 0.24\, N_f\,\alpha_s^2\, \lambda_g \,
T \, \ln \left( 1.65/\alpha_s\, \lambda_g \right) \,\,, 
\,\,\,\, {\rm R}_3 \simeq 2.1 \,\alpha_s^2 \, T \, \sqrt{2\lambda_g - 
\lambda_g^2} \ \ ,
\end{equation}
with $\alpha_s$ being the strong coupling constant. 

We take $\alpha_s = 0.3$ throughout this analysis, 
unless otherwise stated. Note that, while the densities on the
left-hand side of eqs.\ (\ref{r1}) -- (\ref{r3}) in general contain the
full phase-space distribution function (\ref{1}), the right-hand sides
of these equations and the terms
(\ref{Rs}) are derived assuming the factorized distribution (\ref{fac})
\cite{Biro}. This is in principle inconsistent. At least
close to equilibrium, however, the right-hand sides of the rate 
equations (\ref{r1}) -- (\ref{r3}) are small, and we are allowed to 
neglect this inconsistency.
On the other hand, far away from equilibrium we expect
other contributions to the chemical reaction rates to be more
important than the difference between factorized and full
distribution functions. Such contributions are, for
instance, multi-gluon processes $gg \rightarrow n\, g$ \cite{shuryakxiong}.
In this work, we only include the lowest-order rates (\ref{Rs}).

In chemical equilibrium, the right-hand sides of the rate equations
(\ref{r1}) -- (\ref{r3}) vanish by definition, and we obtain
\begin{equation}
\partial_\mu ( n_i^{\rm eq} u^\mu ) = 0 \,\,, \,\,\,\, i = g, q, \bar{q}
\,\, . \label{ra}
\end{equation}
We can now derive a relation for the time evolution
of the ratio of the parton number density $n_i$ to its corresponding
equilibrium value $n_i^{\rm eq}$.
By writing $n_i = n_i^{\rm eq} \, n_i/n_i^{\rm eq}$ on the left-hand
sides of the rate equations (\ref{r1}) -- (\ref{r3}), we
derive with eq.\ (\ref{ra}) the condition
\begin{equation} \label{comoving}
u^\mu \partial_\mu \left( \frac{n_i}{n_i^{\rm eq}} \right) = 
\frac{R_i}{n_i^{\rm eq}} \,\, .
\end{equation}
This means that, for $R_i/n_i^{\rm eq} > 0$, the comoving time derivative
of the ratio of density to equilibrium density is positive,
and thus this ratio is bound to grow with increasing proper time in each fluid
element. Vice versa, this ratio decreases with time if 
$R_i /n_i^{\rm eq} < 0$.
In other words, in the rest frame of a fluid element, $n_i/n_i^{\rm eq}$
cannot decrease (increase) if $R_i/n_i^{\rm eq} > 0 \, (< 0)$.

\subsection{Entropy}

Local thermodynamical equilibrium implies the conservation of entropy.
If the system is not chemically equilibrated, the entropy is bound
to grow, until chemical equilibrium is reached \cite{MMS,EntropyProd}. 
To see this, we make the following observation: 
the form of the distribution function (\ref{1}) or (\ref{fac})
is exactly that of a distribution function in 
local thermodynamical equilibrium, except that
the fugacities, $\lambda_i$, assume 
values which are {\em not\/} the same as
in chemical equilibrium, $\lambda_i^{\rm eq}$. 
Note that in this case there are {\em no\/} 
further dissipative terms \cite{thankstoJR} in the tensor
decomposition of (\ref{tmunu}) and (\ref{nmu}) \cite{DRproc}.
Nevertheless, the system is {\em not\/} in thermodynamical equilibrium, and
the actual values of $\epsilon, \, p$, and $n_i$ will differ
from the equilibrium values, 
$\epsilon^{\rm eq},\, p^{\rm eq}$, and $n_i^{\rm eq}$.

Since the distribution function has the same form as
in thermodynamical equilibrium, we can 
imagine each fluid element to actually {\em be\/} in chemical equilibrium
with a (local) particle bath which determines the value of the
fugacities instead of the rate equations (\ref{5}).
In other words, we can imagine local changes of particle number
to be induced by a change of the parameters of the particle bath
instead by chemical reactions within the fluid element itself.
As long as these changes happen faster than those of the
macroscopic fluid variables, one can view the fluid element to
remain in (local) thermodynamic equilibrium, and
apply thermodynamical relationships.

The assumption that microscopic reaction rates are much larger
than the rate of change of macroscopic fluid variables
is certainly an over-idealization in view of
the characteristic time scales in relativistic heavy-ion collisions.
So is, in fact, the ideal fluid limit which assumes
complete (local) thermodynamical equilibrium. Nevertheless, our
point of view is that ideal fluid dynamics still offers valuable information 
about the collective behavior of the system. Our approach
is the simplest possible extension of ideal fluid dynamics.
A better approximation would be to solve the
equations for dissipative relativistic fluid dynamics in the
presence of chemical reactions \cite{raju}. In general, chemical
non-equilibrium gives rise to additional diffusion terms
which do not appear in our treatment. So far, the
complexity of these equations has discouraged attempts to apply them
to the description of nuclear collision dynamics.

Accepting these {\it caveats}, we now proceed by contracting (\ref{2}) 
with $u_\nu$,
\begin{equation}
0= u_\nu \partial_\mu T^{\mu \nu} = u \cdot \partial \epsilon +
(\epsilon + p) \partial \cdot u \,\, ,
\end{equation}
and use the first law of thermodynamics,
\begin{equation}
{\rm d} \epsilon = T\, {\rm d} s + \sum_i \mu_i\, {\rm d}n_i\,\, ,
\end{equation}
($s$ is the entropy density),
as well as the fundamental relation of thermodynamics,
\begin{equation}
\epsilon + p = T \, s + \sum_i \mu_i\, n_i\,\,,
\end{equation}
to derive
\begin{equation}
0 = T \, (u \cdot \partial s  + s \, \partial \cdot u)
+ \sum_i \mu_i (u \cdot \partial n_i + n_i \, \partial \cdot u)\,\, .
\end{equation}
Equations (\ref{5}) and (\ref{nmu}), together with the definition of 
$\lambda_i$, can be used to conclude that
\begin{equation} \label{noneqS}
\partial_\mu S^\mu = - \sum_i \ln \lambda_i \, R_i\,\, ,
\end{equation}
where   $S^\mu = s\, u^\mu$.
In chemical equilibrium, $R_i$ vanishes, and entropy is conserved.

Consider now the case of gluons, or quarks when the net-baryon number is
zero. In this case, the equilibrium value of the
fugacity $\lambda_i^{\rm eq} = 1$, {\it i.e.},
$\mu_i^{\rm eq} = 0$. 
Now assume that the {\em actual\/} 
particle number density $n_i$ is {\em smaller\/} than
the equilibrium value, corresponding to $\lambda_i < \lambda_i^{\rm eq} =1$. 
The rate equations drive $n_i$ towards equilibrium, 
{\it i.e.}, the right-hand side has to be positive, $R_i >0$, in order
to produce particles of species $i$. 
Since $\ln \lambda_i <0$, the change in entropy is positive.
On the other hand, if $n_i$ is larger than in equilibrium, $\lambda_i > 
\lambda_i^{\rm eq} =1$,
the rate equations reduce the number of particles of species
$i$, {\it i.e.}, 
the right-hand side is negative, $R_i<0$. Again, entropy increases,
since $\ln \lambda_i > 0$.

Now suppose that $\lambda_i^{\rm eq} >1$, for instance for quarks
when the net-baryon number is positive. Naively, one would think
that the previous argument fails in this case, and one
might worry whether this could lead to a situation where
entropy actually {\em decreases\/} towards its equilibrium value,
which contradicts the second law of thermodyamics. This is, however,
not the case. In a relativistic system, one can create quarks 
only in pairs with antiquarks. In order to conserve the net-baryon
number, $\partial_\mu (N_q^\mu - N_{\bar{q}}^\mu) \equiv 0$,
the right-hand side of the rate equation for the quark
number density, $R_q$, {\em must\/} be equal to the right-hand
side of the rate equation for the antiquark number density, $R_{\bar{q}}$,
see eqs.\ (\ref{r2a}) -- (\ref{r3a}) below.

In eq.\ (\ref{noneqS}), these two terms in the sum over $i$ can then
be combined to $- (\mu_q + \mu_{\bar{q}}) R_q/T$.
In equilibrium, $\mu_q^{\rm eq} = - \mu_{\bar{q}}^{\rm eq}$, such
that $\mu_q + \mu_{\bar{q}} = \delta \mu_q + \delta \mu_{\bar{q}}$,
where $\delta \mu_i \equiv \mu_i - \mu_i^{\rm eq}$. If the
quark (and consequently, the antiquark) number density is {\em smaller\/}
than its equilibrium value, $\delta \mu_q+ \delta \mu_{\bar{q}}< 0$, the
right-hand side of the rate equation is positive, $R_q > 0$. Consequently,
the sum of the quark and antiquark term on the right-hand side of 
(\ref{noneqS}) is positive, leading again to an increase in entropy. 
The argument is similar in the case when the quark number density is 
larger than in equilibrium, or when $\lambda_i^{\rm eq} < 1$. In all cases, the
entropy increases.

\section{Initial conditions}

In order to solve the macroscopic transport equations discussed in
the last section, one has to specify the initial conditions.
In this paper, we use results from two different approaches, 
both for RHIC and LHC energies.

The first approach is the SSPC model employed in \cite{MMS,woundedmms}.
We decided to use these initial conditions for two reasons. First,
the values for initial energy and parton densities obtained in this
approach constitute an upper bound
of what is expected to be created in ultrarelativistic nuclear collisions
at RHIC energies. Second, it allows us to directly compare the
results of \cite{MMS,woundedmms} with ours and point out
possible discrepancies.

The second approach is that of \cite{eskola2}. This approach is
purely based on pQCD, and thus does not contain the contribution from
the soft background, which constitutes about half of the total 
transverse energy at RHIC energies. 
In this sense, at RHIC this second approach
constitutes a lower bound for the initial energy and parton number densities.

Table \ref{table1} contains the initial values for energy and parton number
density for each approach and collision energy.
The SSPC values are taken from \cite{MMS}, 
while the values shown for the pQCD approach
\cite{eskola2} are computed 
from the ${\rm d}N/{\rm d}\eta$ and ${\rm d}E_T/{\rm d}\eta$ 
values near midrapidity ($-0.5\leq \eta \leq 0.5$) 
using a volume element $\Delta V = \pi R^2 \tau_0 \Delta \eta$ \cite{Bjorken},
such that
\begin{equation} \label{dV}
\frac{{\rm d}E_T}{{\rm d}\eta} \equiv \epsilon_0 \,\frac{{\rm d}V}{{\rm d}\eta}
= \epsilon_0 \tau_0 \pi R^2 \;\;\;\; , \;\;\;\;\;\;
\frac{{\rm d}N_i}{{\rm d}\eta} \equiv n_i^0 \, \frac{{\rm d}V}{{\rm d}\eta}
= n_i^0 \tau_0 \pi R^2 \;\;\;\; , \;\;\;\;\; i = g,\, q,\,\bar{q} \,\, ,
\end{equation}
where $R=1.12\, A^{1/3}$ fm. 
From the four values $\epsilon_0,\, n_g^0,\, n_q^0$, and $n_{\bar{q}}^0$,
using eqs.\ (\ref{e}) and (\ref{ng}) -- (\ref{nbarq})
one can unambiguously extract $T_0, \, \lambda_g^0,\, \lambda_q^0$, and
$\lambda_{\bar{q}}^0$, and, from the latter, the values
$n_i^0/\tilde{n}_i^0$ given in the last three columns of Table \ref{table1}.

The initial time $\tau_0$ in the SSPC and pQCD approaches is chosen
as $\tau_0 \equiv 1/p_0$, where $p_0$ is the infrared momentum cutoff 
required to regularize the mini-jet production cross sections.
The values for $\tau_0$ given in Table \ref{table1} arise from the cutoffs
$p_0 \sim 0.8$ GeV for the SSPC approach and $p_0 \sim 2$ GeV for 
the pQCD approach. The transverse momentum $p_T \geq p_0$ of a 
produced mini-jet
determines the time scale for the respective parton to come on mass-shell.
Only {\em after\/} the partons are on their respective mass-shell, 
elastic scattering processes between them will drive the
system towards kinetic equilibrium. How the system approaches
kinetic equilibrium has to be studied in the framework of
kinetic theory; our approach does not make any statements about
this pre-equilibrium stage of a heavy-ion collision. In the following,
we take $\tau_0= 1/p_0$, which is certainly the
{\em earliest\/} possible time for the system to reach kinetic equilibrium.
In the next section, we shall also use larger (and thus more conservative)
values for $\tau_0$, in order to test the sensitivity of our results
on this important parameter.

For RHIC initial conditions, the energy density in pQCD
is much smaller than for the SSPC model, despite the smaller initial
kinetic equilibrium time $\tau_0$. The reason is the absence of the 
soft background contribution in the former approach. For LHC, pQCD gives larger
values for the initial energy density than the SSPC model. If
scaled to the same initial proper time, however, the initial energy densities
are approximately equal. This reflects the fact that the soft
background contribution is rather small at LHC energies.
Nevertheless, despite the similarity of the initial energy densities, in pQCD
the gluons are close to chemical equilibrium at
LHC energies, while this is not the case for the SSPC approach.
Note that there is a small initial net-baryon number density
$(n_q^0 - n_{\bar{q}}^0)/3$ in the pQCD approach, as opposed to 
the SSPC model where vanishing net-baryon number is assumed.

\begin{table}
\begin{center}
\caption {Initial conditions for the fluid-dynamical expansion phase in
central $Au+Au$ collisions at BNL RHIC and $Pb+Pb$ collisions at
CERN LHC energies from the SSPC model and pQCD.}
\bigskip\bigskip
\begin{tabular}{|c|c|c|c|c|c|c|c|}  \hline
\multicolumn{1}{|c|}{ } &
\multicolumn{1}{c|}{ } &
\multicolumn{1}{c|}{ } &
\multicolumn{1}{c|}{ } &
\multicolumn{1}{c|}{ } &
\multicolumn{1}{c|}{ } &
\multicolumn{1}{c|}{ } &
\multicolumn{1}{c|}{ } \\
\multicolumn{1}{|c|}{ Approach } &
\multicolumn{1}{c|}{  Energy } &
\multicolumn{1}{c|}{  ${\tau}_0$ } &
\multicolumn{1}{c|}{  ${\epsilon}_0$ } &
\multicolumn{1}{c|}{  $T_0$ } &
\multicolumn{1}{c|}{  $n_g^0/\tilde{n}_g^0$ } &
\multicolumn{1}{c|}{  $n_q^0/\tilde{n}_q^0$ } &
\multicolumn{1}{c|}{  $n_{\bar q}^0/\tilde{n}_{\bar q}^0$ } \\
\multicolumn{1}{|c|}{  } &
\multicolumn{1}{c|}{   } &
\multicolumn{1}{c|}{  (fm/c) } &
\multicolumn{1}{c|}{  (GeV/fm$^{3}$) } &
\multicolumn{1}{c|}{  (GeV) } &
\multicolumn{1}{c|}{   } &
\multicolumn{1}{c|}{   } &
\multicolumn{1}{c|}{   } \\
\multicolumn{1}{|c|}{ } &
\multicolumn{1}{c|}{   } &
\multicolumn{1}{c|}{ } &
\multicolumn{1}{c|}{ } &
\multicolumn{1}{c|}{ } &
\multicolumn{1}{c|}{ } &
\multicolumn{1}{c|}{ } &
\multicolumn{1}{c|}{ } \\ \hline
 &  &  &    &    &  &  &  \\           
SSPC & RHIC   & $0.25$ & $61.4$ & $0.66$ & $0.34$ & $0.064$ & $0.064$ \\
SSPC & LHC    & $0.25$ & $425$  & $1.01$ & $0.43$ & $0.082$ & $0.082$ \\
 &   & &    &    &  &  &  \\ \hline
 &   & &    &    &  &  &  \\ 
pQCD & RHIC   & $0.10$ & $23.9$ & $0.889$ & $0.042$ & $0.0077$ & $0.0048$ \\
pQCD & LHC    & $0.10$ & $1057$ & $1.09$ & $0.99$ & $0.064$ & $0.061$ \\
 &   &   &  &    &  &  &  \\ \hline
\end{tabular}
\label{table1}
\end{center}
\end{table}

\section{Boost-invariant longitudinal expansion}

In this section, we study purely longitudinal expansion 
(in the $z$-direction) with boost-invariant initial conditions 
\cite{Bjorken}. In this case,
physics is constant along the space-time hyperbolae
$\tau = \sqrt{t^2-z^2} =const.$.
Our aim is to establish how the parton equilibration process is affected
when the full phase-space distribution (\ref{1}) is used instead
of the factorized distribution (\ref{fac}). We will furthermore
compare the time evolution of the parton densities and the
entropy in the SSPC and pQCD scenarios, and
investigate the dependence of equilibration on the initial proper time
$\tau_0$ and the strong coupling constant $\alpha_s$.

\subsection{Factorized vs.\ full phase-space distribution}

For boost-invariant longitudinal expansion, 
energy-momentum conservation (\ref{2}) reads 
\begin{equation} \label{enmomcons}
\dot{\epsilon} +  \frac{\epsilon+p}{\tau}  =  0 \,\, ,
\end{equation}
with $\dot{\epsilon} \equiv {\rm d}\epsilon/{\rm d}\tau$, 
$\tau$ being the proper time. The rate equations
(\ref{r1}) -- (\ref{r3}) assume the form
\begin{eqnarray} \label{r1a}
\dot{n_g} + \frac{n_g}{\tau}   & = &  {\rm R}_3 \, n_g 
\left( 1 - \frac{n_g}{\tilde{n}_g} \right) - 
2 \, {\rm R}_2\, n_g \left( 1 - \frac{n_q \,n_{\bar{q}} \, \tilde{n}_g^2}{ 
\tilde{n}_q \, \tilde{n}_{\bar{q}}\, n_g^2} \right)  \,\,, \\
\dot{n_q} + \frac{n_q}{\tau} & = & {\rm R}_2 \, n_g 
\left( 1 - \frac{n_q \, n_{\bar{q}} \, \tilde{n}_g^2 }{\tilde{n}_q\,
\tilde{n}_{\bar{q}} \, n_g^2} \right)   \,\, , \label{r2a} \\
\dot{n_{\bar{q}}} + \frac{n_{\bar{q}}}{\tau} & = &
{\rm R}_2\, n_g \left(1 - \frac{n_q\, n_{\bar{q}}\, \tilde{n}_g^2 }
{\tilde{n}_q\, \tilde{n}_{\bar{q}}\, n_g^2} \right)  \,\,. \label{r3a}
\end{eqnarray}
Equations (\ref{enmomcons}) -- (\ref{r3a}) are
four ordinary differential equations in the variable $\tau$,
containing four unknowns, 
$T,\, \lambda_g,\, \lambda_q$, and $\lambda_{\bar{q}}$. 
They are numerically solved with a standard Runge--Kutta
integration routine.
At this point note that, while the evolution of the fluid energy  
and momentum densities are completely decoupled from the evolution of the 
parton densities on account of the equation of state (\ref{pe3}), energy
used up in parton production will be reflected by the temperature 
falling faster than in complete (local) thermodynamical
equilibrium. In the latter case, $T \sim {\tau}^{-1/3}$ 
\cite{Bjorken}.

For the factorized distribution, the evolution equations 
(\ref{enmomcons}) -- (\ref{r3a}) can be further simplified \cite{Biro}:
\begin{eqnarray}
\frac{\dot{\lambda}_g + b(\dot{\lambda}_q + \dot{\lambda}_{\bar{q}})}{
\lambda_g + b(\lambda_q  + \lambda_{\bar{q}} ) }  +
4\, \frac{\dot{T}}{T} + \frac{4}{3 \tau} & = & 0\,\, ,\\
\frac{\dot{\lambda}_g}{\lambda_g} +
\frac{3\, \dot{T}}{T} + \frac{1}{\tau} & = & 
{\rm R}_3 \, (1-\lambda_g) - 2\, {\rm R}_2 \left( 1 - 
\frac{\lambda_q\, \lambda_{\bar{q}}}{\lambda_g^2}\right) \,\,, \\
\frac{\dot{\lambda}_q}{\lambda_q}   +
\frac{3\, \dot{T}}{T} + \frac{1}{\tau} & = & R_2 \, \frac{a_1}{b_1} 
\left( \frac{\lambda_g}{\lambda_q} - 
\frac{\lambda_{\bar{q}}}{\lambda_g}\right) \,\, , \label{r2b} \\
\frac{\dot{\lambda}_{\bar{q}}}{\lambda_{\bar{q}}}     +
\frac{3\, \dot{T}}{T} + \frac{1}{\tau} & = &R_2 \, \frac{a_1}{b_1}
\left( \frac{\lambda_g}{\lambda_{\bar{q}}} - 
\frac{\lambda_q}{\lambda_g}\right) \,\, , \label{r3b}
\end{eqnarray}
with $b=b_2/a_2=21N_f/64$. 

\begin{figure}
\vspace*{10cm}
\caption{Parton densities for the SSPC RHIC (left) and 
LHC scenarios (right), normalized to the corresponding
values for $\lambda_i = 1$ (denoted as $\tilde{n}_i$ in the text). 
The solid line shows results for the full phase-space distribution function, 
while the dashed line those for the factorized 
distribution function $f^{\rm fac}$.}
\includegraphics{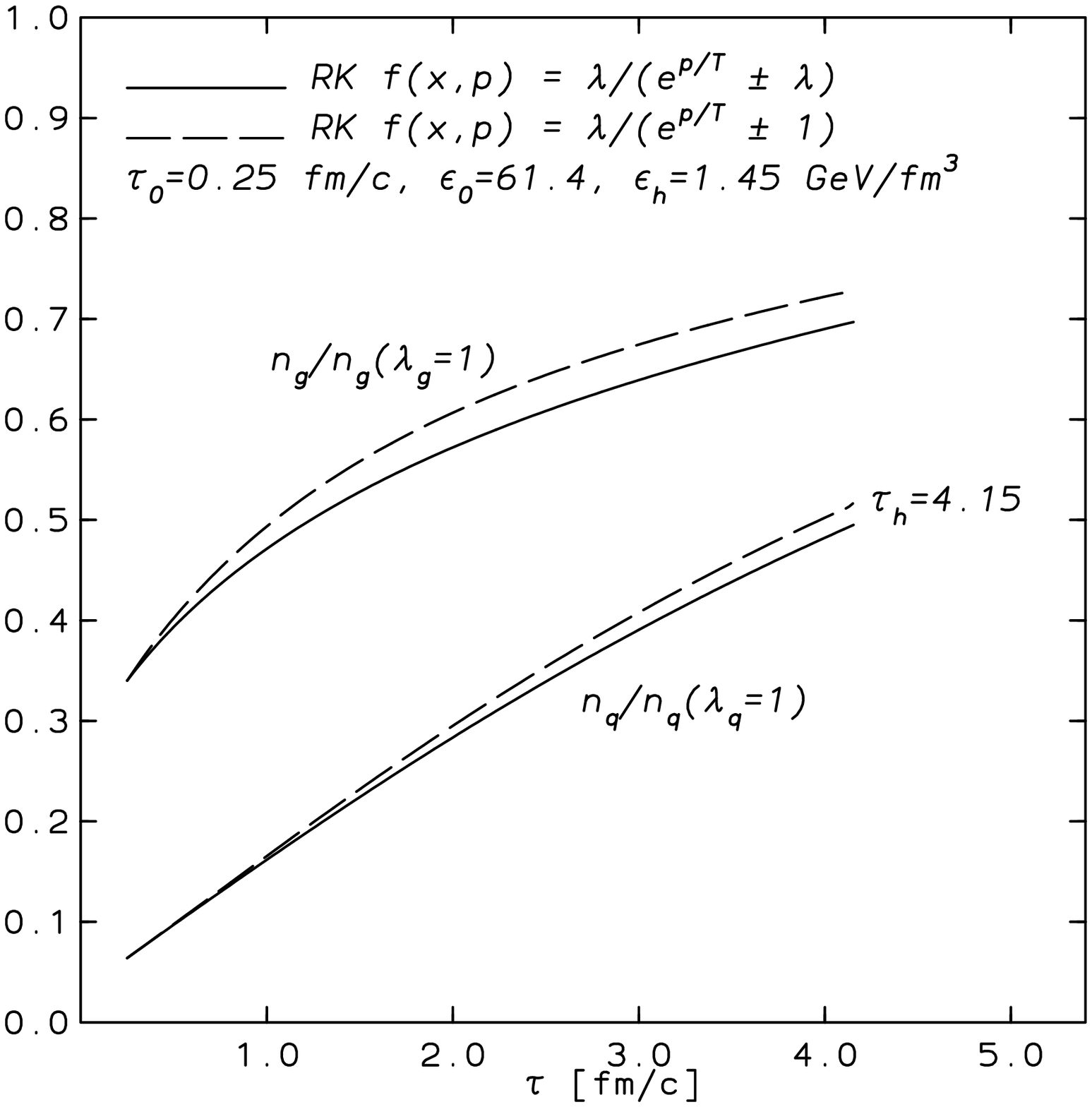}
\includegraphics{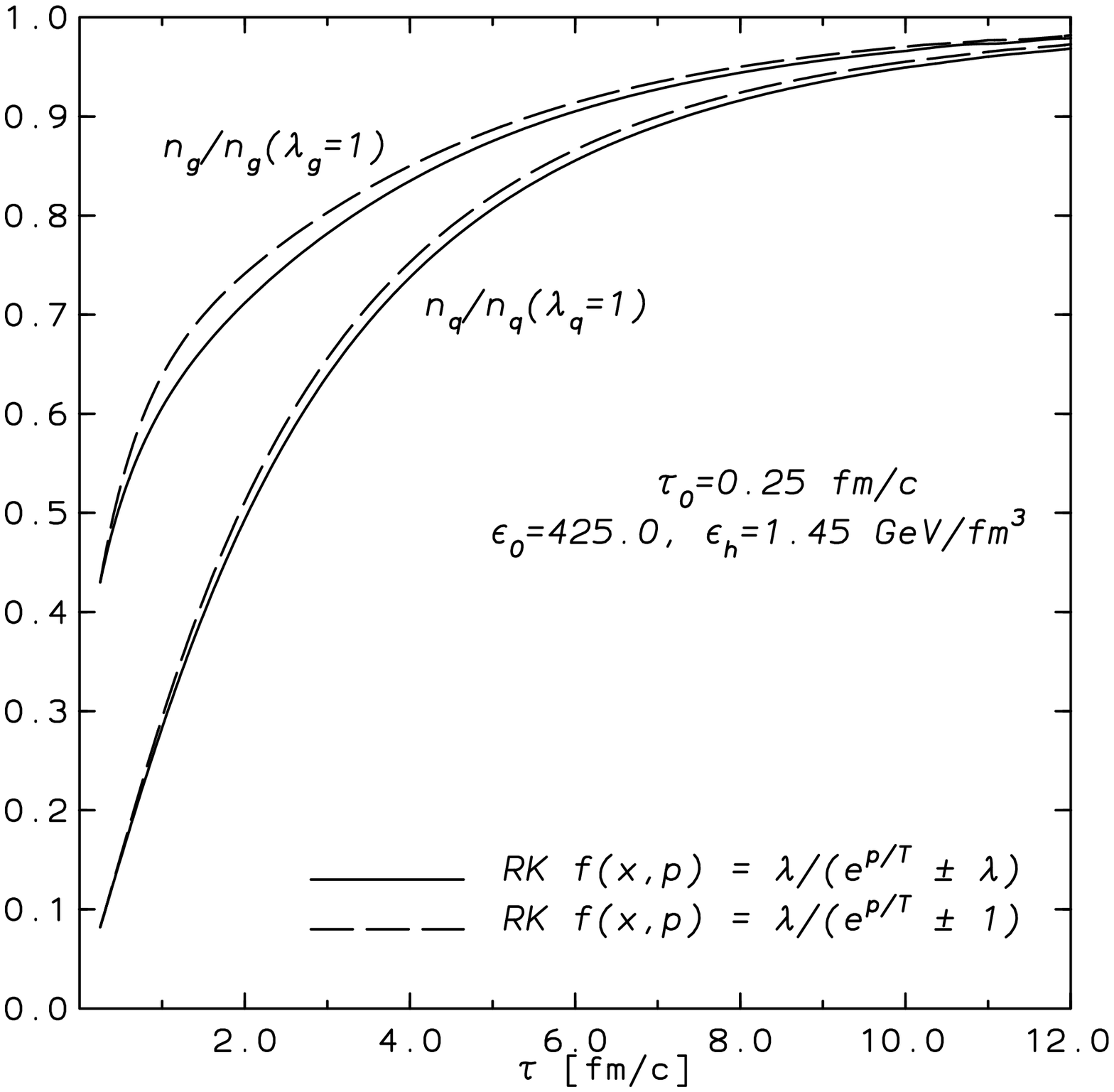}
\label{fig3}
\end{figure}

We are now in a position to check how the equilibration process
differs when using the 
factorized phase-space distribution function (\ref{fac})
as compared with the full distribution function (\ref{1}).
Fig.\ \ref{fig3} shows the proper time evolution of
quark and gluon number densities, $n_q$ and $n_g$,
normalized to the corresponding
values for $\lambda_i = 1$, $\tilde{n}_q$ and $\tilde{n}_g$, 
for SSPC initial conditions at RHIC and LHC
(cf.\ Table \ref{table1}). 
The results for the factorized distribution function
tend to slightly overestimate the degree of equilibration, but the
deviation is of the order of a few percent only. This was to be expected
from Fig.\ \ref{fig2}.

\subsection{Comparison of SSPC and pQCD scenarios}

We now compare the proper time evolution of the parton densities and the 
entropy in the SSPC and pQCD scenarios.
We exclusively use the full distribution function (\ref{1}) 
in the following. The entropy in a given rapidity unit is
${\rm d}S/{\rm d} \eta = s \, \tau \,\pi R^2$, where $\pi R^2$
is the transverse area of the expanding system. Since this area is constant
in a purely longitudinal expansion, we may consider the product of
entropy density and proper time, $s \, \tau$, as a measure for the
entropy per rapidity unit.

\begin{figure}
\vspace*{15.5cm}
\includegraphics{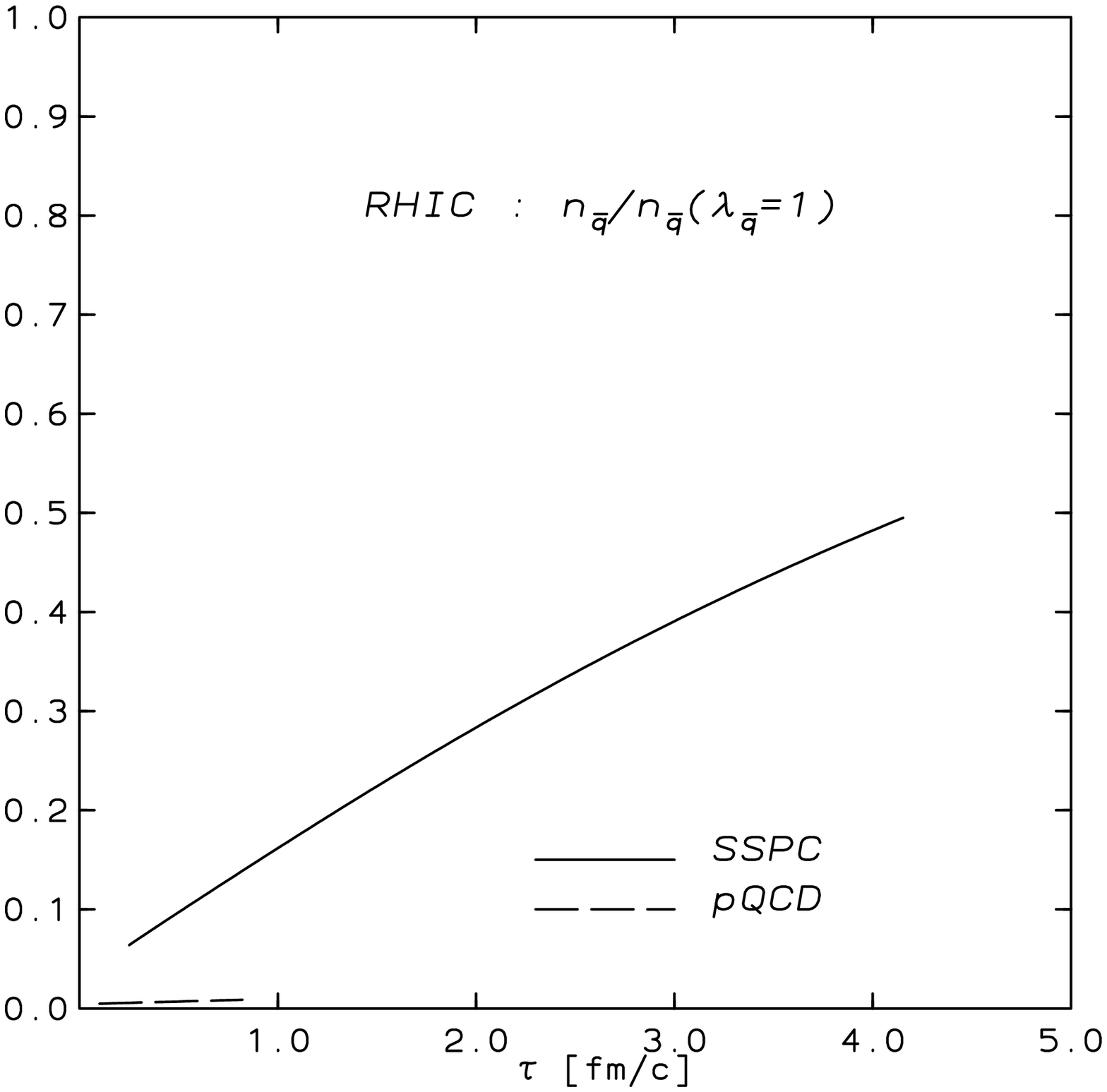}
\includegraphics{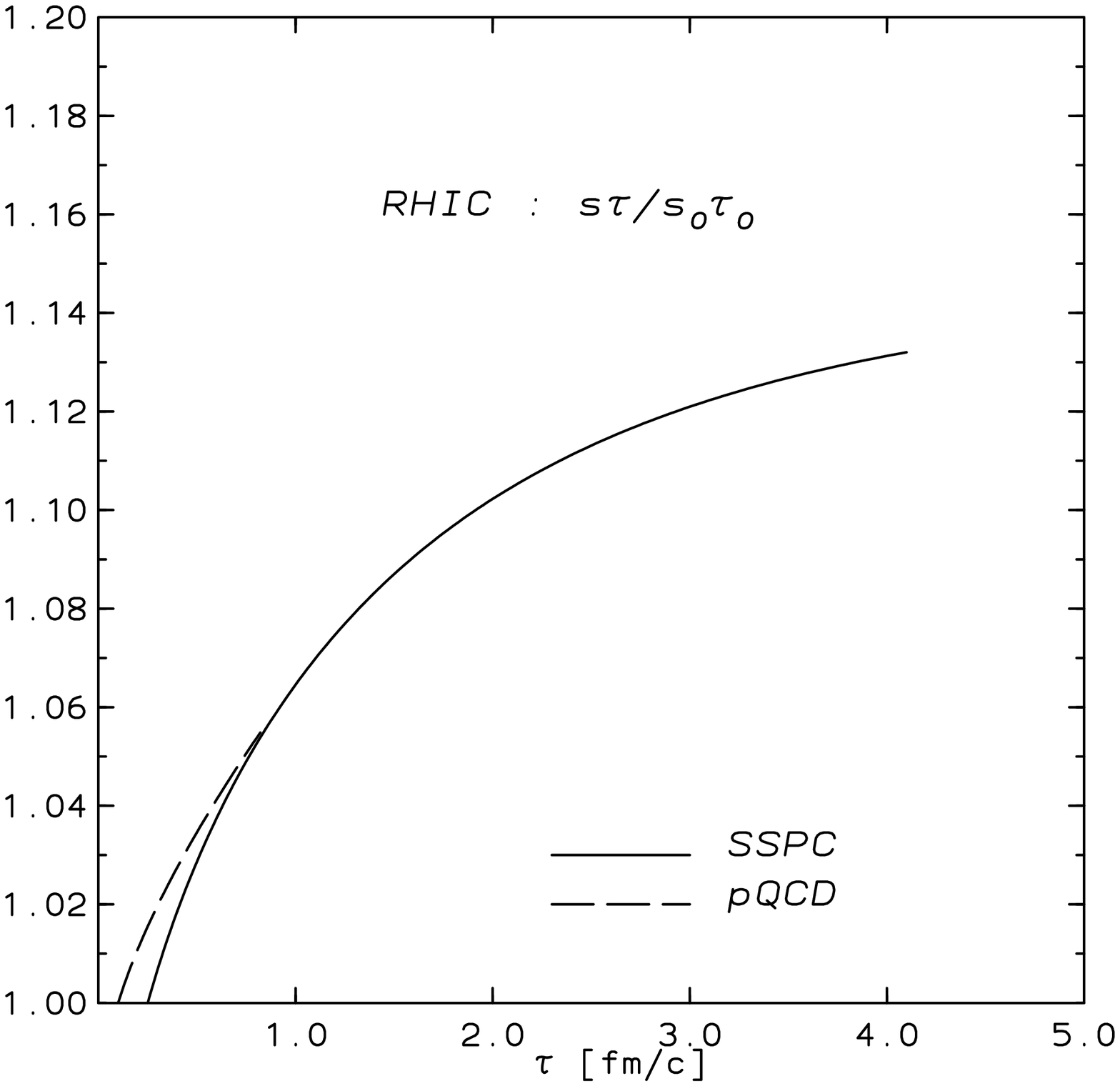}
\includegraphics{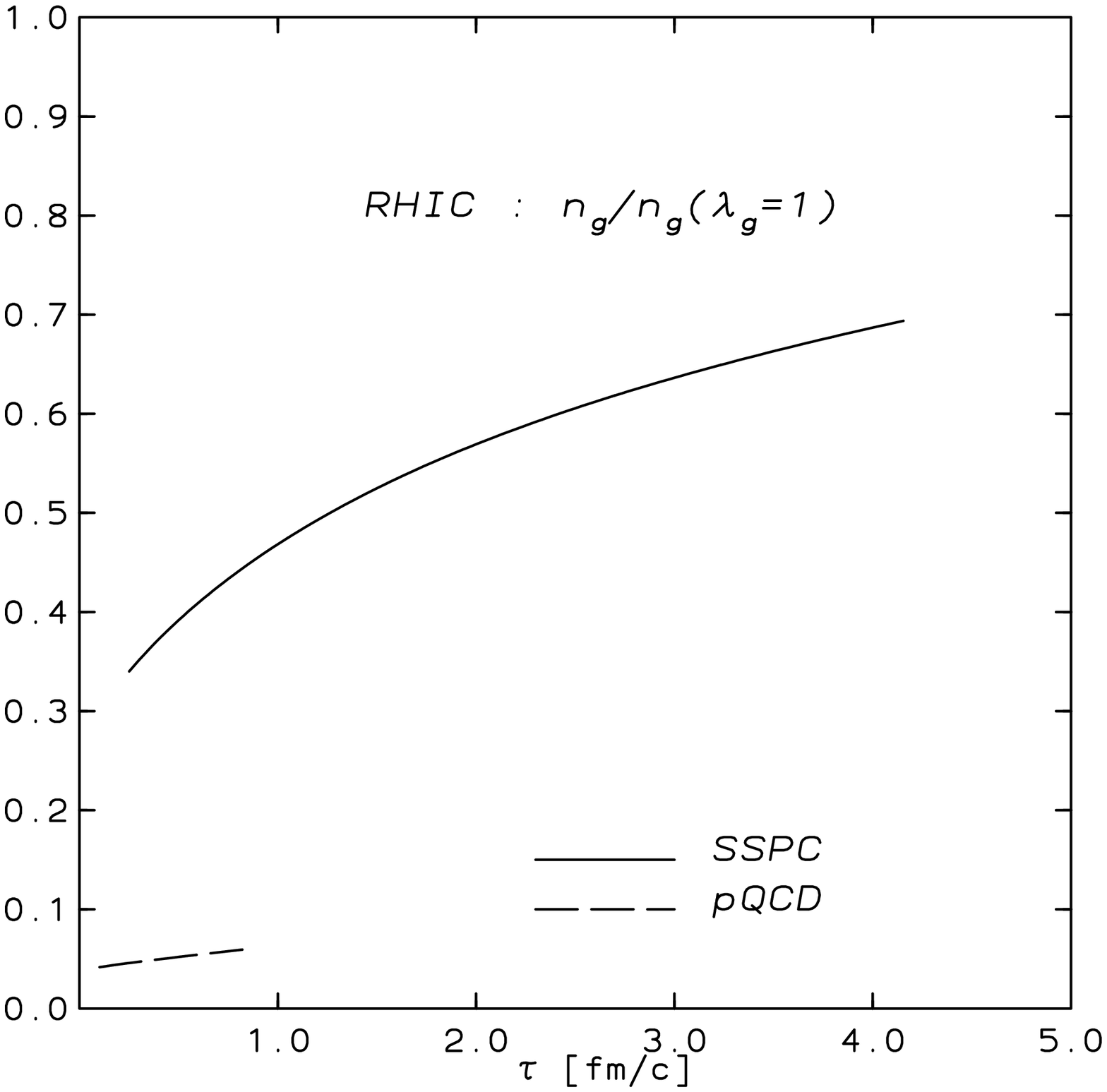}
\includegraphics{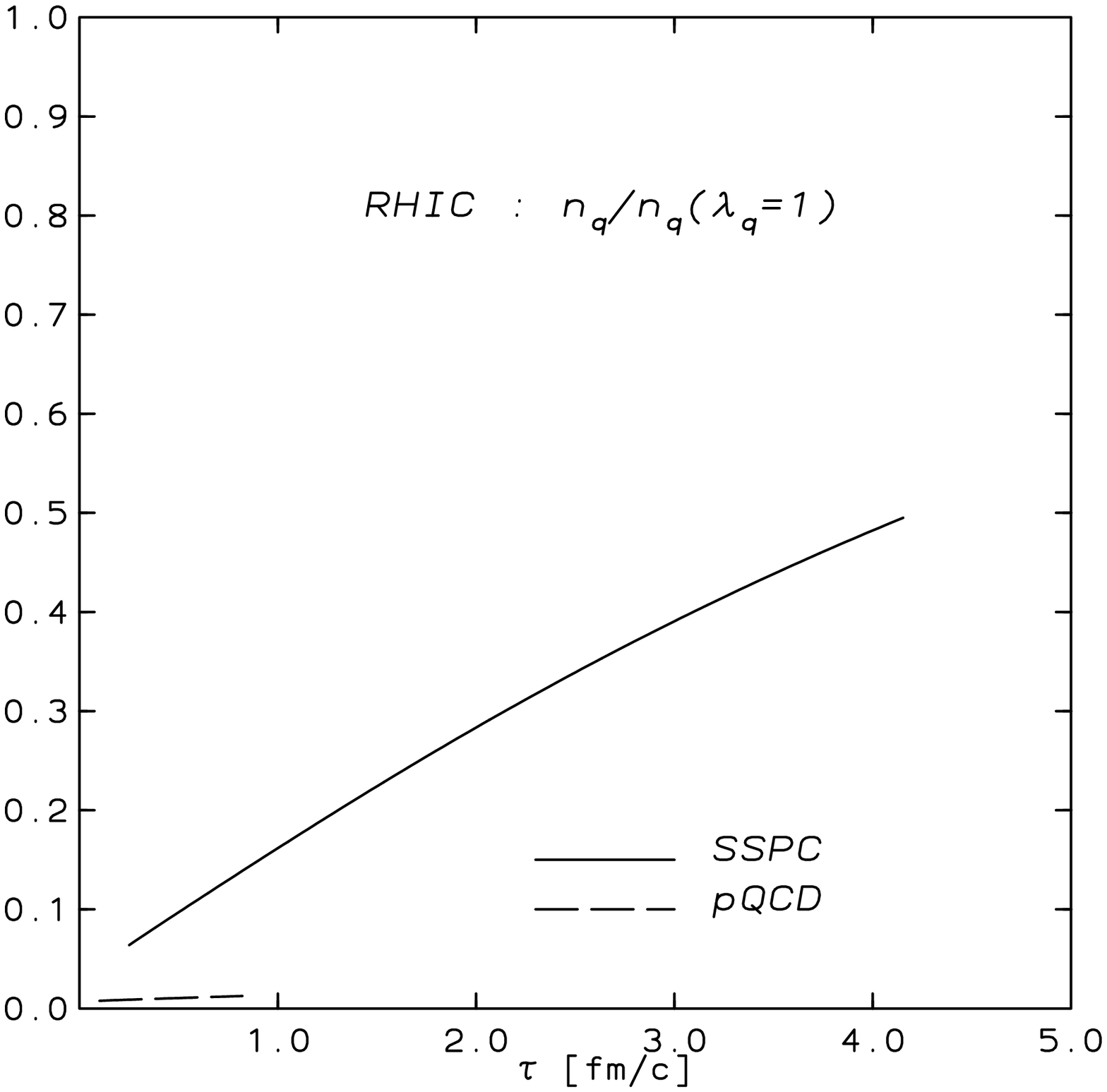}
\vspace*{-2cm}
\caption{RHIC SSPC and pQCD scenarios (see Table \ref{table1}): 
proper time evolution of the gluon (upper left panel), 
quark (upper right panel), and antiquark densities 
(lower left panel), and the entropy (lower right panel). The densities
are normalized to their corresponding values for 
$\lambda_i=1$, the entropy ($\sim s \tau$) to its initial value. }
\label{fig4}
\end{figure}

In Fig.\ \ref{fig4} we show the parton equilibration process for
RHIC initial conditions (see Table \ref{table1}), for the
SSPC model (solid lines) and pQCD (dashed lines). For the SSPC case,
the hadronization energy density $\epsilon_h = 1.45$ GeVfm$^{-3}$ is
reached at a proper time $\tau_h=4.15$ fm/c, at which we stop
the time evolution. The relative gluon density, $n_g/\tilde{n}_g$, reaches 
about $0.7$, and the relative quark density, $n_q/\tilde{n}_q$, about
$0.5$. The entropy increases $13$ \% before hadronization.  
For pQCD initial conditions, the QGP phase has little
time to evolve, hadronizing only about $0.7$ fm/c 
after thermalization. The partons have no time to equilibrate, and the 
entropy increases only $\sim 5$ \%. 

While the SSPC scenario was considered to be net-baryon free, 
the pQCD scenario shows a slight difference in the quark and 
antiquark initial densities, corresponding to an initial baryon number 
density of about $0.12$ f${\rm m}^{-3}$. We checked that
our Runge-Kutta solver respects baryon number conservation, which in the 
purely longitudinally expanding geometry implies that the product 
$n_B\, \tau = (n_q - n_{\bar{q}})\tau /3$ is constant throughout the 
expansion. As a consequence of nonzero net-baryon number, the
equilibrium ratio of $n_q /\tilde{n}_q$ will be larger than
1, while that of $n_{\bar{q}}/\tilde{n}_{\bar{q}}$ will be 
smaller than 1, because $\tilde{n}_i$ is computed with $\lambda_i = 1$
instead of the correct equilibrium value $\lambda_q^{\rm eq}=
1/\lambda_{\bar{q}}^{\rm eq} > 1$. For a small initial baryon
number density of about $0.12\, {\rm fm}^{-3}$, however, the deviation
of $\lambda_i^{\rm eq}$ from 1 is negligible.

\begin{figure}
\vspace*{15.5 cm}
\includegraphics{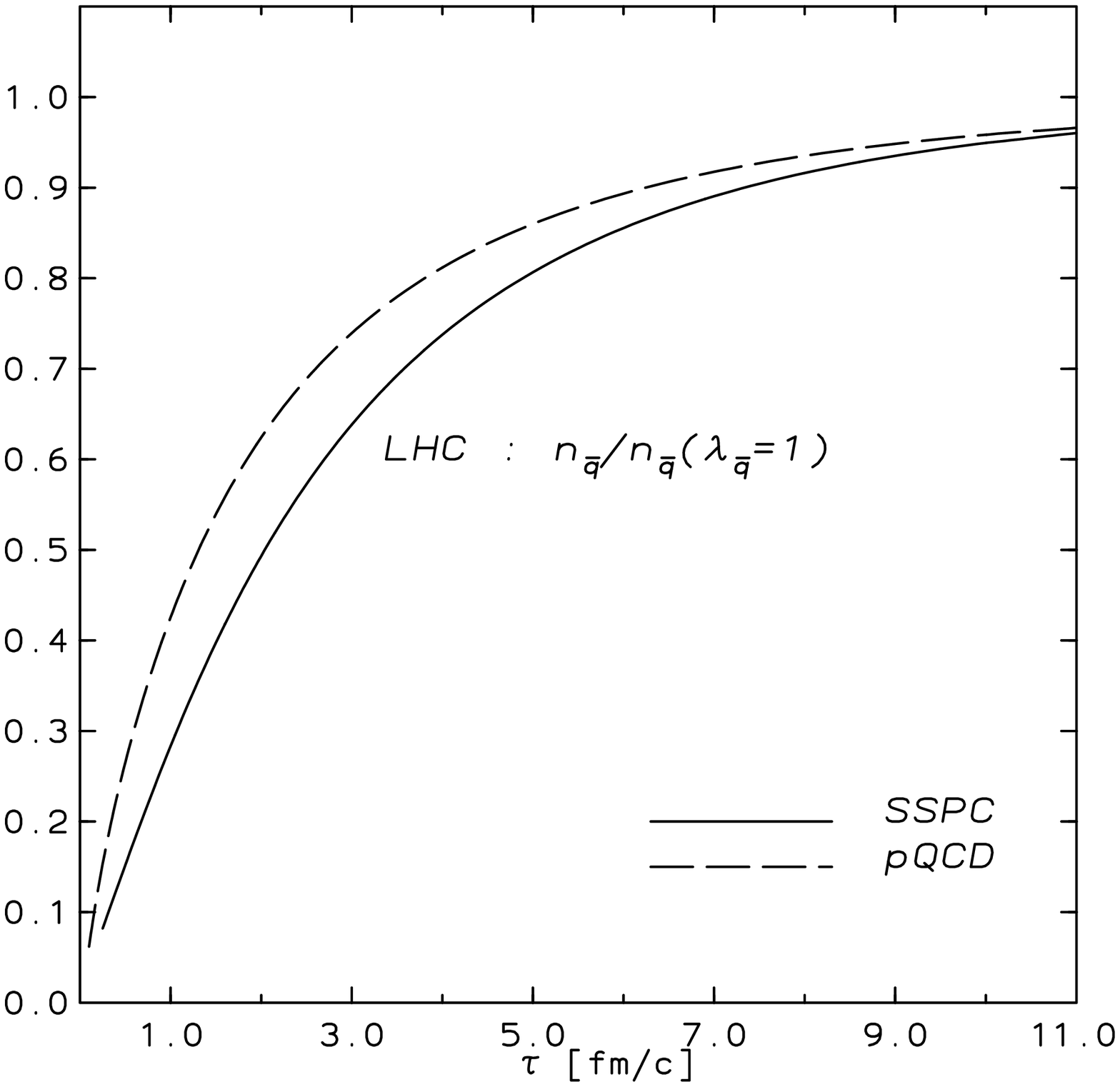}
\includegraphics{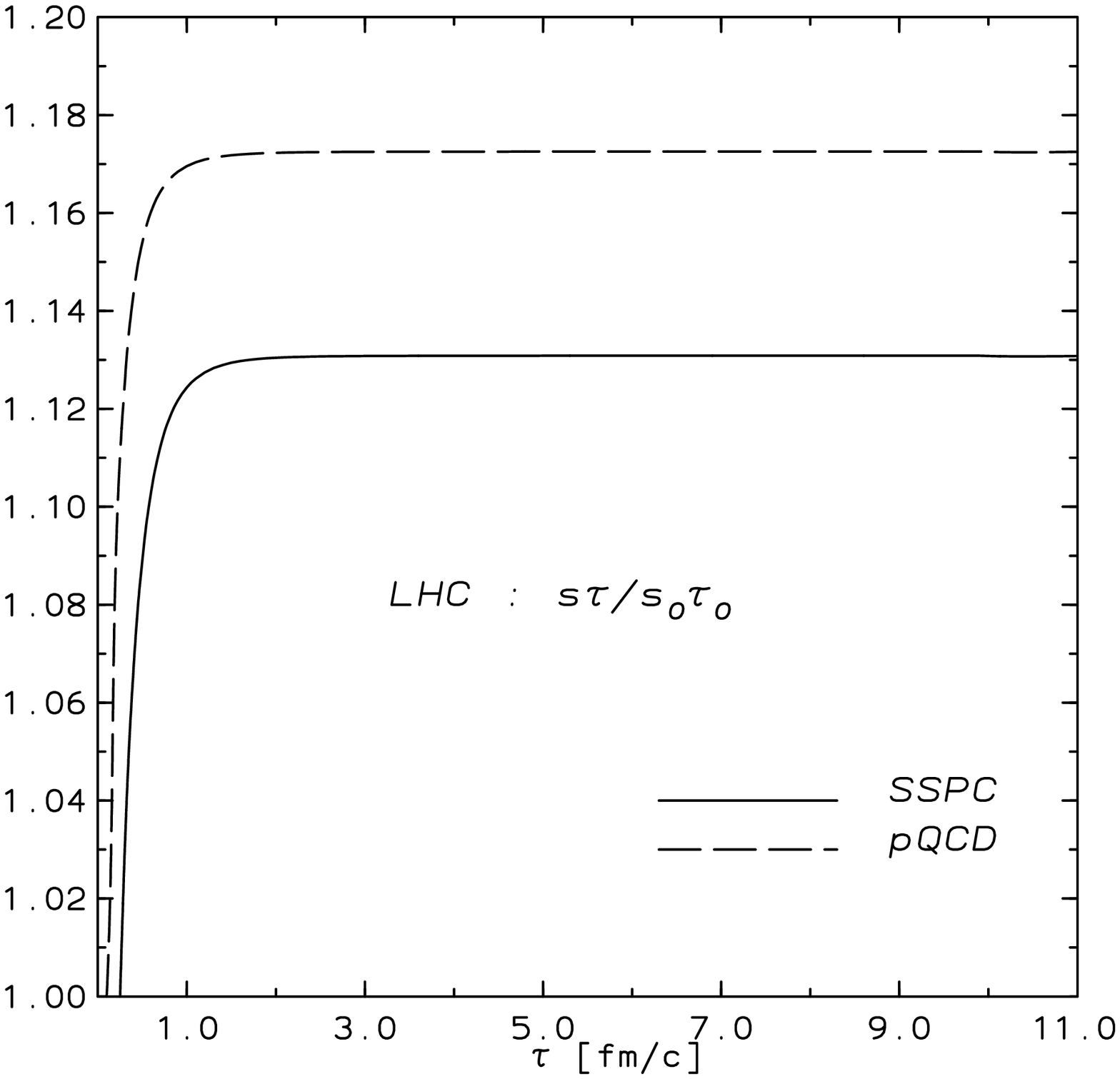}
\includegraphics{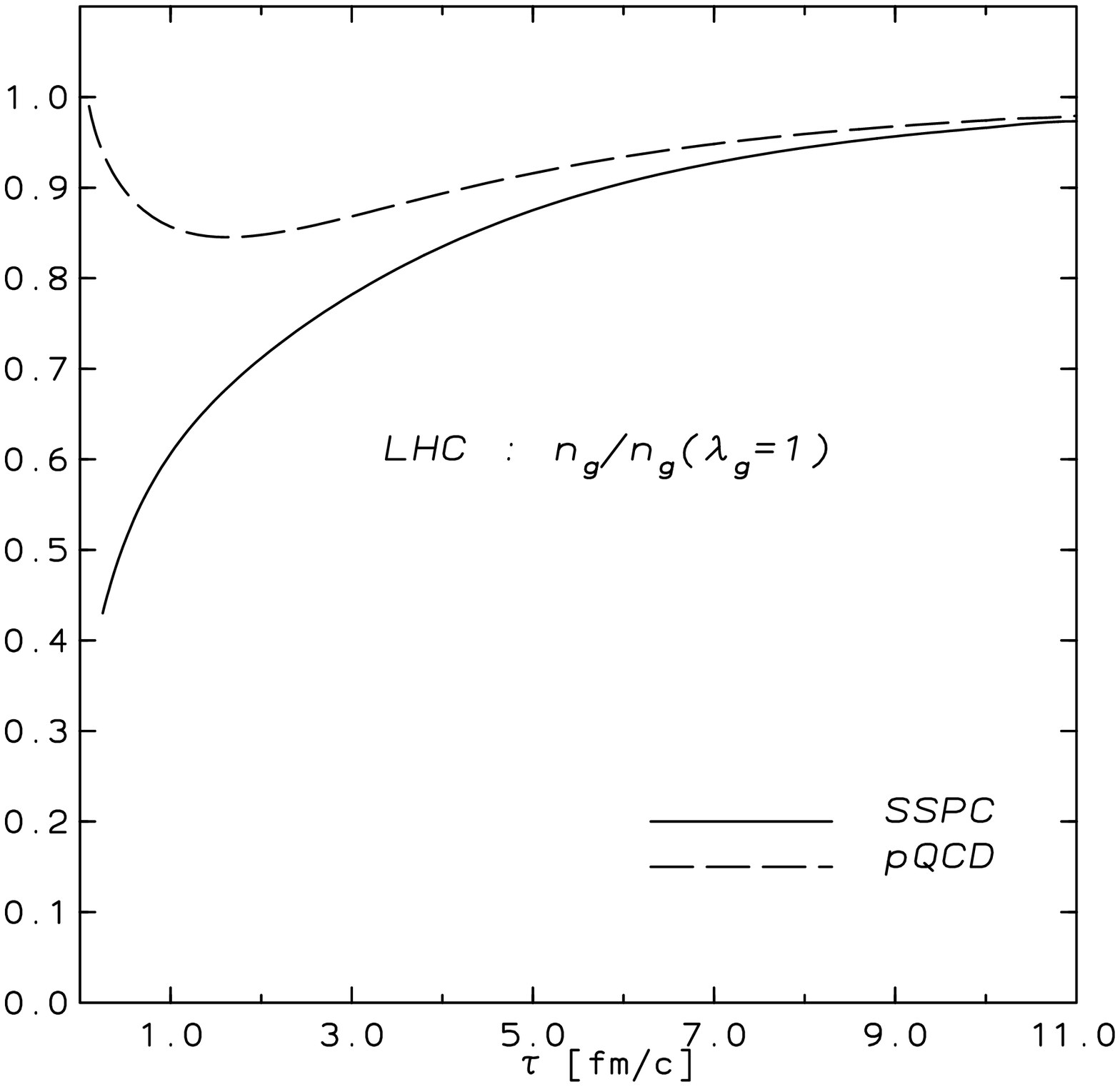}
\includegraphics{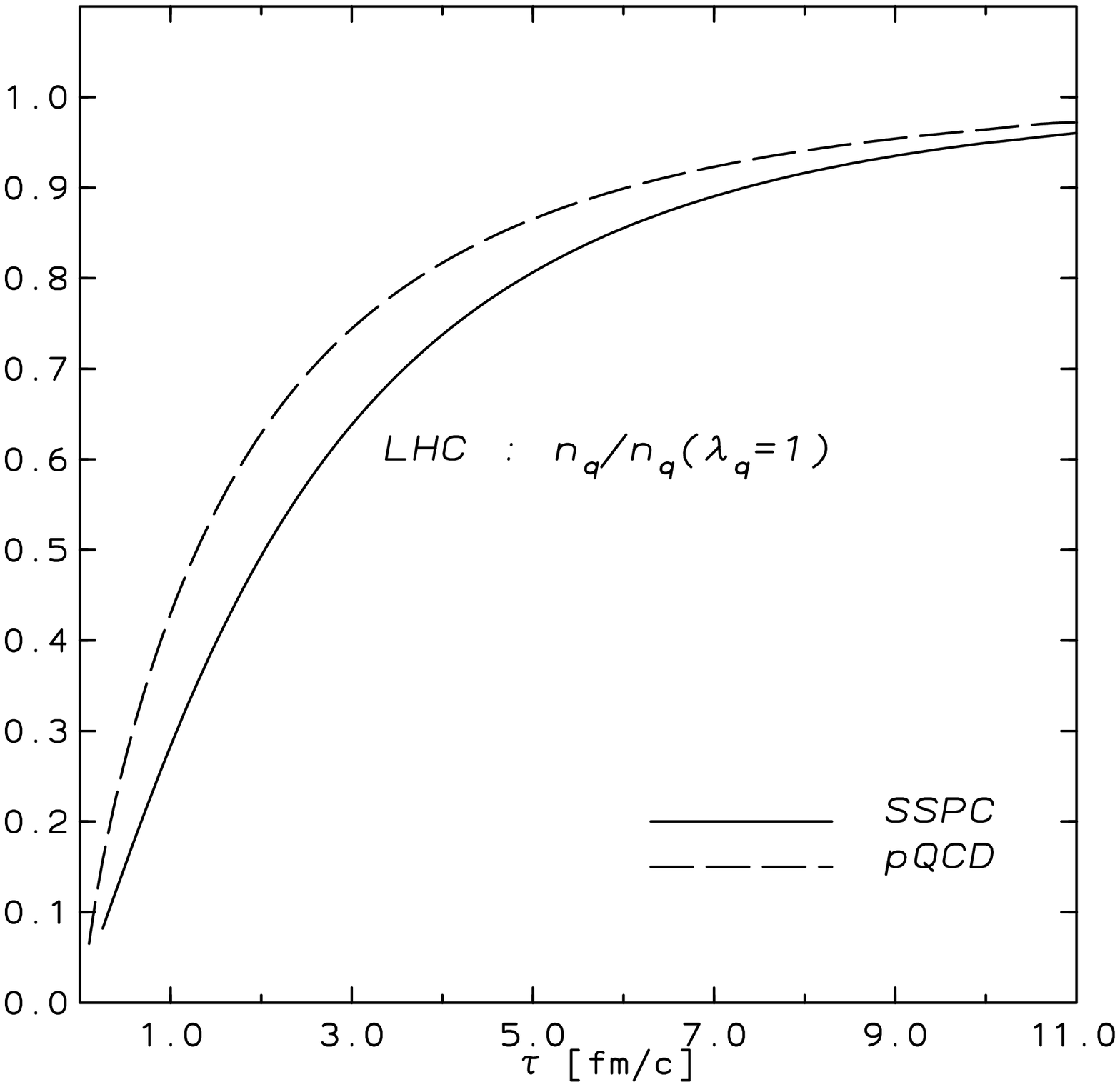}
\vspace*{-2cm}
\caption{As in Fig.\ \ref{fig4}, for LHC initial conditions.}
\label{fig5}
\end{figure}

The LHC case is shown in Fig.\ \ref{fig5}.
The parton species are seen to approach equilibrium after about
10 fm/c. We note that this time is well before a transverse rarefaction 
front can penetrate to the center of the QGP (see discussion in
section \ref{transverse}). The reason why partons equilibrate is
that the system starts at much higher temperature and higher initial values
for the fugacities than at RHIC (see Table \ref{table1}).
Particularly noteworthy is that, in the pQCD case, 
the gluons are almost completely equilibrated already at $\tau_0$, 
while the quarks and antiquarks are not. Consequently, quark--antiquark 
production processes drive the gluons temporarily {\em out\/} of equilibrium.
Entropy production is of the order of $13 \%$ (SSPC) to $17 \%$ (pQCD).

\subsection{Sensitivity to {\boldmath {\bf $\tau_0\ {\rm and} \ \alpha_s$}}}

We now investigate the sensitivity of the equilibration 
process to variations of the initial proper time, $\tau_0$, and
the strong coupling constant, $\alpha_s$.
As discussed above, the values for the initial time $\tau_0$ 
in Table \ref{table1} are most certainly {\em lower\/} bounds. 
We therefore consider values $\tau_0 \geq 0.25$ fm/c for the 
SSPC scenario and $\tau_0 \geq 0.1$ fm/c for the pQCD approach.
In varying $\tau_0$, we keep the produced transverse energy,
${\rm d}E_T/{\rm d}\eta$, and the parton numbers, ${\rm d}N_i/{\rm d}\eta$, 
constant. By eq.\ (\ref{dV}), the energy density and parton number
densities must then decrease like $\sim 1/\tau_0$. In
essence, this means that we allow the system to
evolve {\em without\/} doing longitudinal work. This is certainly
an idealization: only when the pressure vanishes,
the system does not perform any longitudinal work, but in our case
pressure builds up as the system approaches kinetic equilibrium.

To facilitate the presentation, we only give the {\em final\/}
values for the parton densities and the entropy at the end of the
expansion of the QGP phase. The evolution of the QGP is terminated either 
at the {\em hadronization time}, $\tau_h$, when
the longitudinal expansion has cooled the system down to an energy
density $\epsilon_h = 1.45$ GeV/fm$^{3}$, or
at the {\em rarefaction time}, $\tau_{\rm rarefac}$, when
a transverse rarefaction wave reaches the center of the system
(see discussion in section \ref{transverse}).

The time evolution equation for the 
energy density, (\ref{enmomcons}), with the equation of state (\ref{pe3}), 
has the solution $\epsilon/\epsilon_0 = (\tau_0/\tau)^{4/3}$.
According to eq.\ (\ref{dV}), the product $\epsilon_0 \tau_0 =
({\rm d}E_T/{\rm d} \eta)/(\pi R^2)\,$ is constant for constant 
transverse energy per unit rapidity and constant transverse area.
Consequently, the hadronization time $\tau_h$ grows with the initial proper
time $\tau_0$ according to
$\tau_h = \tau_0^{1/4}\, \left[ ({\rm d}E_T/{\rm d} \eta)/(\pi R^2 \epsilon_h)
\right]^{3/4}$. The time spent in the QGP phase is therefore
\begin{equation} \label{tauh}
\Delta \tau \equiv
\tau_h - \tau_0= \tau_0^{1/4}\, \left(  \frac{{\rm d}E_T/{\rm d} \eta}{
\pi R^2 \epsilon_h} \right)^{3/4}  - \tau_0\,\,.
\end{equation}
This time increases for small $\tau_0$, has a maximum at
$\tau_0^* = ({\rm d}E_T/{\rm d} \eta)/( 4^{4/3}\pi R^2 \epsilon_h)$, and
then decreases again.
For the SSPC model at RHIC, $\tau_0^* \simeq 1.67$ fm/c, while for
pQCD at RHIC, $\tau_0^* \simeq 0.27$ fm/c. At LHC energies, the values
for $\tau_0^*$ are quite similar, for the SSPC model,
$\tau_0^* \simeq 11.54$ fm/c, and for pQCD,
$\tau_0^* \simeq 11.48$ fm/c.

The hadronization time grows proportional to the initial 
transverse energy. For LHC energies, the transverse energy is so 
large that a transverse rarefaction
wave (see section \ref{transverse}), 
travelling with sound velocity $c_s = 1/\sqrt{3}$ into matter at rest,
reaches the center of the system {\em before\/} the
longitudinal expansion has cooled matter down to $\epsilon_h$.
At $z=0$, this transverse rarefaction wave reaches the center at time
$\tau_{\rm rarefac} = \tau_0 + R/c_s$. 
For a $Pb$ nucleus,
$R = 1.12\, A^{1/3} \simeq 6.6$ fm and 
$ \tau_{\rm rarefac}\simeq \tau_0 + 11.5$ fm/c.
While we use $\tau_h$ to terminate the time evolution at RHIC
energy, for the LHC case, the time evolution is terminated
at $\tau_{\rm rarefac}$ instead of $\tau_h$.

\begin{figure}
\vspace*{15cm}
\includegraphics{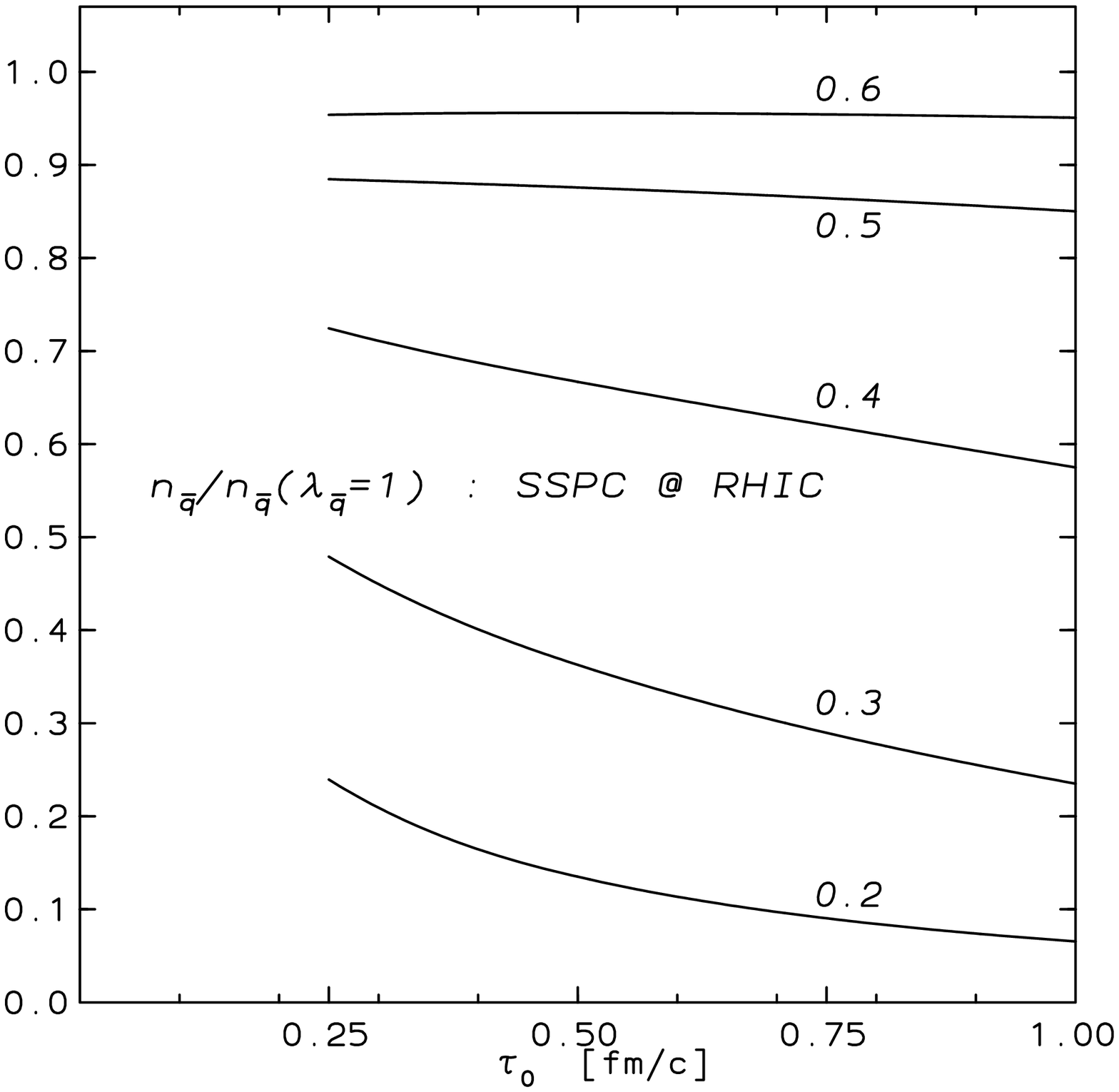}
\includegraphics{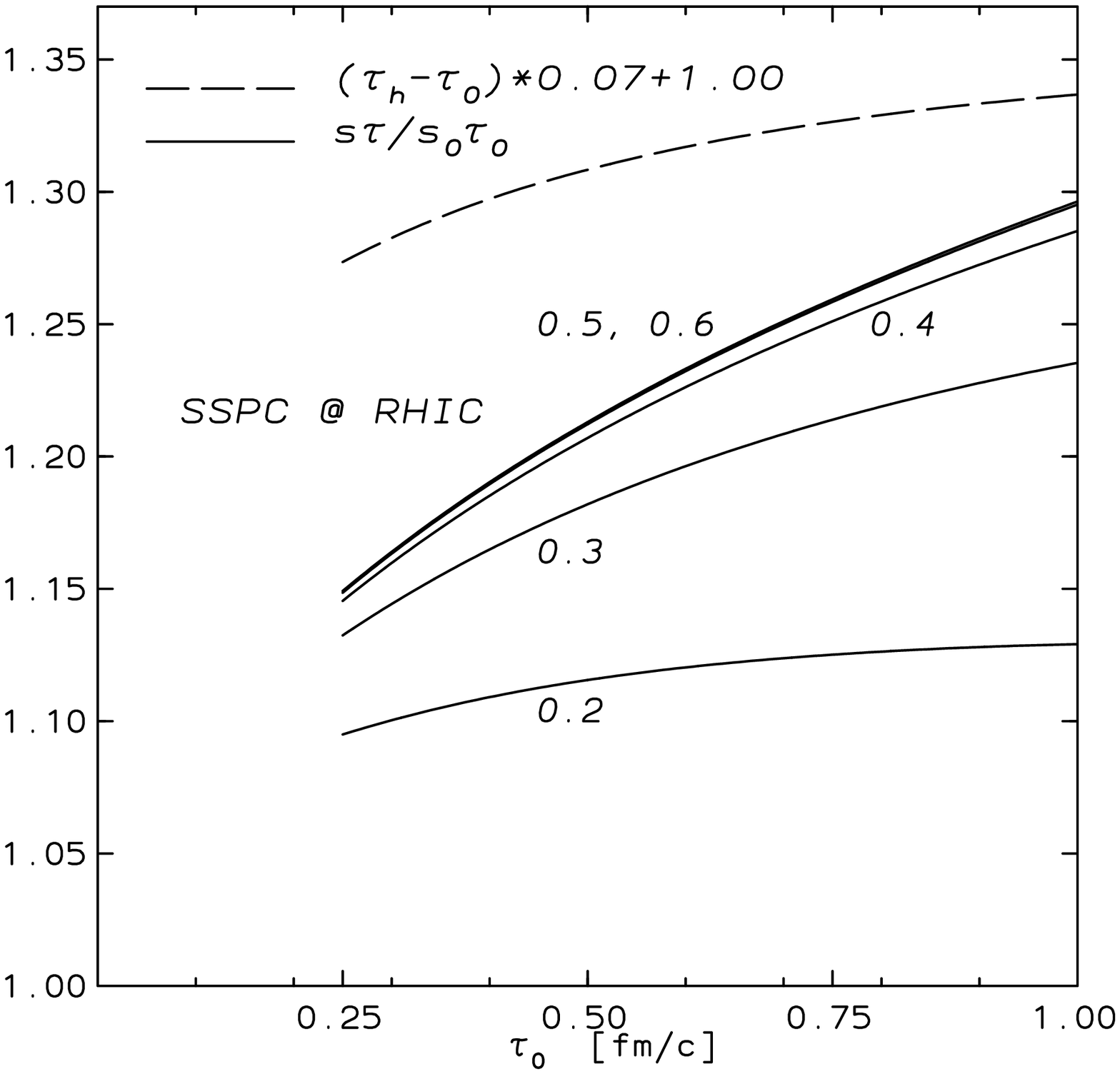}
\includegraphics{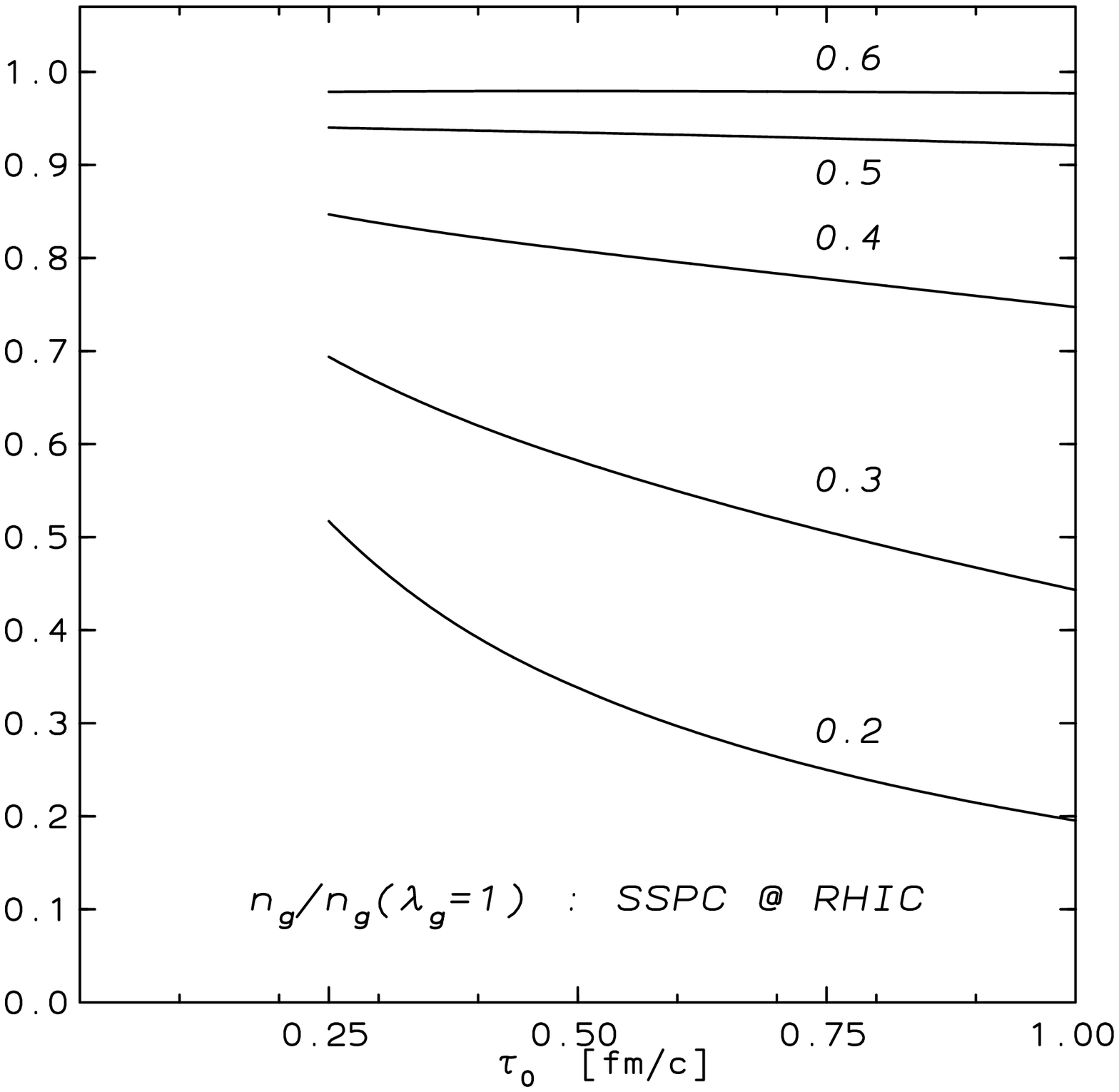}
\includegraphics{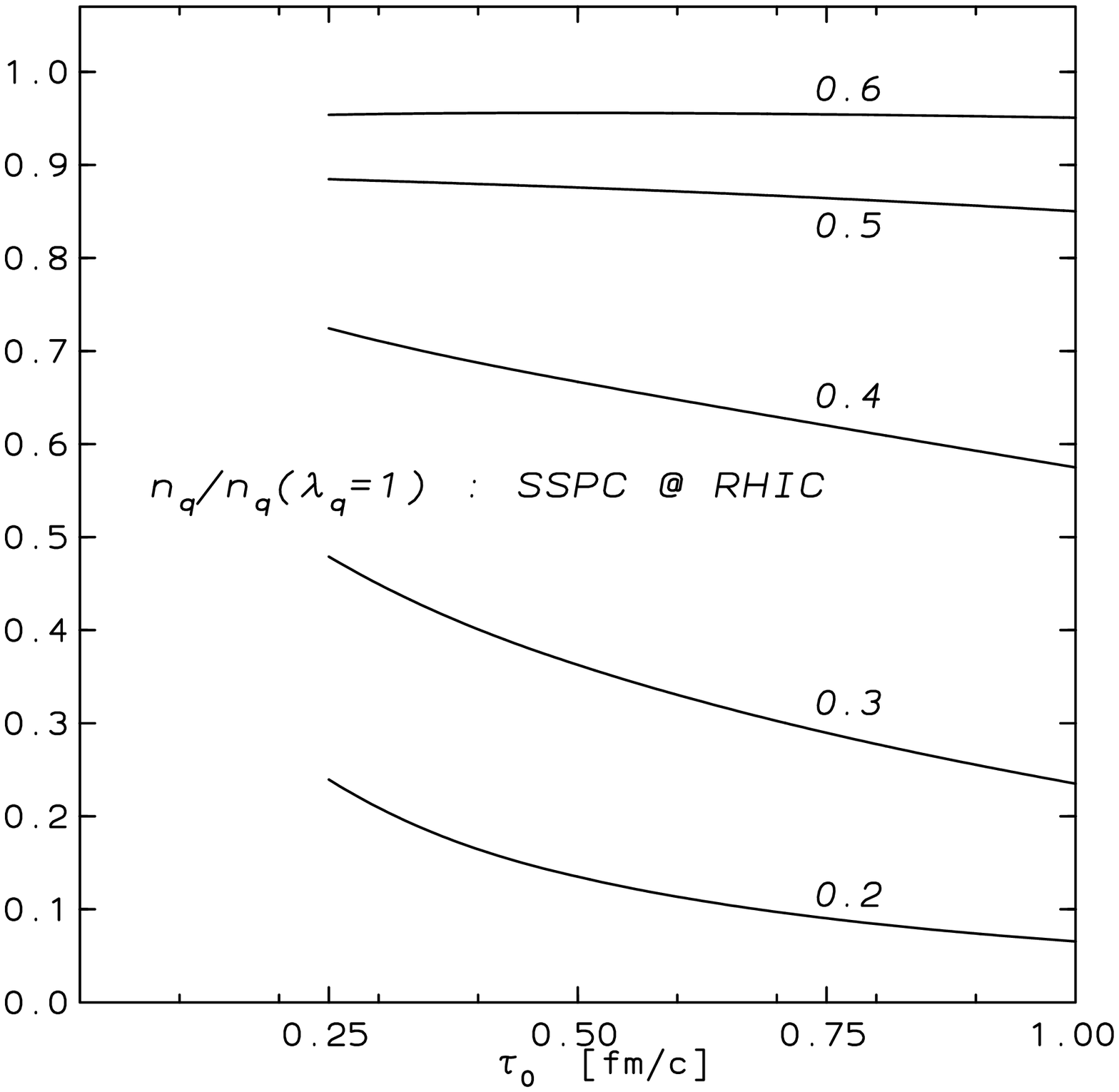}
\vspace*{-1cm}
\caption{Hadronization values of the relative densities and the
entropy for ${\alpha}_s = 0.2,\, 0.3, \ldots,\, 0.6$, for the SSPC model 
with RHIC initial conditions.
Upper left: gluons, upper right: quarks, lower left: antiquarks, lower right: 
$s\tau /s_0{\tau}_0$. The dashed line is the time the system spends
in the QGP phase. }
\label{fig6}
\end{figure}

Fig.\ \ref{fig6} shows the case for the SSPC model with
RHIC initial conditions. 
The hadronization values of the densities are closer to equilibrium 
for larger values of $\alpha_s$. This is obvious, 
since then the right-hand sides of the 
rate equations (\ref{r1a}) -- (\ref{r3a}) are larger, cf.\ eq.\ (\ref{Rs}),
driving the system faster towards equilibrium. Consequently, also
entropy production increases with $\alpha_s$.

On the other hand, the hadronization values of the densities are
further away from equilibrium 
for increasing values of the initial time $\tau_0$. The reason
is that, according to eq.\ (\ref{dV}), the values for the initial
energy density and the parton densities decrease for increasing $\tau_0$
(the parton number rapidity density ${\rm d}N_i/{\rm d}\eta$ and 
transverse energy rapidity density ${\rm d}E_T/{\rm d}\eta$ are kept
constant). Surprisingly, this does not have an effect on the initial 
temperature which is to all intents and purposes independent of $\tau_0$. 
To understand this,
consider the factorized expressions (\ref{efac}) and (\ref{nfac});
reducing energy and parton densities by the same factor is
achieved by reducing the fugacities only, keeping the temperature constant.
However, the reduction of the fugacities
leads to smaller initial values for the parton densities. 
This puts the initial
densities further away from their equilibrium values. This difference is, 
with increasing $\tau_0$, increasingly harder
to overcome during the lifetime $\Delta \tau$, eq.\ (\ref{tauh}), 
of the QGP phase.
The amount of entropy produced in the RHIC scenarios
is proportional to the lifetime of the QGP phase, {\it i.e.}, the
time during which chemical reactions drive the system towards
thermodynamical equilibrium and thus increase the entropy. 
Note that equilibration is never complete at RHIC, unless one uses
rather large values of $\alpha_s$.

\begin{figure}
\vspace*{15.5cm}
\includegraphics{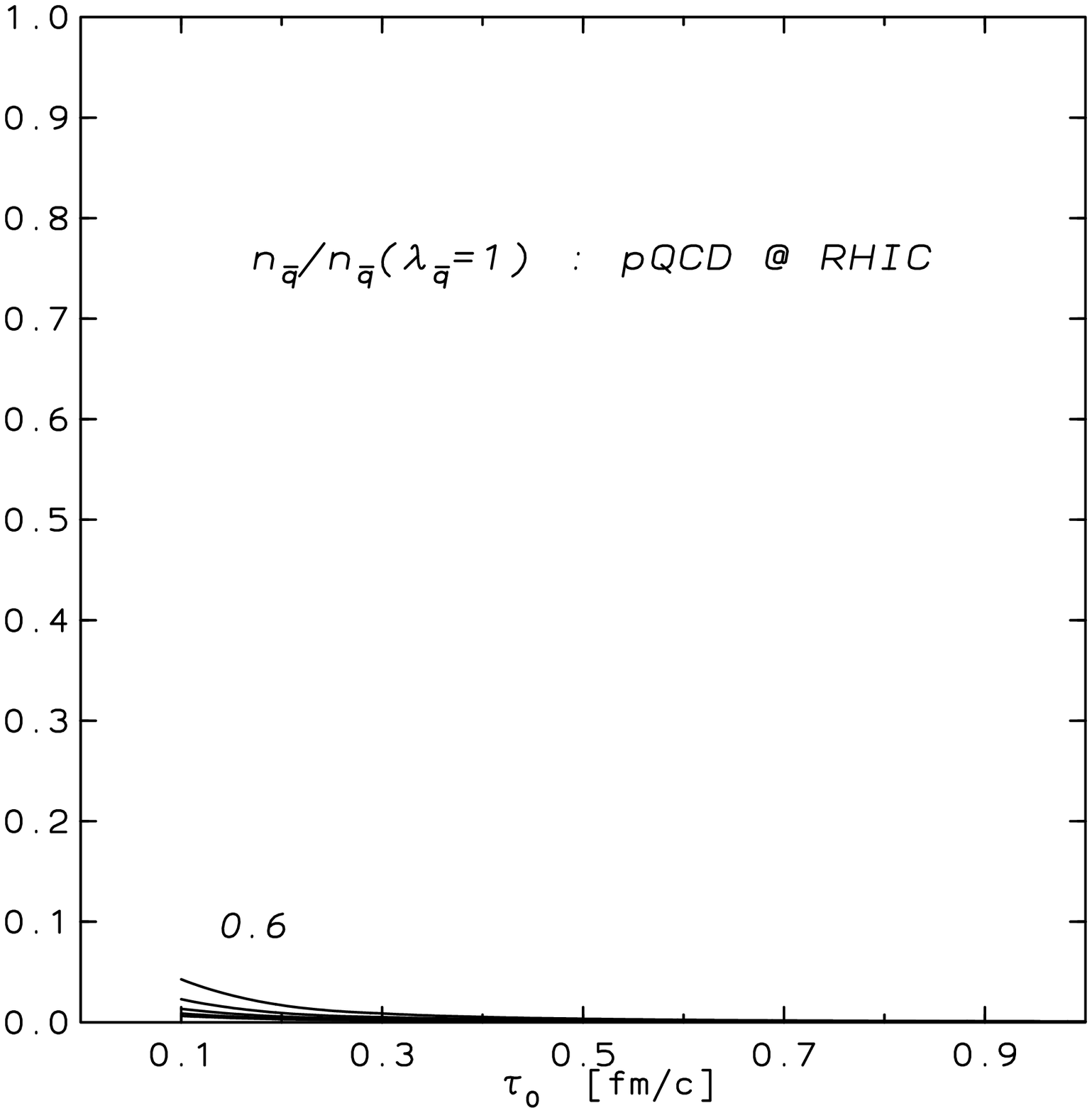}
\includegraphics{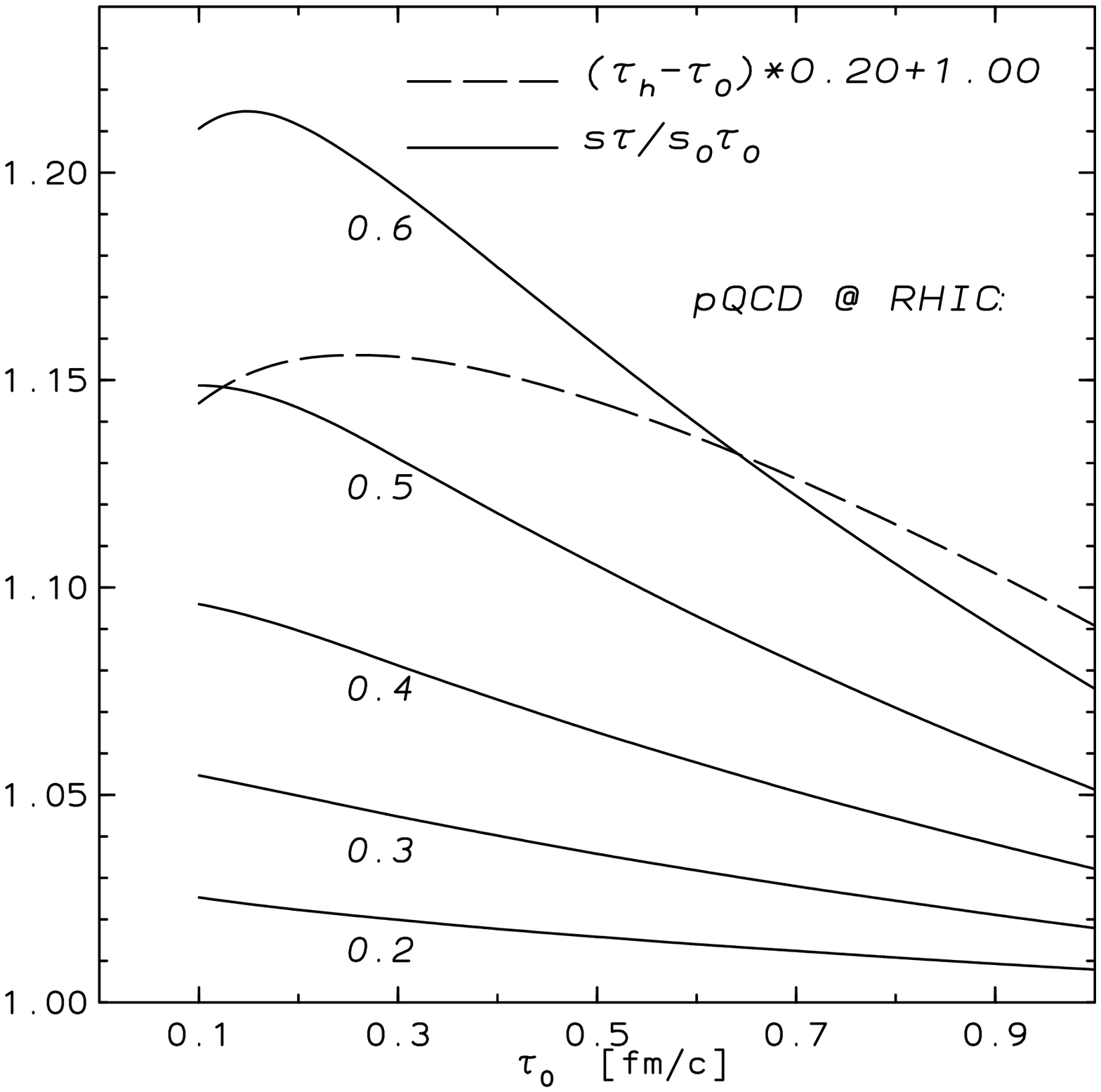}
\includegraphics{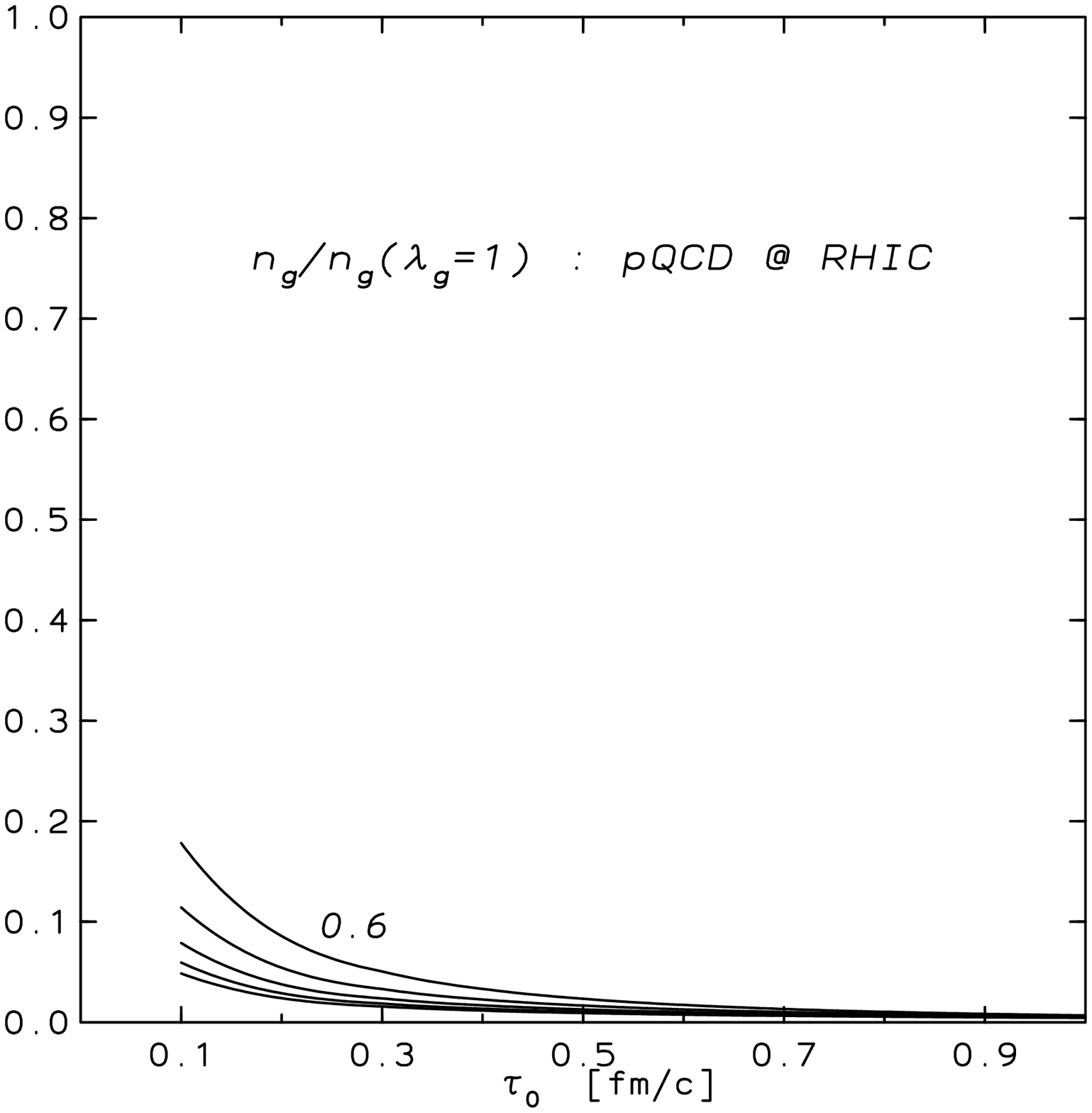}
\includegraphics{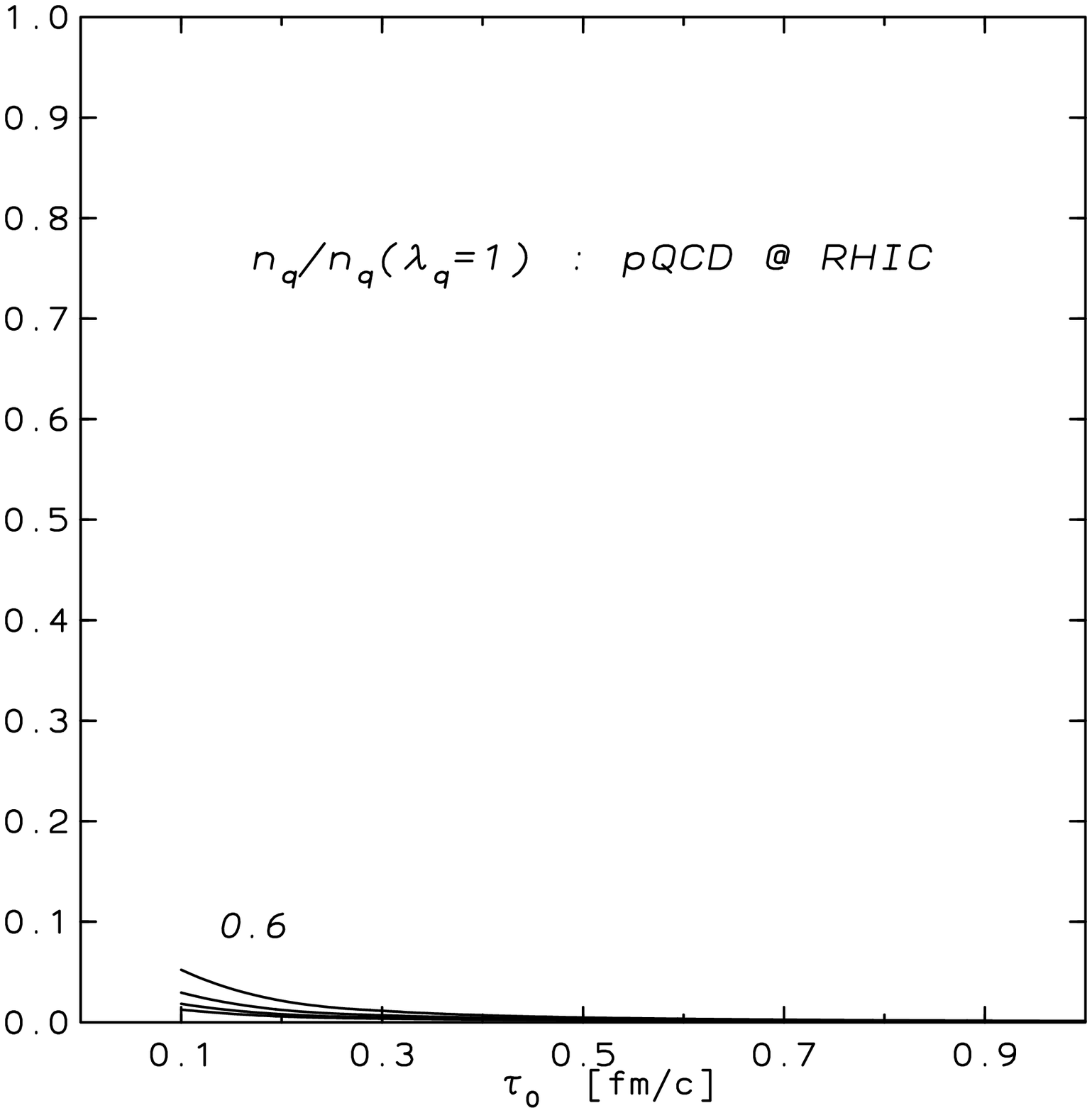}
\vspace*{-2cm}
\caption{As in Fig.\ \ref{fig6}, for pQCD initial conditions.}
\label{fig7}
\end{figure}

The pQCD RHIC scenario is depicted in Fig.\ \ref{fig7}. 
The parton densities behave as in Fig.\ \ref{fig6}.
The main quantitative difference is that, due to the small initial values of
the energy and parton densities, even for large values 
of $\alpha_s$ the system {\em never\/} reaches chemical equilibrium.
The behavior of the entropy follows again that of the
lifetime of the QGP.

\begin{figure}
\vspace*{15.5cm}
\includegraphics{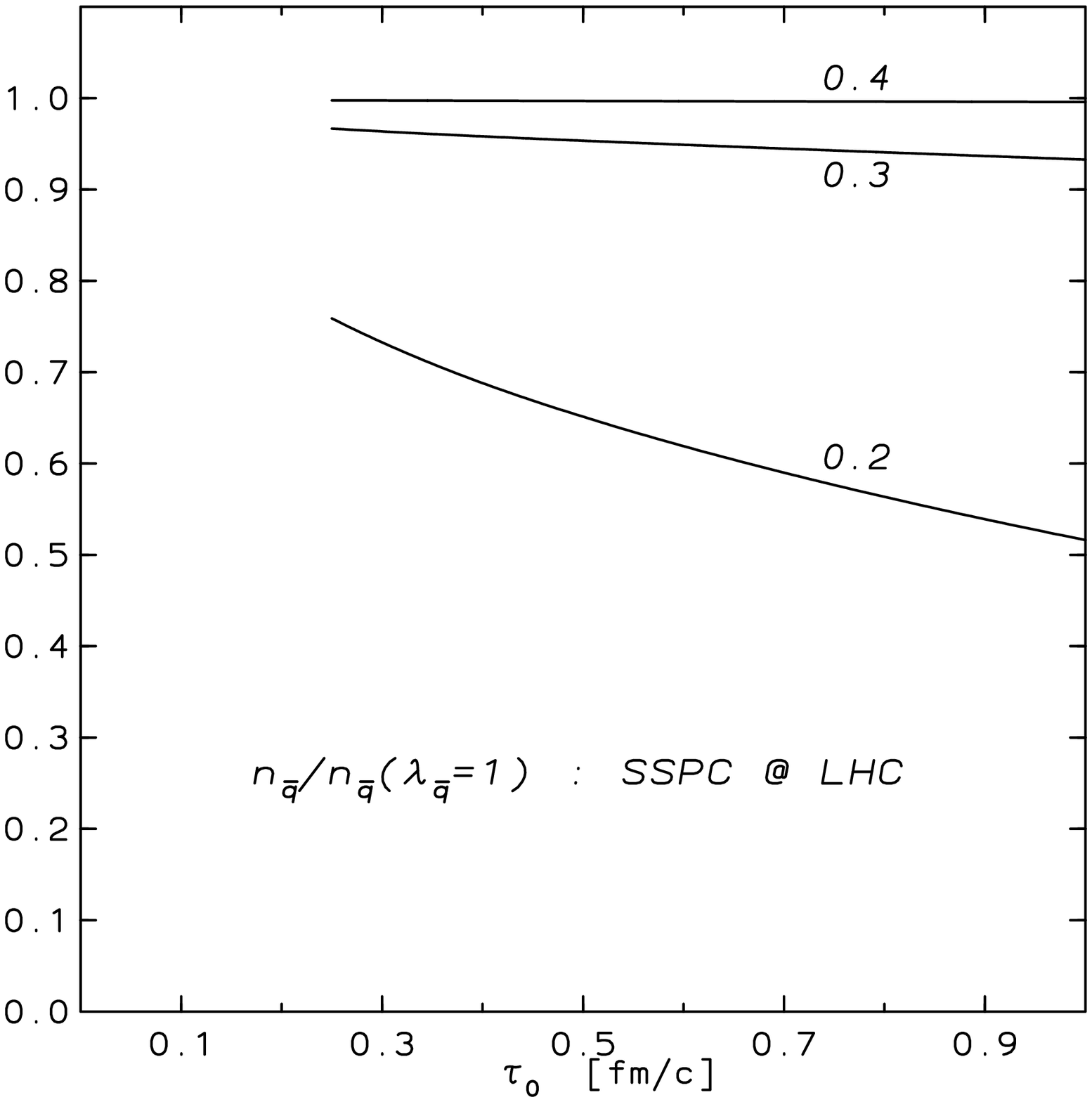}
\includegraphics{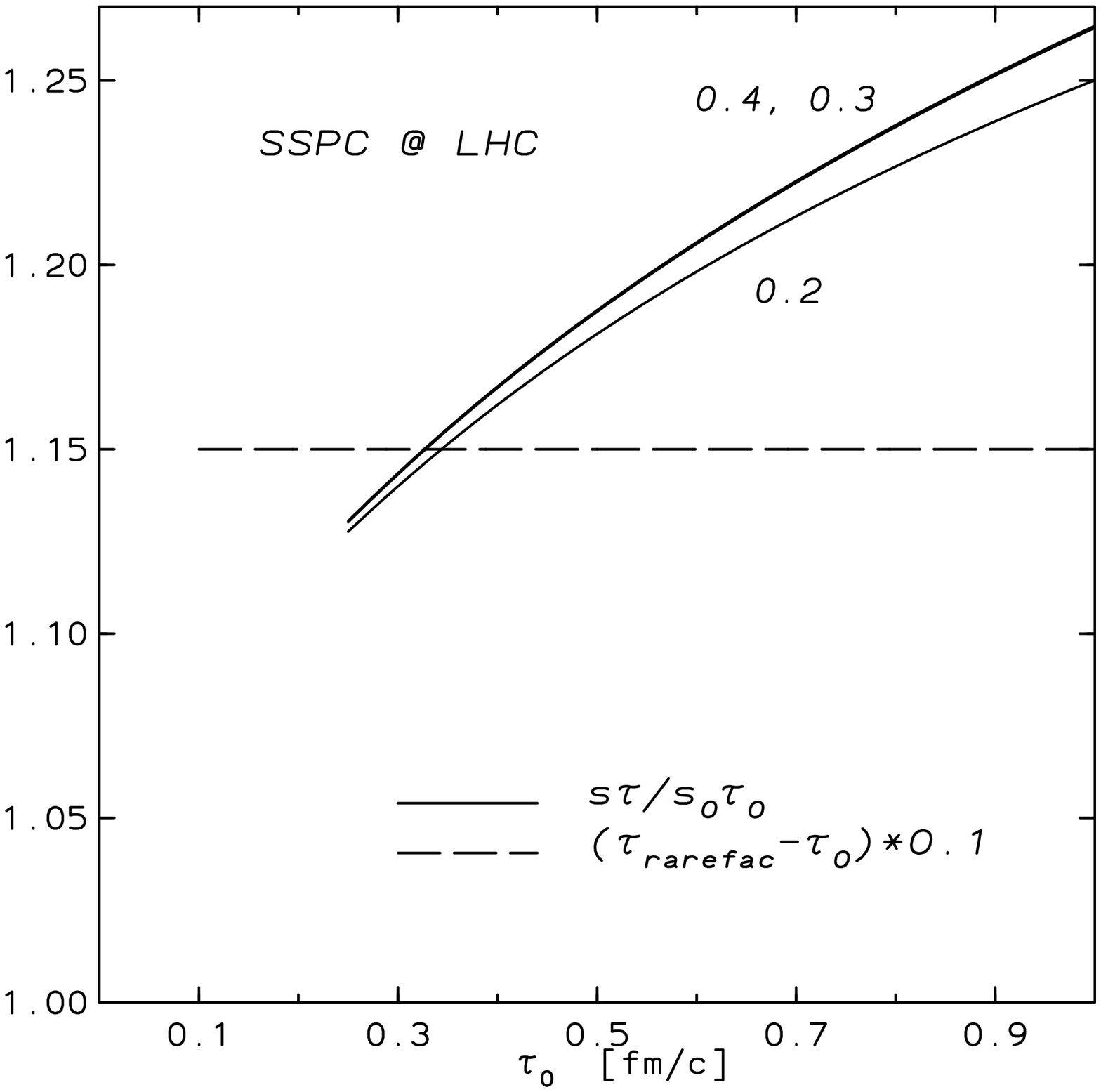}
\includegraphics{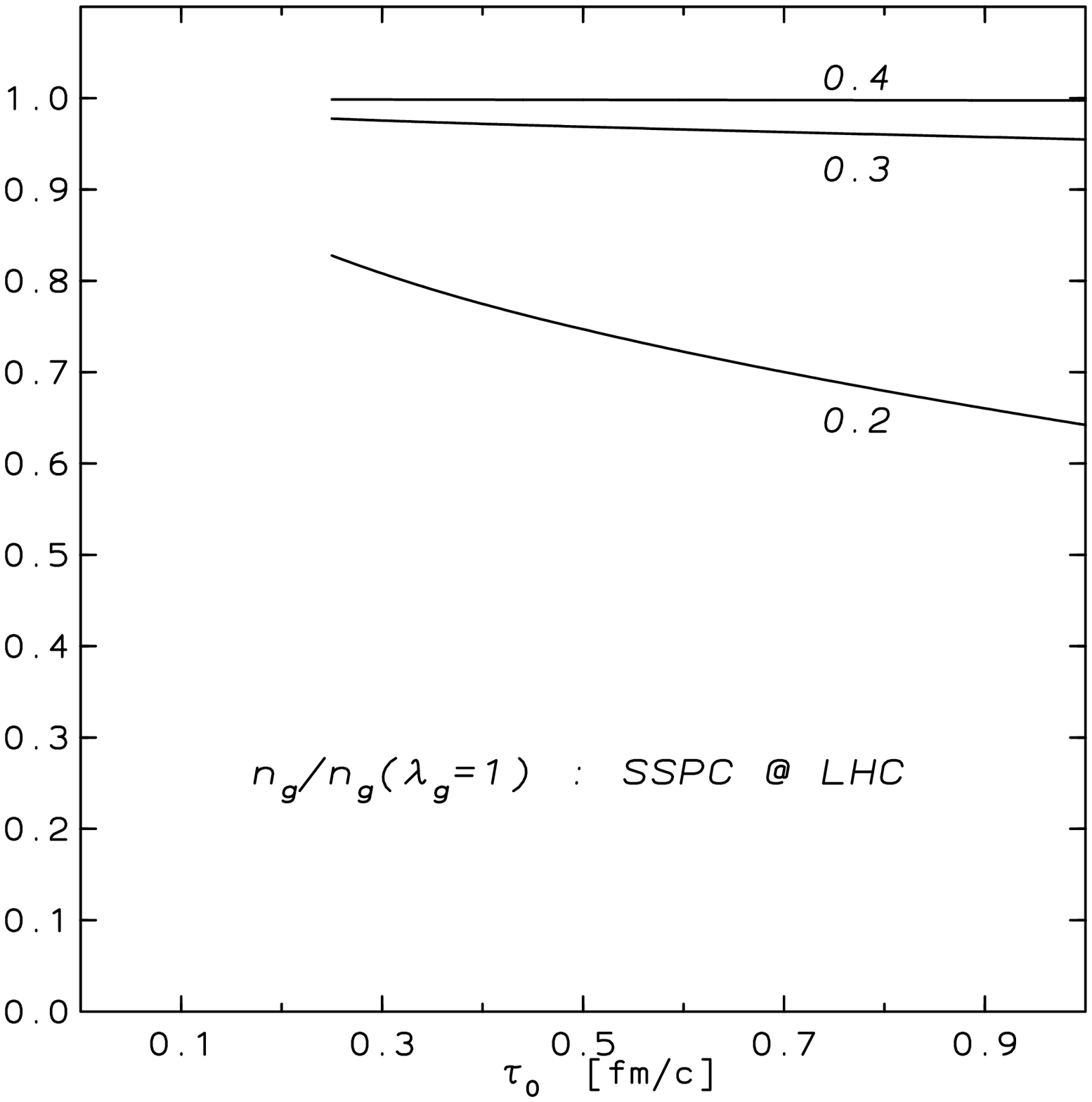}
\includegraphics{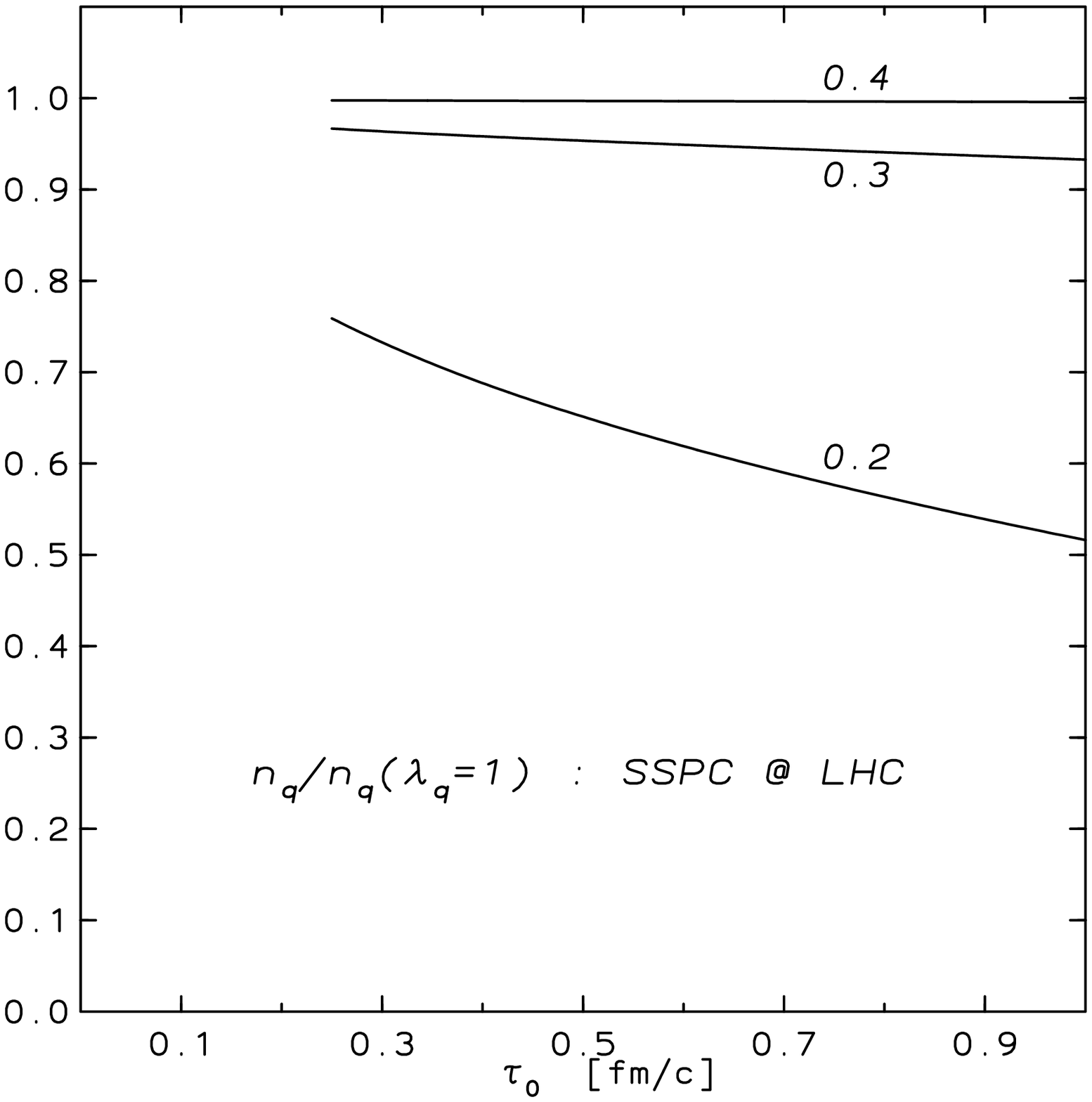}
\vspace*{-2cm}
\caption{As in Fig.\ \ref{fig6}, for LHC initial conditions.
The largest value of $\alpha_s$ taken here is $0.4$; the curves
for larger values are indistinguishable from this case.}
\label{fig8}
\end{figure}

\begin{figure}
\vspace*{14.8 cm}
\caption{As in Fig.\ \ref{fig8}, for pQCD initial conditions.}
\includegraphics{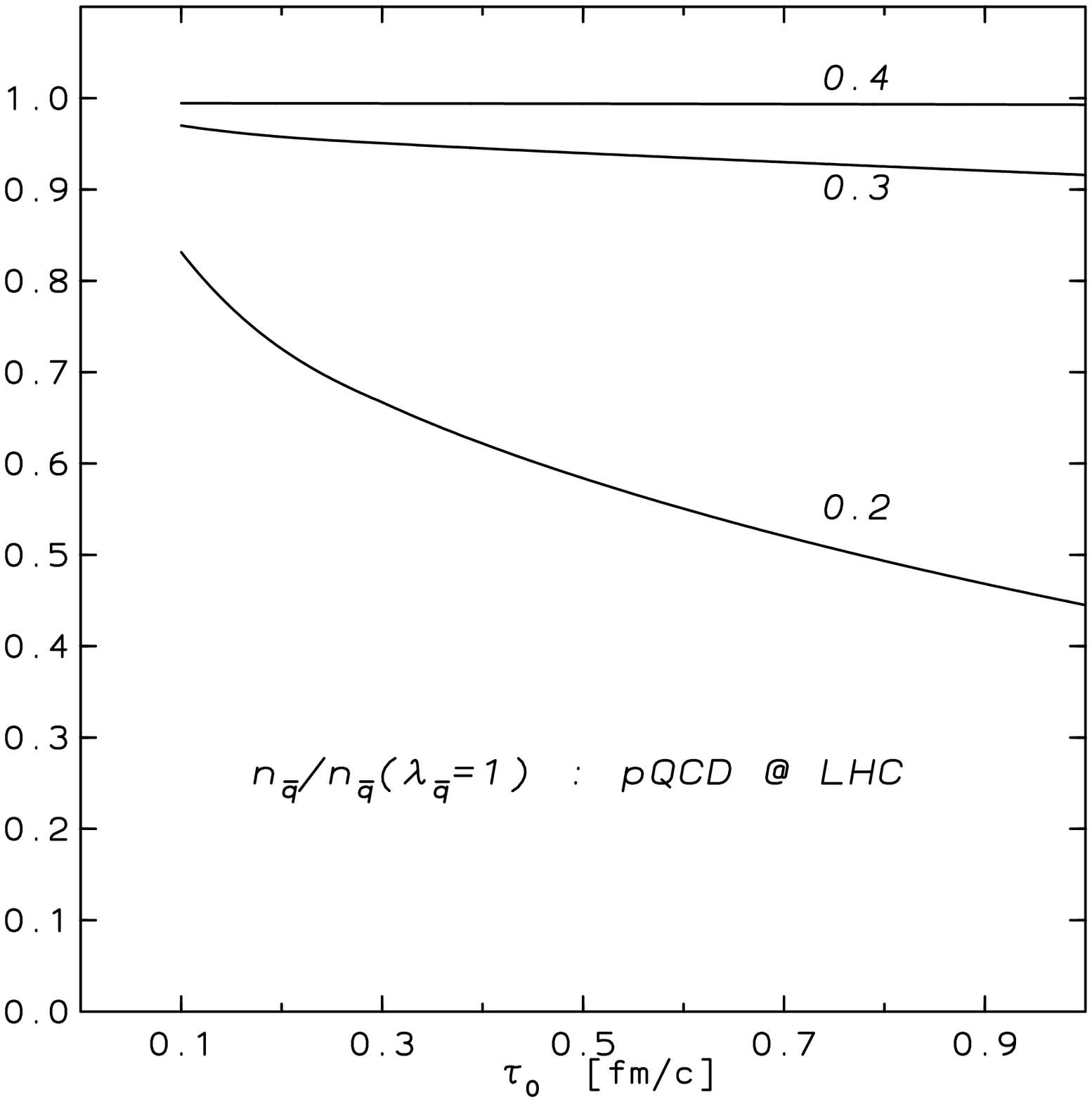}
\includegraphics{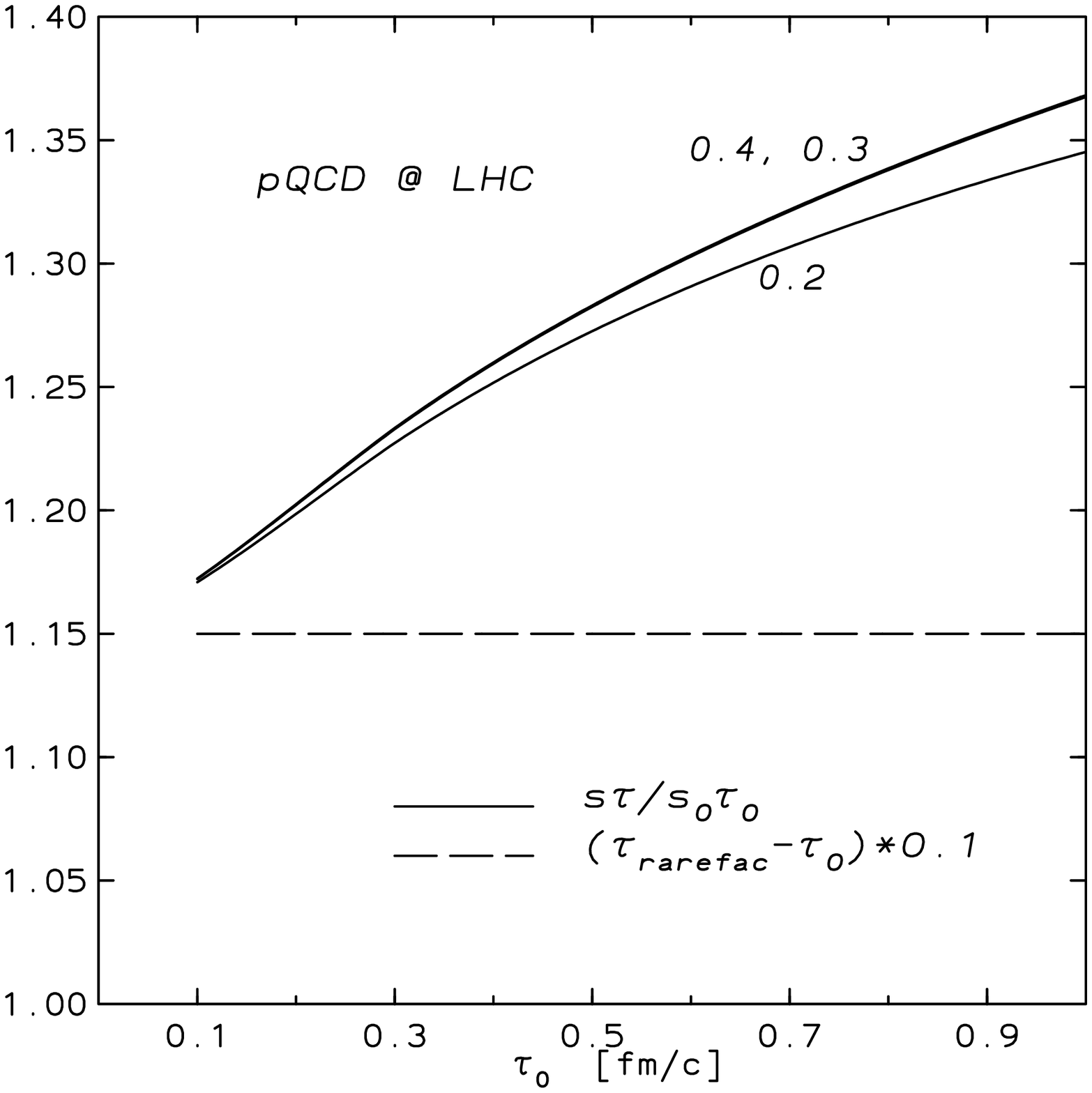}
\includegraphics{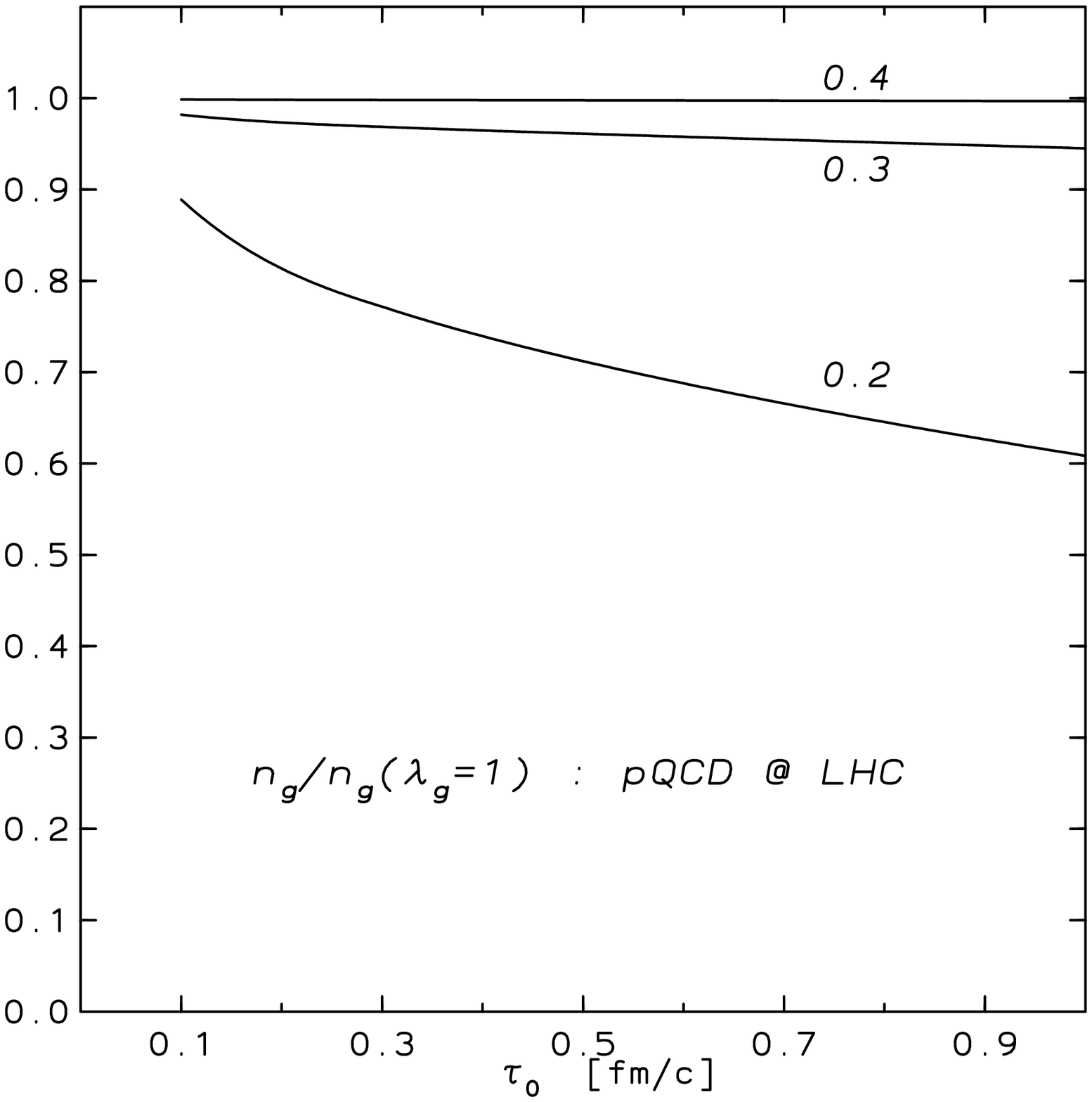}
\includegraphics{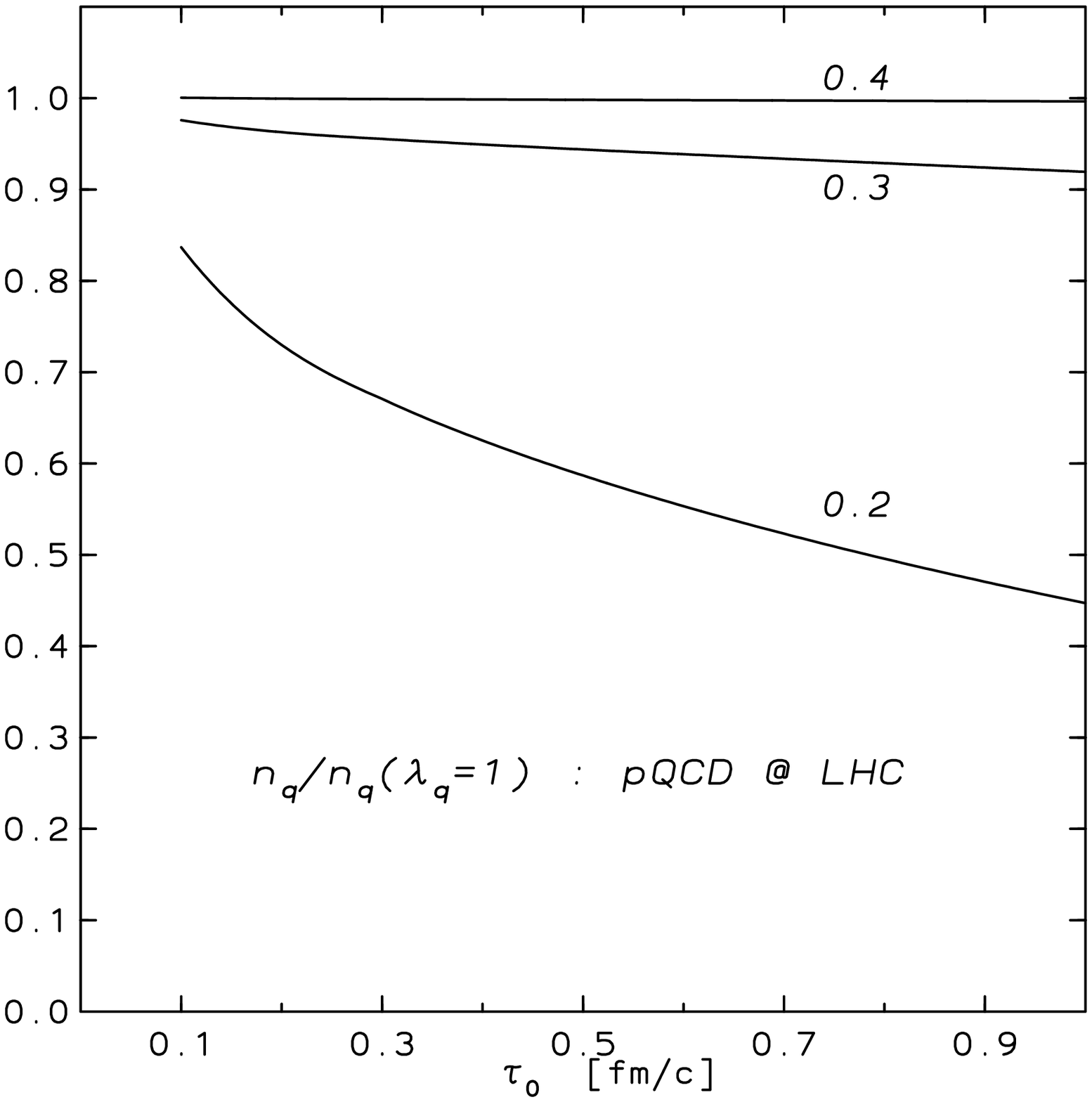}
\label{fig9}
\end{figure}

The LHC case is shown in Figs.\ \ref{fig8} and \ref{fig9}.
Despite different initial conditions, the relative
densities at $\tau_{\rm rarefac}$ are remarkably 
similar in both the SSPC model and pQCD. 
Again, this ratio increases with increasing $\alpha_s$ and decreasing
$\tau_0$. The entropy increases with $\alpha_s$ and $\tau_0$. 
However, in this case 
the time the system spends in the QGP phase is constant,
$\Delta \tau \equiv \tau_{\rm rarefac} - \tau_0 = R/c_s \simeq 
11.5$ fm/c for $Pb$.
The increase in entropy can only be explained by the fact that
the time integral over the right-hand side of 
eq.\ (\ref{noneqS}) is larger, if the system is further away
from equilibrium (which is the case for larger values of $\tau_0$).
As already seen in Fig.\ \ref{fig5}, entropy
production is stronger in the pQCD case. 
Equilibration is nearly complete for large values of $\alpha_s$, 
independent of the value of $\tau_0$.
The only scenario where the QGP will not reach full chemical
equilibration is when $\alpha_s$ is small, and the initial 
time is large.

\section{Longitudinal and transverse expansion}\label{transverse}

In this section, we consider (cylindrically symmetric)
transverse expansion in addition to boost-invariant longitudinal
expansion. In this case, the time evolution is no longer given
by ordinary differential equations, and one has to resort to numerical
algorithms which solve the equations of fluid dynamics, eqs.\
(\ref{2}), and the rate equations (\ref{r1}) -- (\ref{r3}).
The numerical scheme to solve the fluid-dynamical equations
used in this analysis is the relativistic 
Harten--Lax--Van Leer--Einfeldt (RHLLE) algorithm \cite{RHLLE3}, 
with geometrical corrections performed using Sod's method
\cite{RHLLE4}. Since the rate equations are 
conservation equations with source terms, they can also
be solved with the RHLLE algorithm, with the right-hand sides 
being treated with Sod's method. Because of boost-invariance, 
it is sufficient to solve the equations in the central plane, $z=0$,
where $\tau \equiv t$.

Our results are exclusively for the SSPC model and for the
initial values given in Table \ref{table1}; 
we did not vary $\tau_0= 0.25$ fm/c or $\alpha_s=0.3$.
We first show results for a box profile $\sim \Theta(R-r)$, and then
for a wounded-nucleon profile $\sim 3/2 \sqrt{1-r^2/R^2}$.
To facilitate comparison with \cite{MMS}, for the box profile
we use $R=7$ fm both at RHIC and LHC energies. For the wounded-nucleon
profile we use $R=1.12\, A^{1/3}$ to compute the actual values for 
the radii of $Au$ nuclei (at RHIC) and $Pb$ nuclei (at LHC).

\subsection{Box profile}

\begin{figure}
\vspace*{15.5cm}
\includegraphics{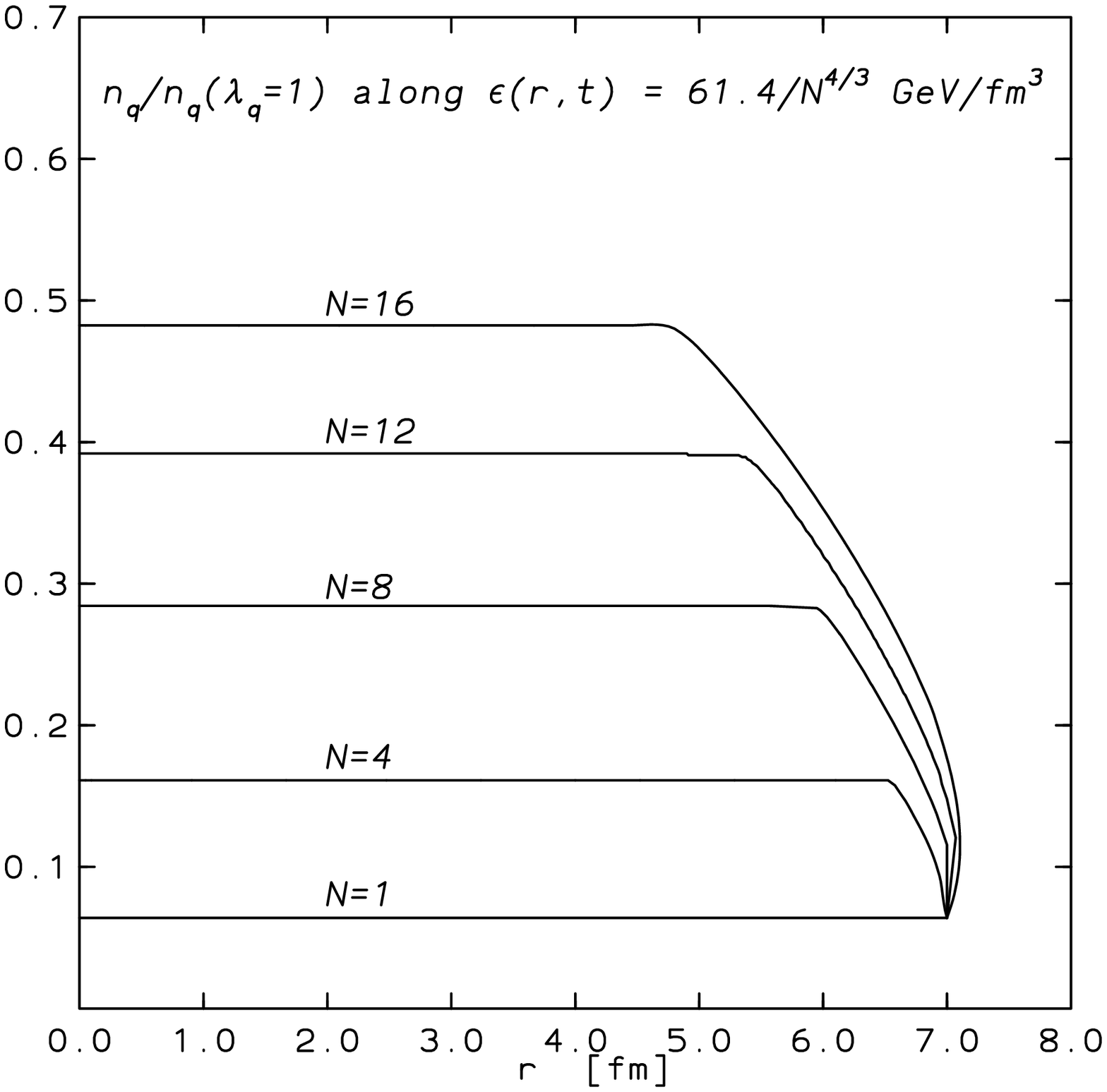}
\includegraphics{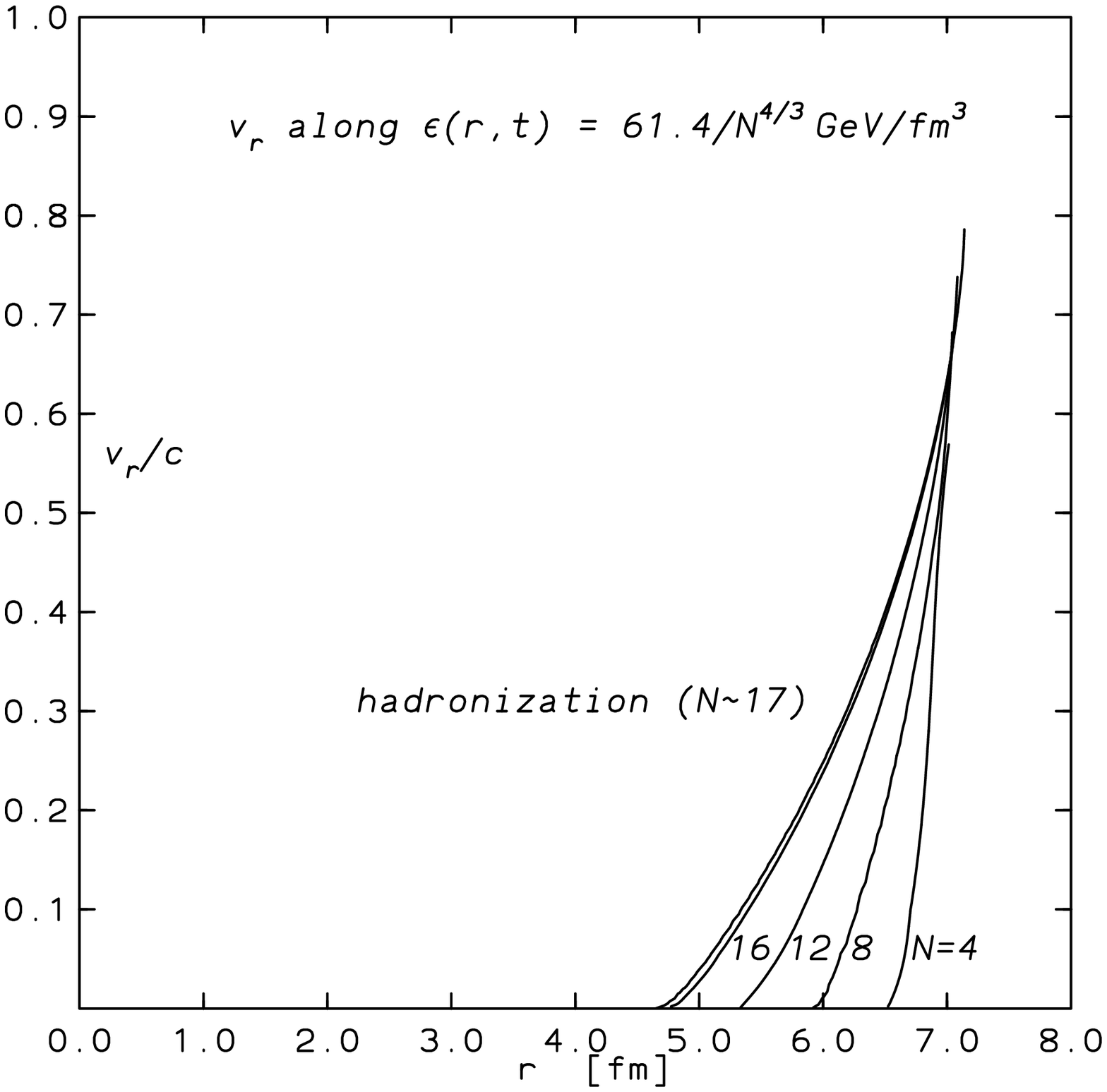}
\includegraphics{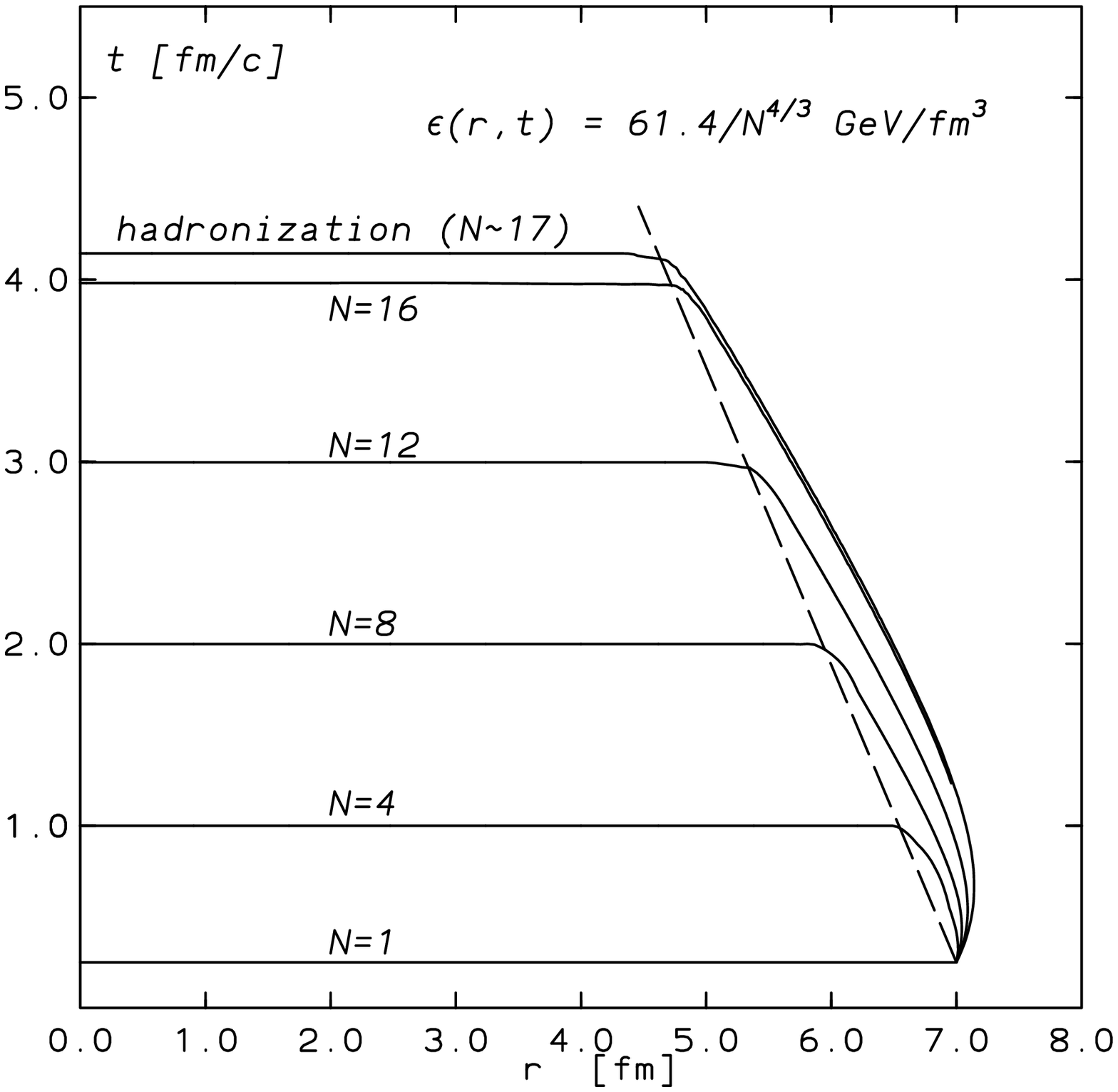}
\includegraphics{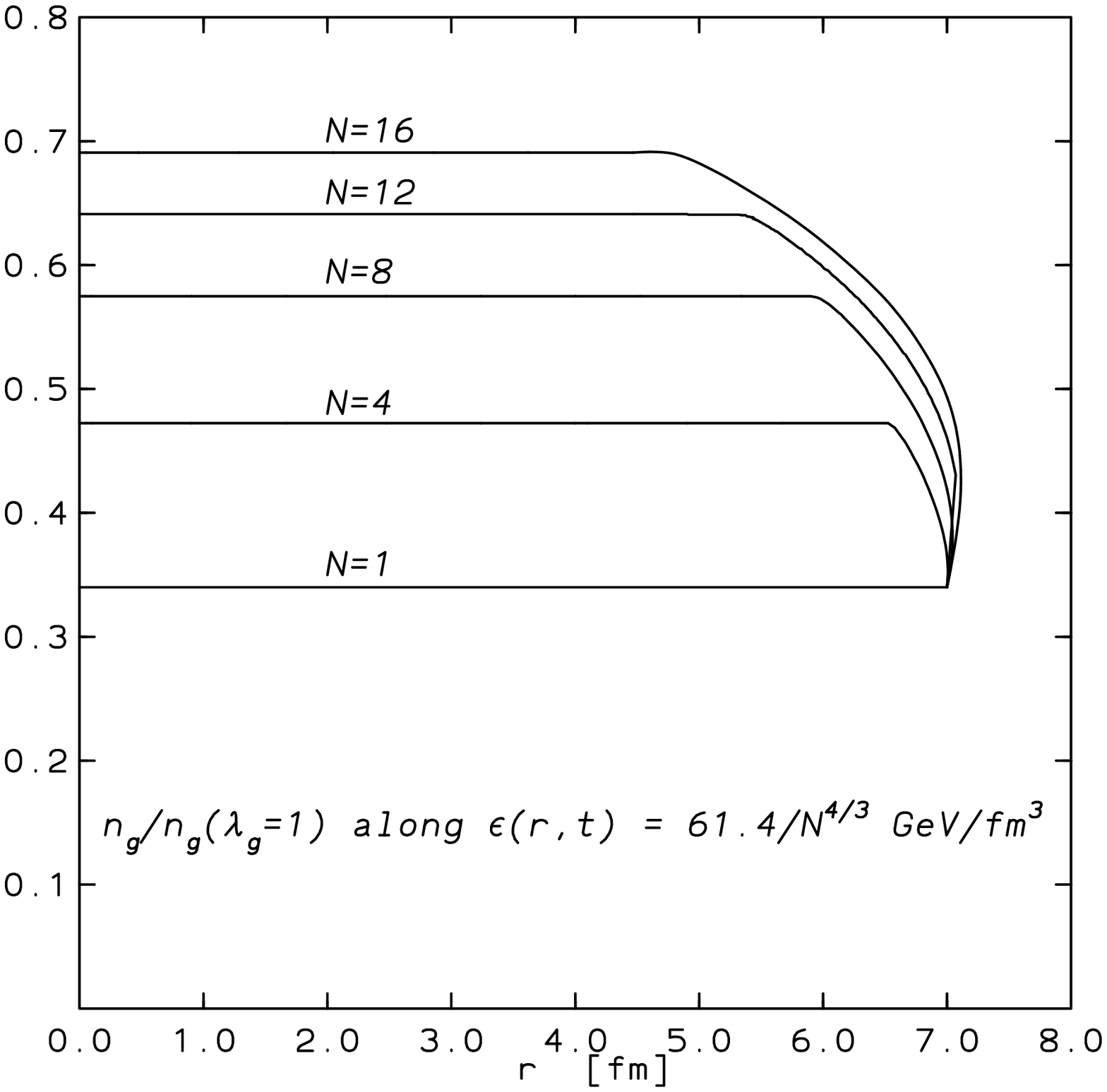}
\vspace*{-2cm}
\caption{SSPC RHIC scenario: evolution in the $t-r$ plane. 
Upper left panel: energy density 
contours. The dashed line represents the advance of the rarefaction front.
Upper right panel: relative density of gluons. Lower left
panel: relative density of quarks (antiquarks identical). Lower right panel: 
transverse velocity. }
\label{fig10}
\end{figure}

In Fig.\ \ref{fig10} we show the evolution of the system in
the $t-r$ plane for RHIC initial conditions. In order to
highlight the consequences of transverse expansion and facilitate
comparison with \cite{MMS,woundedmms}, we plot 
constant energy contours $\epsilon (r,t) = {\epsilon}_0/N^{4/3}$. If 
there were no transverse expansion of the system, the energy 
contours for $N\geq 2$ would be parallel to the $N=1$ contour. We see 
that a rarefaction wave travels into the system from the surface, but 
does not reach the center before the interior cools below the
hadronization energy density $\epsilon_h$ at $\tau_h=4.15$
fm/c. This agrees with the Runge-Kutta analysis of the purely
longitudinal expansion presented above. 
It disagrees with Fig.\ 1 of \cite{MMS}, where the hadronization time
is of the order of 5 fm/c. Upon hadronization, the relative gluon 
density, $n_g/\tilde{n}_g$, has reached roughly $0.7$, while the relative 
quark (and antiquark) density is $n_q/\tilde{n}_q \simeq 0.5$. 
Modest transverse velocities develop near the surface of the system. 
We find them to be smaller than those in Fig.\ 1 of \cite{MMS}. This 
difference is due to the different numerical scheme used in \cite{MMS}
(the SHASTA algorithm of \cite{Kataja}; we use the RHLLE algorithm
\cite{RHLLE3}). 
In contrast to \cite{MMS}, Fig.\ 1, we do not see that these velocities drive 
the fluid locally away from chemical equilibrium; in our case, the
respective fluid elements continue to equilibrate. 

Let us give a simple argument as to why this must be the case.
Consider a fluid element on a given energy density contour in 
the $t-r$ plane (upper left panel of Fig.\ \ref{fig10}).
Then, as the system expands, $\partial \cdot u >0$, 
the energy density drops as a function of proper time,
$u \cdot \partial \epsilon = - (\epsilon + p) \partial \cdot u < 0$.
Since any other contour lying outside the original contour 
(relative to the origin) corresponds to a smaller energy density, 
it also corresponds to later proper times. Therefore, in going from the 
first contour to the second, the proper time in the rest frame of a 
fluid element increases. Since the actual parton number densities
are smaller than their equilibrium values, and consequently the
right-hand sides of the rate equations positive, $R_i >0$,
we conclude from eq.\ (\ref{comoving}) that
$n_i/n_i^{\rm eq}$ must increase in going from the first contour
to the second. In order to compare
with $n_i/\tilde{n}_i$, note that $\tilde{n}_i$ is computed 
at the same temperature $T$ as $n_i$, with the fugacity $\lambda_i$ set
to 1. On the other hand, while $\lambda_i^{\rm eq}$ is also equal to 1 
for the SSPC case considered here (on account of vanishing net-baryon density),
the temperature $T^{\rm eq}$ in $n_i^{\rm eq}$ is in general {\em different\/}
from $T$. Let us assume that the initial temperature for
an evolution in complete local equilibrium is the same as that for
the evolution including chemical equilibration. This has the
consequence that initially $\tilde{n}_i = n_i^{\rm eq}$, and thus
$n_i/\tilde{n}_i = n_i/n_i^{\rm eq}$ at $\tau_0$. As noted earlier in
section 4.1, in complete local equilibrium the temperature falls
less rapidly than in chemical non-equilibrium, $T^{\rm eq} > T$ at all 
times $\tau > \tau_0$. Therefore,
$n_i/\tilde{n}_i = (n_i/n_i^{\rm eq})(T^{\rm eq}/T)^3 \geq n_i/n_i^{\rm eq}$,
with the equality holding at $\tau_0$.
Thus, not only does $n_i/\tilde{n}_i$ not decrease, it increases even 
faster than $n_i/n_i^{\rm eq}$ during the evolution. 
Consequently, in the $n_i/\tilde{n}_i-r$ diagram
(upper right and lower left panels of Fig.\ \ref{fig10}), the
second contour must lie outside the first contour (with respect to
the origin). This is exactly what we find numerically.

\begin{figure}
\vspace*{15.5cm}
\includegraphics{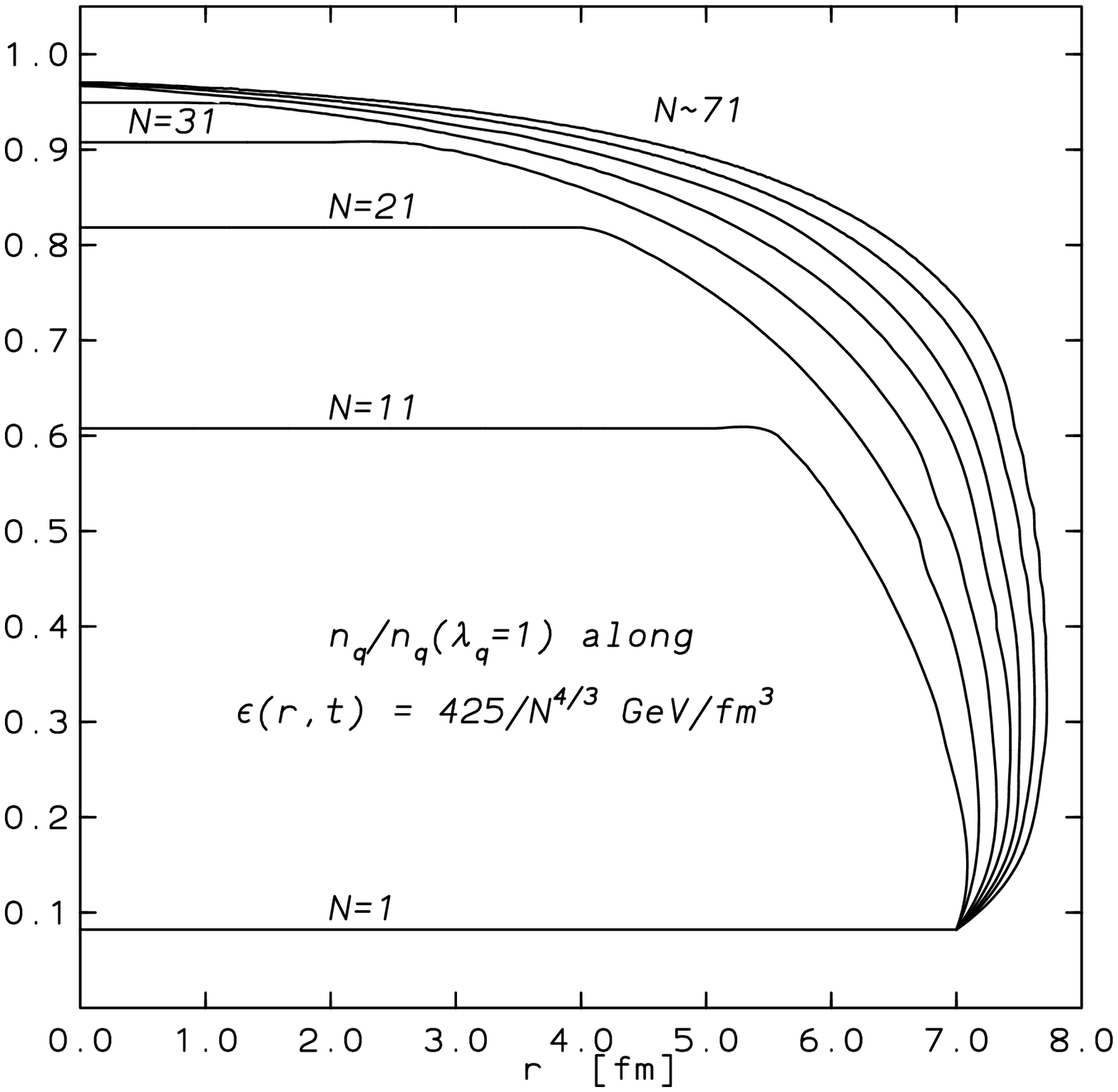}
\includegraphics{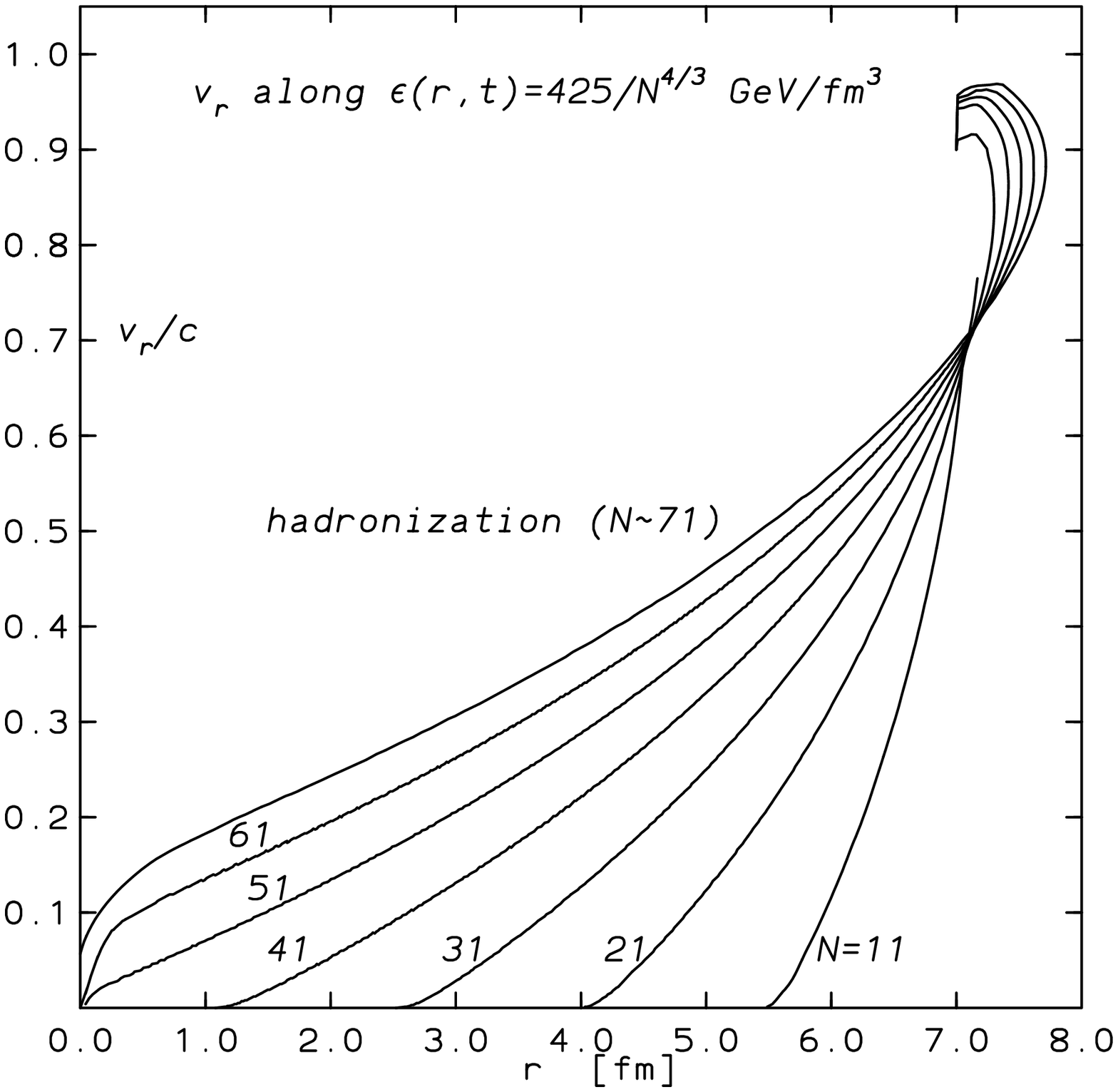}
\includegraphics{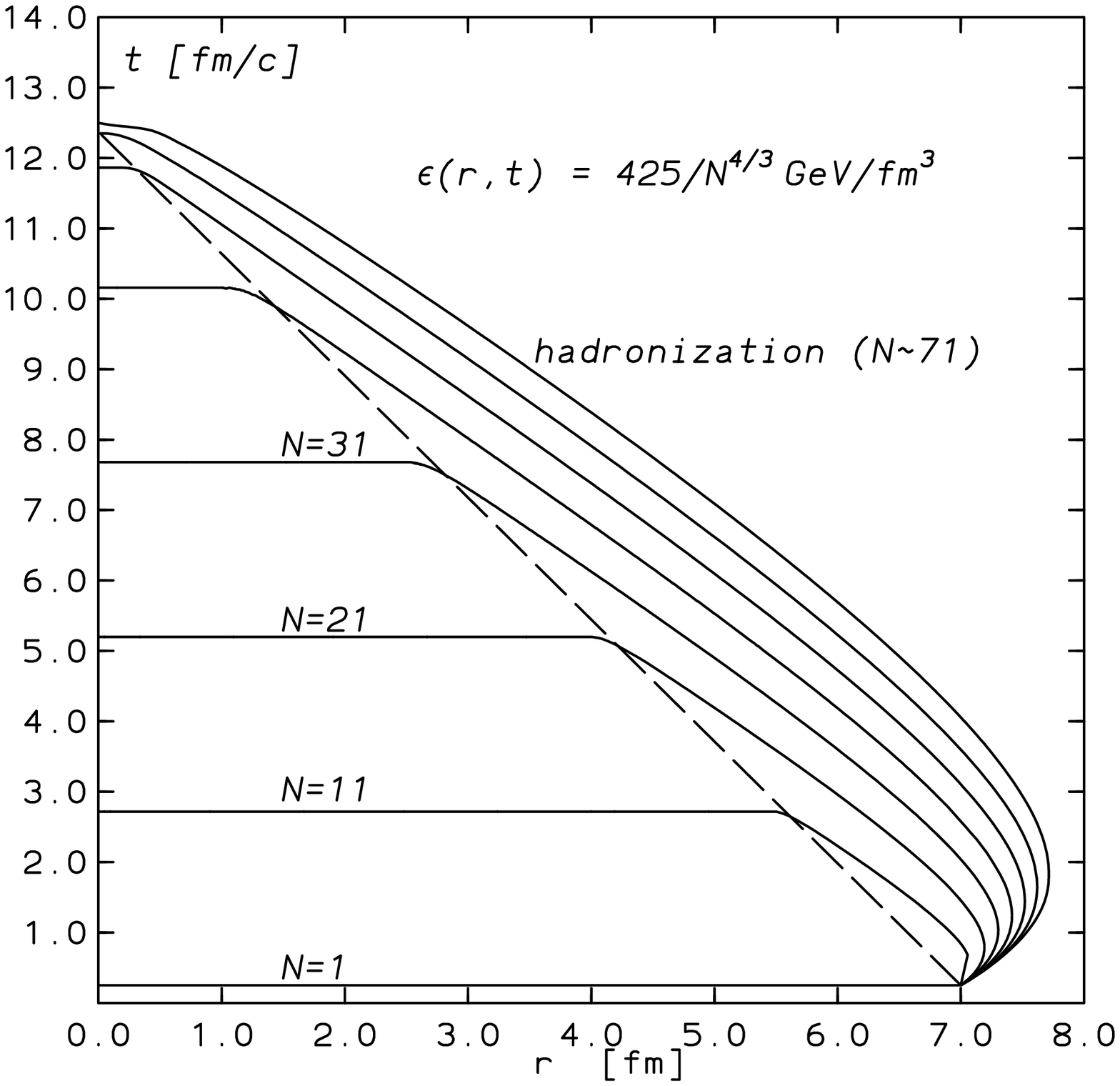}
\includegraphics{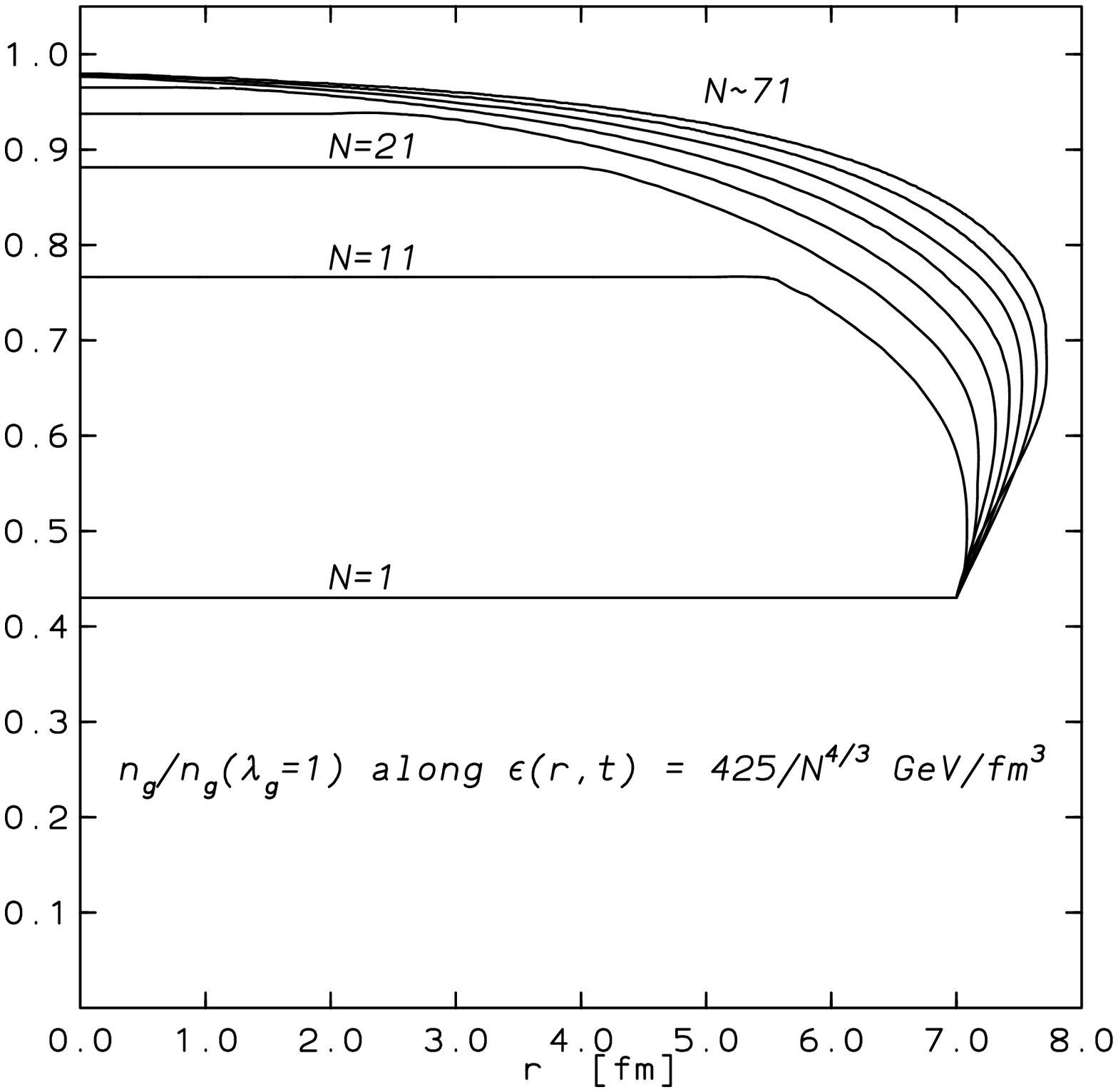}
\vspace*{-2cm}
\caption{As in Fig. \ref{fig10}, for the SSPC LHC scenario.}
\label{fig11}
\end{figure}

Fig.\ \ref{fig11} shows the SSPC LHC scenario. In this case,
a rarefaction wave has enough time to penetrate into the center
of the system, albeit only shortly before the energy density
falls to $\epsilon_h=1.45$ GeV/fm$^{3}$.
Until the rarefaction wave reaches the center 
at proper time $\tau = \tau_0 + R/c_s$, 
the evolution of the system at $r=0$ is given by a purely
longitudinal expansion, as discussed in the previous section.
Once again, the results obtained with the RHLLE algorithm agree well with the
analysis using the Runge--Kutta method. As seen above,
equilibration of the parton species is nearly complete.

The hadronization time agrees well with that found in \cite{MMS},
Fig.\ 3. However, in contrast to that analysis we
again find that the transverse velocities
developing in the rarefaction front do not impede chemical equilibration.
In particular, we do not find the tendency to drive the
system out of equilibrium at larger transverse distances $r$, and
to overshoot the equilibrium values close to the origin, as in 
Fig.\ 3 of \cite{MMS}. As explained above, in the $n_i/\tilde{n}_i -r$
diagram, contours corresponding to smaller energy density (larger proper
times) must lie outside those corresponding to larger energy density
(smaller proper times) relative to the origin. 

\subsection{Wounded-nucleon profile}

Throughout the analysis presented above 
a box-profile function has been assumed.  
While this is instructive and facilitates
direct comparison between the results from a purely longitudinal
expansion and those including transverse expansion (as long as the
transverse rarefaction wave has not yet reached the center of the
system), a box profile is clearly an idealization.

In this section we repeat the analysis of the SSPC RHIC and 
LHC scenarios with a more realistic 
nuclear profile function, the so-called wounded-nucleon profile
\begin{equation} \label{ta}
T_A(r)  = \frac{3}{2}\,  \sqrt{1-\frac{r^2}{R^2}}\,\, ,
\end{equation}
where $r$ is the transverse coordinate. We generate an $r$--dependence
of our initial energy and parton density profiles by multiplying
the initial values of Table \ref{table1} with eq.\ (\ref{ta}).
Clearly, this implies higher
initial densities at the center of the plasma. However, the 
fluid-dynamical expansion begins with a non-zero density gradient already at 
$\tau_0$, which, without the factor $3/2$ in (\ref{ta}), 
would lead to faster cooling than a constant  
nuclear profile function.

\begin{figure}
\vspace*{15.5cm}
\includegraphics{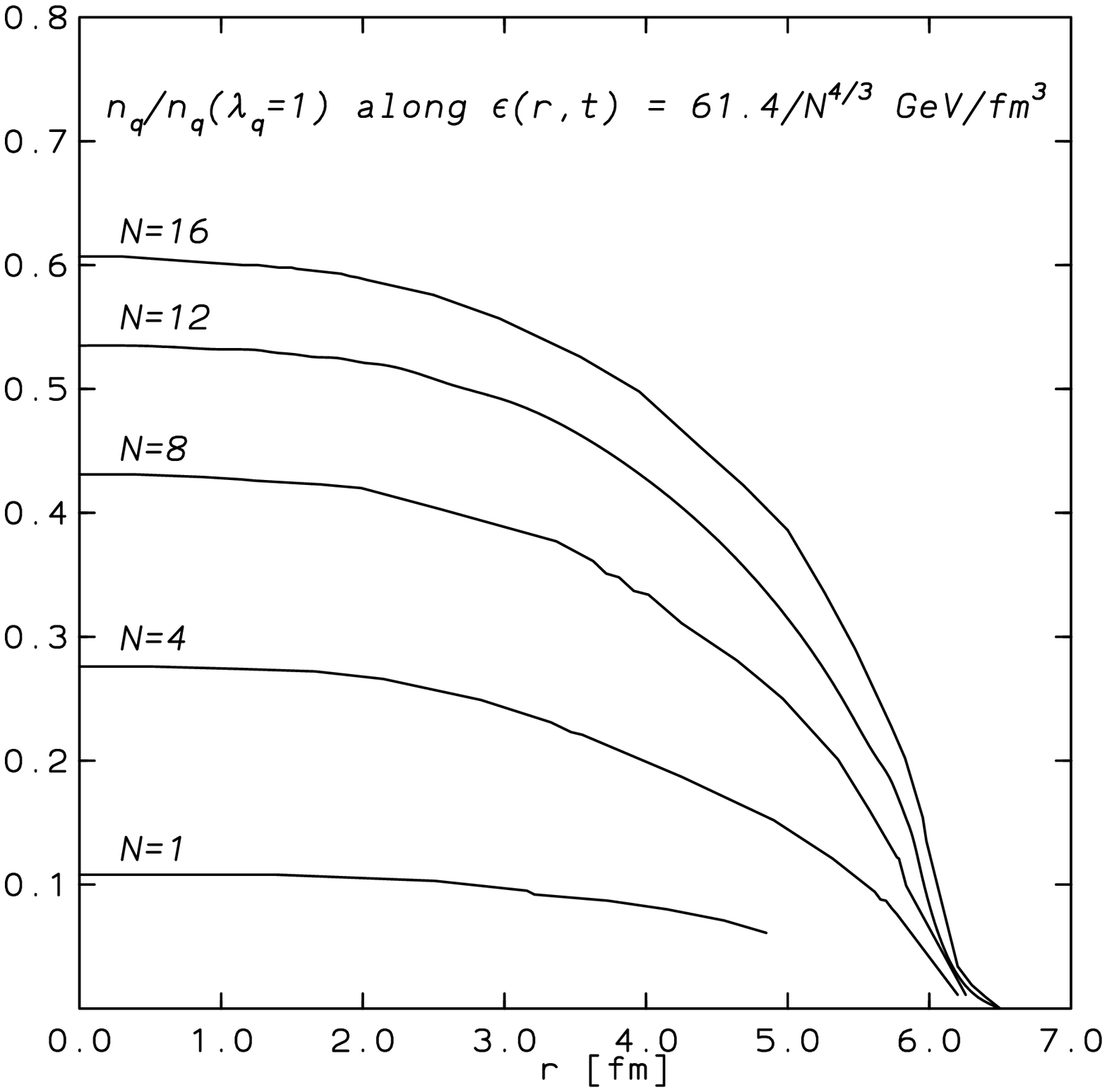}
\includegraphics{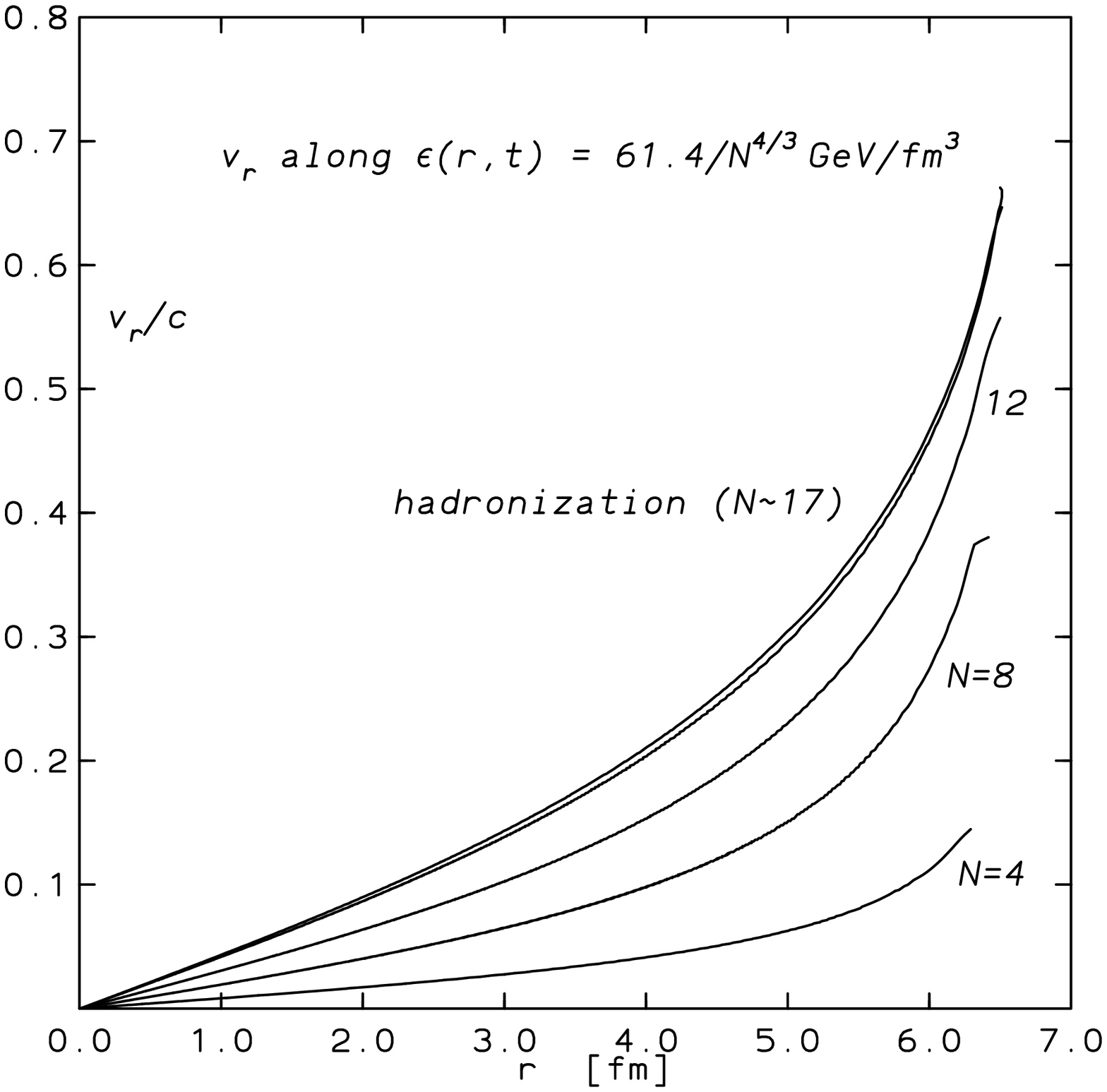}
\includegraphics{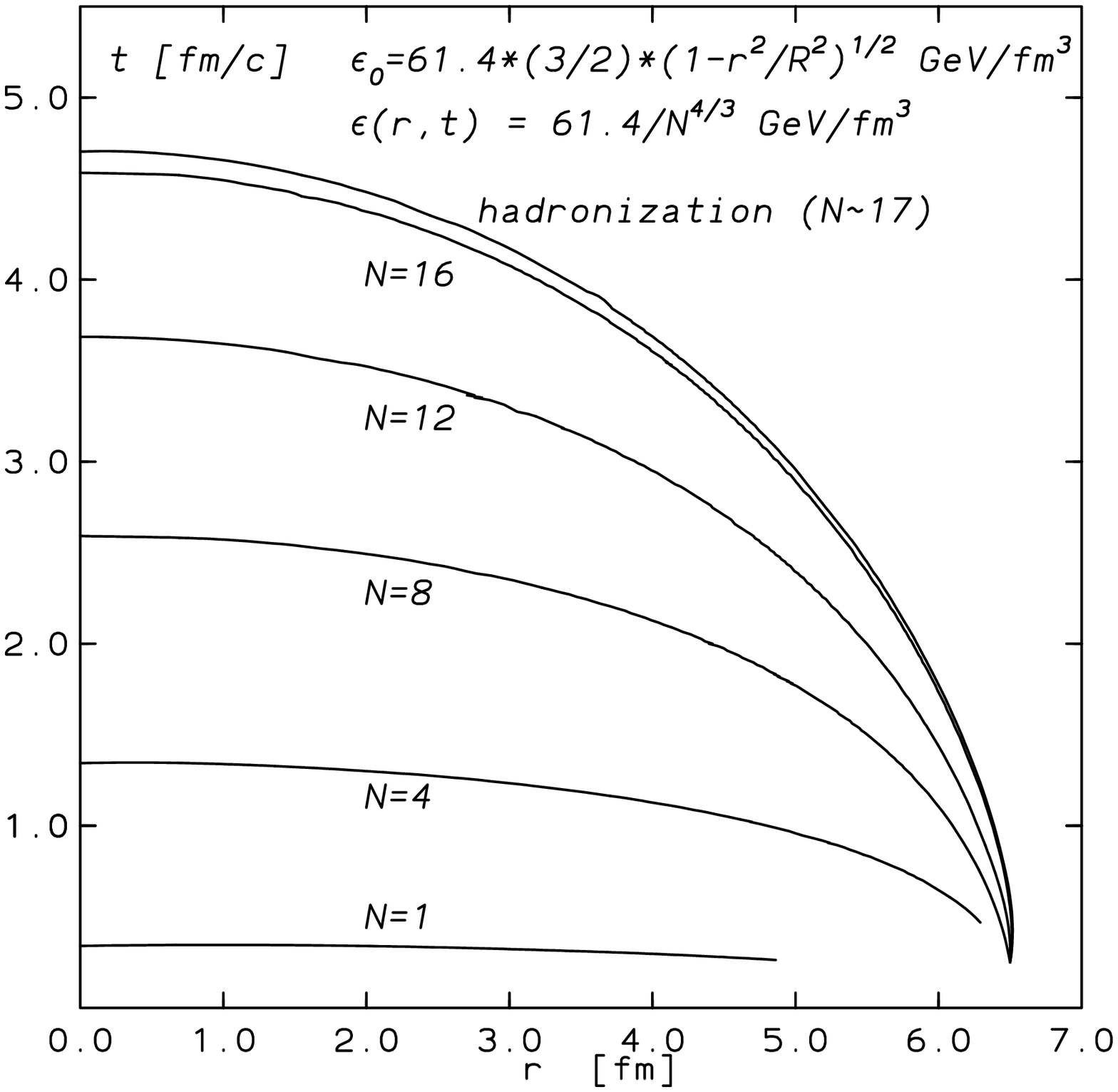}
\includegraphics{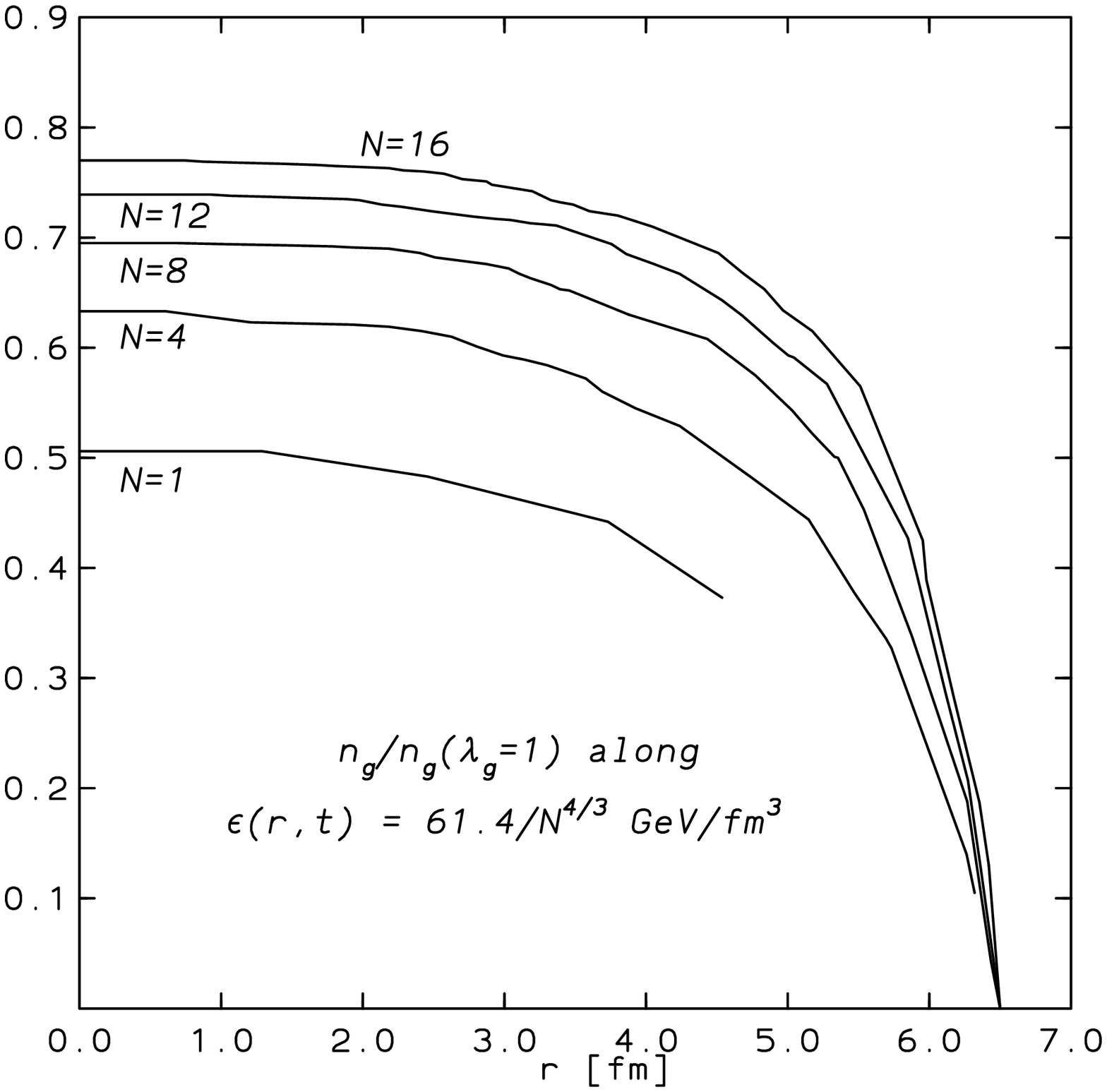}
\vspace*{-2cm}
\caption{As in Fig.\ \ref{fig10}, with an initial wounded-nucleon profile.}
\label{fig12}
\end{figure}

In Fig.\ \ref{fig12} we present the analogue of Fig.\ \ref{fig10}, now
computed with an initial wounded-nucleon density profile.
We observe that the lifetime of the plasma phase is 
actually larger than in Fig.\ \ref{fig10}, $\tau_h=4.70$ fm/c,
instead of $\tau_h=4.15$ fm/c. Due to the initial density gradients,
transverse velocities develop now over the entire volume of the
system.
The cooling effects of the flow become apparent when considering
purely longitudinal expansion with an
initial energy density of $3/2 \times 61.4$ GeV/fm$^{3}$. 
In this case, the QGP lives longer, $\tau_h=5.63$ fm/c. 

The increase of the initial energy and parton densities by a factor
of $3/2$ in the center of the system has the effect that
the partons at small and intermediate $r$ equilibrate a little
further in the wounded-nucleon scenario than in the box-profile
scenario. The relative gluon density reaches values around 
0.8 (0.7 for the box profile)
and the relative quark density values around 0.6 (compared with 0.5 in
the box profile case).  In general, however,
the box-profile and the wounded-nucleon-profile relative parton
density calculations agree to a remarkable extent.

\begin{figure}
\vspace*{15.5cm}
\includegraphics{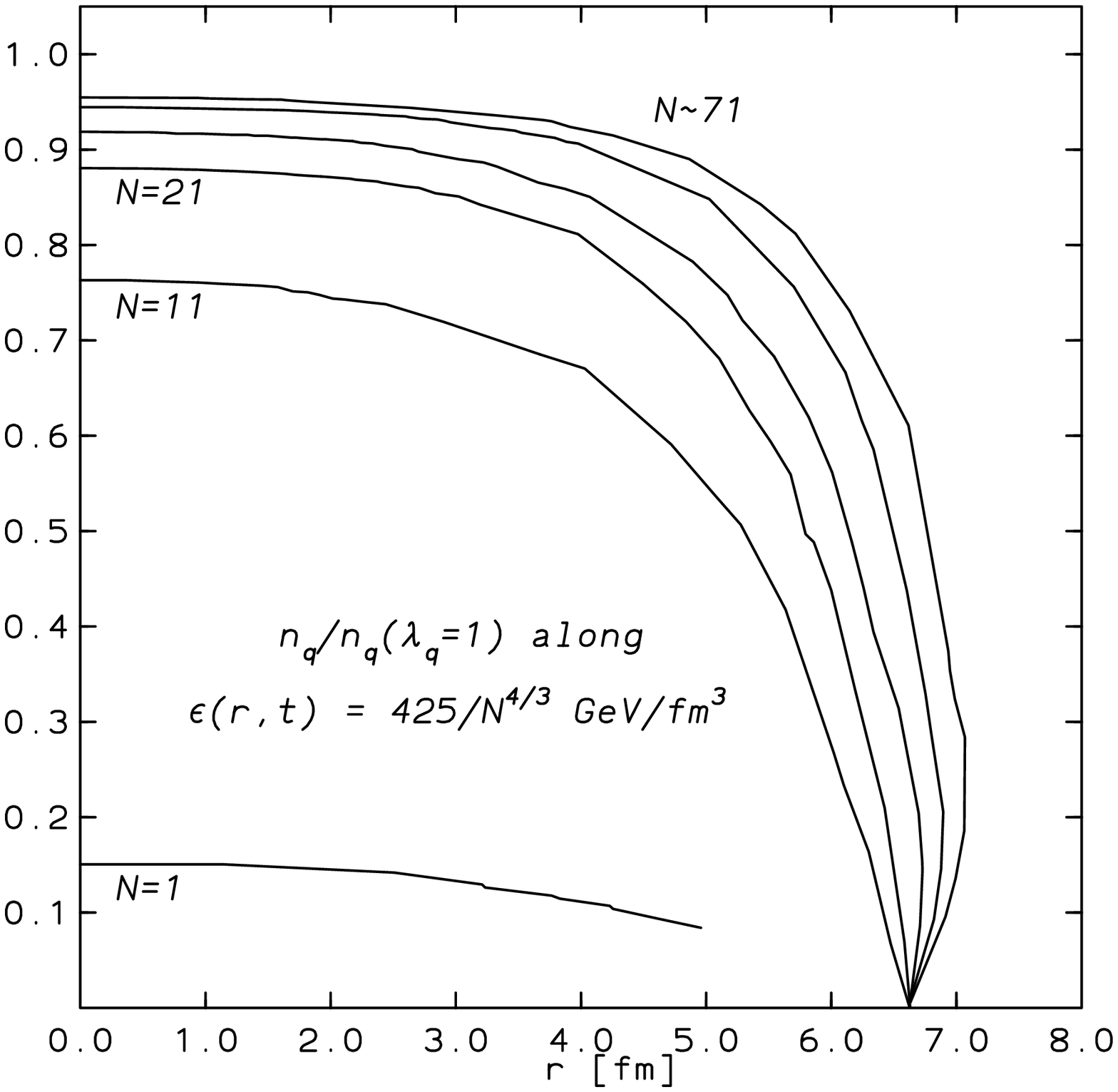}
\includegraphics{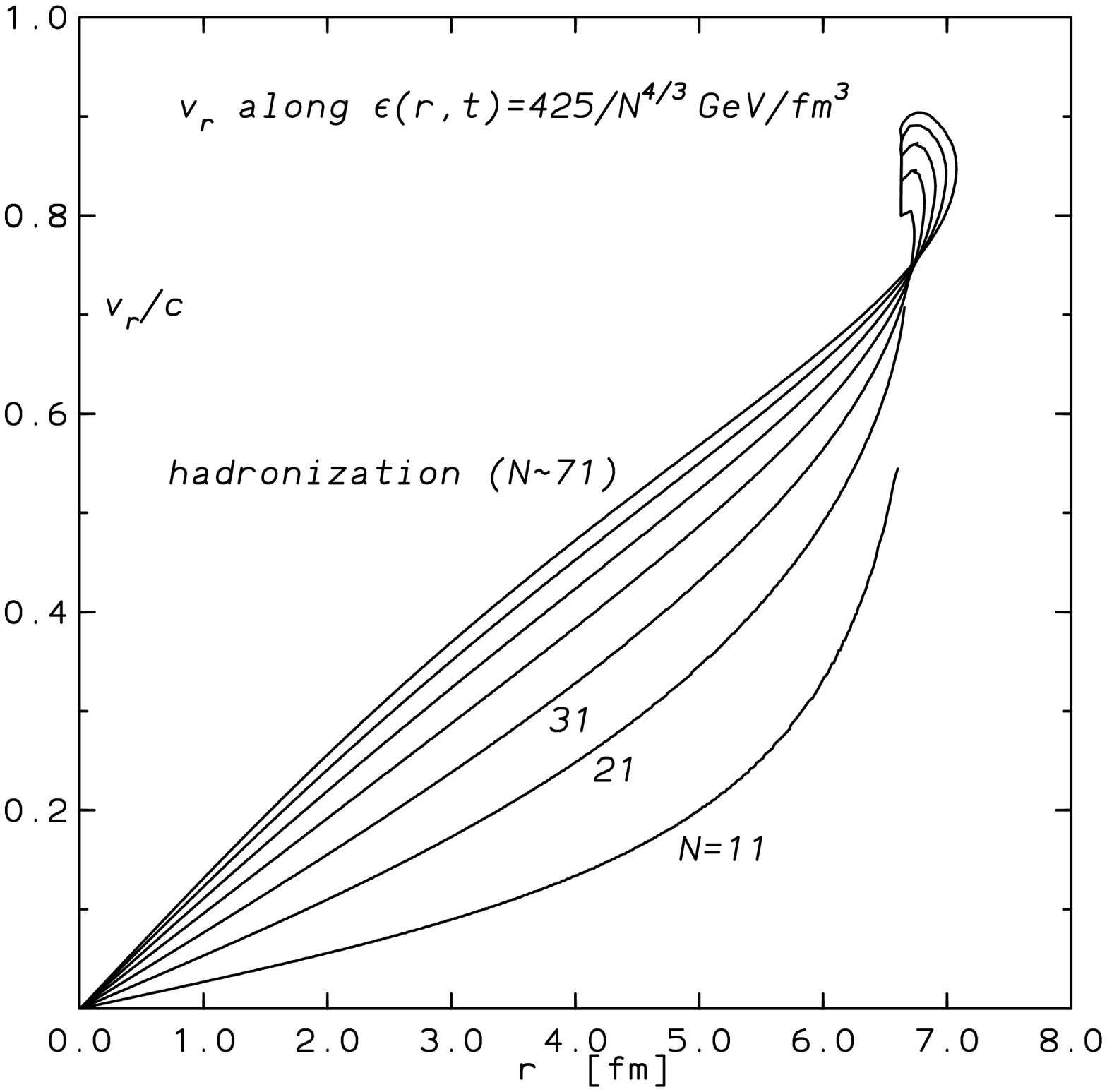}
\includegraphics{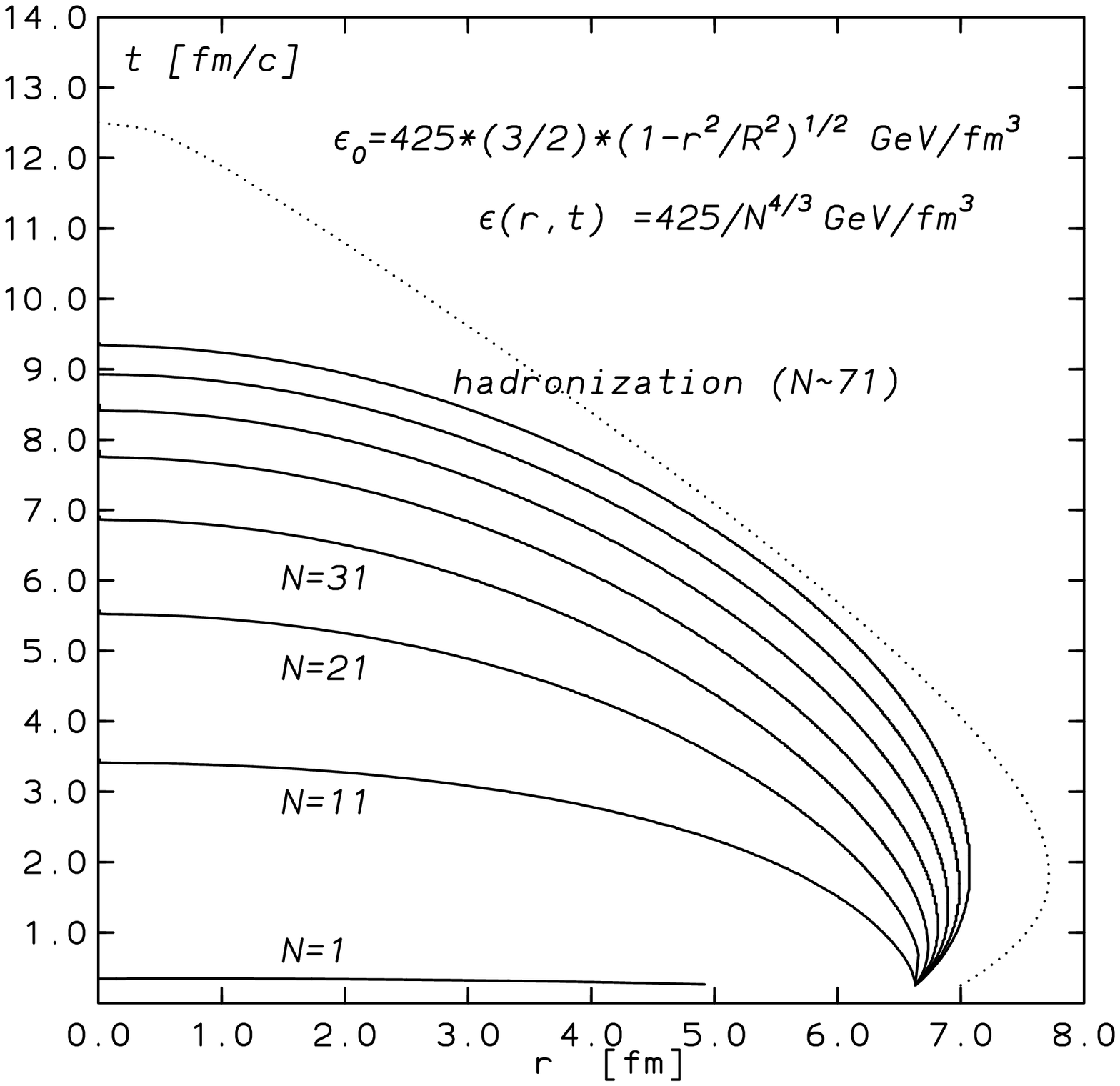}
\includegraphics{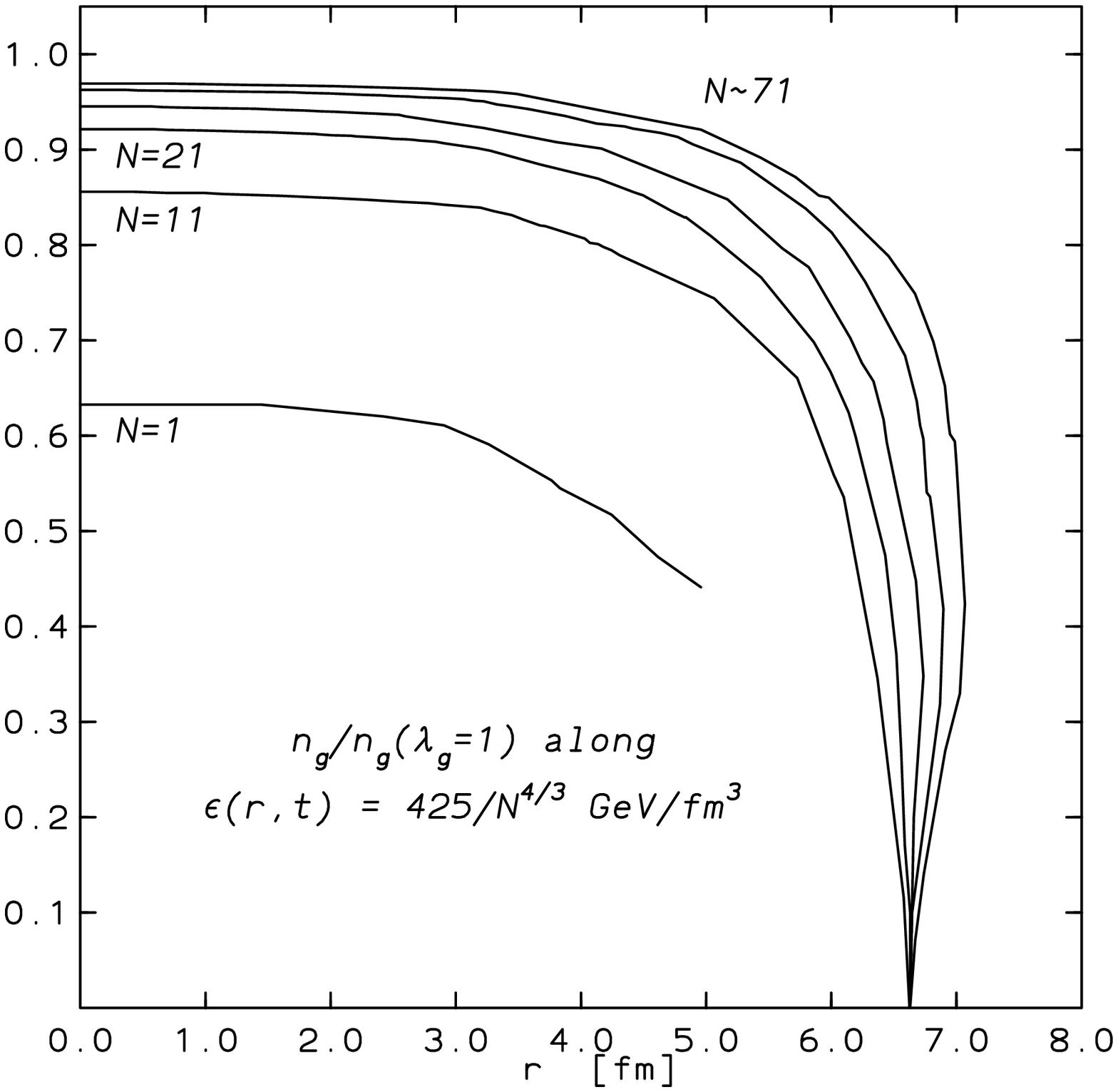}
\vspace*{-2cm}
\caption{As in Fig.\ \ref{fig11}, with an initial wounded-nucleon profile.
The dotted line in the upper left panel depicts the hadronization hypersurface
for an initially box-profiled plasma (of $R=7$ fm), shown 
in Fig.\ \ref{fig11}.} 
\label{fig13}
\end{figure}

In Fig.\ \ref{fig13} we show the LHC case, where the effects of the
wounded-nucleon profile become more apparent. The plasma cools 
markedly more rapidly than in the box-profile case,
hadronizing at a proper time $\tau_h=9.37$ fm/c. 
This is more than $3$ fm/c earlier than for
an initial box profile, in which case $\tau_h$ is essentially
moments after $\tau_{\rm rarefac}$, the arrival of the rarefaction 
wave at $r=0$.

While the plasma's lifetime is significantly reduced, parton chemistry, 
driven by high temperatures and high gluon densities, is
sufficiently rapid to equilibrate the plasma prior to hadronization. 
Qualitatively, the transverse 
velocities along contours of constant energy density in Figs.\ \ref{fig12}
and \ref{fig13} agree
with the analysis given in \cite{woundedmms}. No  
parton evolution plots are shown in \cite{woundedmms}.

\section{Conclusions}

We have investigated chemical equilibration of quarks and
gluons at RHIC and LHC energies. Assuming that a kinetically 
equilibrated quark-gluon plasma is created at these energies, we 
have used ideal fluid dynamics coupled to rate equations for the parton 
densities \cite{Biro,MMS,woundedmms} 
to study chemical equilibration in the further evolution of the system.
Our work is motivated by the
upcoming experiments at RHIC in the fall of this year, and by recent
pQCD estimates on parton production in the initial stage of ultrarelativistic
nuclear collisions \cite{eskola2}. These estimates serve as initial
conditions to the fluid-dynamical as well as the rate equations.
Besides pQCD initial conditions, we also considered 
those given by the so-called SSPC model \cite{MMS,woundedmms}, in order
to compare our results to previous studies.
At RHIC, the SSPC model constitutes an upper limit for
the initial energy and parton densities,
while pQCD sets a lower limit, as the effect of
soft background fields is neglected in this approach.
It was recently argued \cite{keijorecent} that
the transverse momentum cutoff $p_0$ should be chosen to match
the saturation scale, $p_{\rm sat} \sim 1$ GeV for RHIC,
rather than $\sim 2$ GeV chosen here. Then, the initial
conditions obtained from pQCD are rather similar to those in the
SSPC approach.
The produced transverse energy density can also be computed within
classical Yang--Mills theory \cite{rajualex}; the values obtained in this way
support the SSPC estimates.

Our results can be summarized as follows. Chemical equilibration of
partons is never complete at RHIC energies. The highest degree of
equilibration can be reached with SSPC initial conditions and an
(unphysically) large value of the strong coupling constant, 
$\alpha_s \sim 0.6$. With initial conditions from pQCD, the system 
never comes close to equilibration.

The situation changes completely at LHC energies. Here, the initial
energy and parton densities are already so large that chemical
equilibration is complete in almost all scenarios considered.
Only if $\alpha_s \sim 0.2$ or smaller and the initial
proper time $\tau_0$ is large, equilibration was seen to be incomplete.

Within the macroscopic transport approach employed here, 
entropy production due to chemical non-equilibrium processes was found to
be large, in some cases $\sim 30 \%$. This has implications for estimating 
the {\em initial conditions\/} in ultrarelativistic nuclear collisions.
Commonly, one assumes that the evolution of the system 
is entropy conserving. Then, since $s \simeq const. \times n$,
the initial entropy $\sim s_0 \tau_0$ is
estimated from the total multiplicity of hadrons in the final state
by the formula ${\rm d}N_h/{\rm d}\eta \simeq const. \times 
s_0 \tau_0 \pi R^2$. However, if entropy grows due to chemical reactions, 
this formula would severely overestimate the entropy in the initial 
state.

We furthermore established that the use of a factorized phase-space
distribution as in \cite{Biro,MMS,woundedmms} is permissible as
the error is of the order of a few percent only. We also gave an
analytical proof, as well as numerical evidence, that
transverse flow does not drive the system away from chemical equilibrium, in
contrast to the results of \cite{MMS}.

Future studies should focus on the further evolution of the system
into the mixed and hadron phases. To this end, previously developed
non-equilibrium models could be resurrected \cite{barz}.
Further applications of the methods presented here include the study
of strangeness and charm equilibration at RHIC and LHC energies
\cite{dmedhr2}.
Besides this, the reaction rates should be
improved using the full phase-space distribution function instead
of the factorized one. Other improvements of the expressions for
the reaction rates include the
running of the strong coupling constant. As shown in Ref.\ \cite{Wong},
equilibration is faster when the scale of the running coupling constant 
is allowed to vary with the temperature $T$.
\\[1cm]
{\bf Acknowledgements}\\ ~~ \\
We thank S.\ Bass, J.\ Cleymans,  A.\ Dumitru, K.\ Eskola, K.\ Kajantie,
B.\ M\"uller, M.\ Mustafa, J.\ Rafelski,
D.\ Srivastava, and R.\ Venugopalan for valuable discussions. D.H.R.\
thanks RIKEN, Brookhaven National Laboratory, and the U.S.\ Department of 
Energy for providing the facilities essential for the completion of this work.


\end{document}